\documentclass[10pt,journal,compsoc]{IEEEtran}
\usepackage{mathptmx}
\ifCLASSOPTIONcompsoc
  \usepackage[nocompress]{cite}
\else
  \usepackage{cite}
\fi

\usepackage{amsmath,amssymb,amsfonts}
\usepackage{graphicx}
\usepackage{algorithm}
\usepackage{algpseudocode}

\usepackage{array}

\usepackage{mdframed}
\usepackage[draft]{todonotes}
\usepackage{fancyhdr}
\usepackage[normalem]{ulem}
\usepackage[hyphens]{url}
\usepackage{microtype}
\usepackage{caption}
\usepackage{subcaption}
\usepackage{comment}
\usepackage{xspace}
\usepackage{hyperref}

\usepackage{textcomp}
\usepackage{xcolor, soul}
\usepackage{fancyhdr}
\usepackage{multirow}
\usepackage{tikz}
\newcommand*\circled[1]{\tikz[baseline=(char.base)]{
            \node[shape=circle,draw,inner sep=1pt] (char) {#1};}}
\usepackage{pifont}
\newcommand{\cmark}{\ding{51}}%
\newcommand{\xmark}{\ding{55}}%
\usepackage{array}
\newcolumntype{L}[1]{>{\raggedright\let\newline\\\arraybackslash\hspace{0pt}}m{#1}}
\newcolumntype{C}[1]{>{\centering\let\newline\\\arraybackslash\hspace{0pt}}m{#1}}
\newcolumntype{R}[1]{>{\raggedleft\let\newline\\\arraybackslash\hspace{0pt}}m{#1}}

\newcommand{\ignore}[1]{}

\newif\ifcomments
\commentstrue % Comment or set \commentsfalse to disable comments.

\newcommand{\sdc}{SDC-C}
\definecolor{britishracinggreen}{rgb}{0.0, 0.26, 0.1}
\definecolor{cadetblue}{rgb}{0.37, 0.62, 0.63}

\newcommand{\red}[1]{\textcolor{red}{#1}}

\usepackage[hyphens]{url}

\begin{document}
\title{Characterizing and Improving the Resilience of Accelerators in Autonomous Robots}
%\author[]{\rm Deval Shah}
%\author[]{\rm Zi Yu Xue}
%\author[]{\rm Karthik Pattabiraman}
%\author[]{\rm Tor M. Aamodt}
%\affil[]{Electrical and Computer Engineering, University of British Columbia}
%\affil[]{\{devalshah, fzyxue, karthikp, aamodt\}@ece.ubc.ca}
%\author{Deval Shah, Zi Yu Xue, Karthik Pattabiraman,~\IEEEmembership{Senior Member, IEEE,} and Tor M. Aamodt,~\IEEEmembership{Member, IEEE}
%\IEEEcompsocitemizethanks{\IEEEcompsocthanksitem D. Shah, Z. Y. Xue, K. Pattabiraman, and T. M. Aamodt are with the Department of Electrical and Computer Engineering, University of British Columbia.}}
%\author{Michael~Shell,~\IEEEmembership{Member,~IEEE,}
%        John~Doe,~\IEEEmembership{Fellow,~OSA,}
%        and~Jane~Doe,~\IEEEmembership{Life~Fellow,~IEEE}% <-this % stops a space
%\IEEEcompsocitemizethan! Package inputenc Error: Unicode character ̃ (U+0303)
%ks{\IEEEcompsocthanksitem M. Shell was with the Department
%of Electrical and Computer Engineering, Georgia Institute of Technology, Atlanta,
%GA, 30332.\protect\\
%E-mail: see http://www.michaelshell.org/contact.html
%\IEEEcompsocthanksitem J. Doe and J. Doe are with Anonymous University.}% <-this % stops an unwanted space
\author{
\IEEEauthorblockN{Deval Shah, Zi Yu Xue, Karthik Pattabiraman, Tor M. Aamodt} \\
\IEEEauthorblockA{\textit{Electrical and Computer Engineering} \\
\textit{The University of British Columbia}\\
%Vancouver, Canada \\
\{devalshah, fzyxue, karthikp, aamodt\}@ece.ubc.ca}
}
\IEEEtitleabstractindextext{%
\begin{abstract}
  Motion planning is a computationally intensive and well-studied problem in autonomous robots. 
However, motion planning hardware accelerators (MPA) must be soft-error resilient for deployment in safety-critical applications, and blanket application of traditional mitigation techniques is ill-suited due to cost, power, and performance overheads. 
%Motion planning is a computationally intensive and well-studied problem in autonomous robots. 
%Recently, hardware accelerators for real-time motion planning have seen industry adoption.
%However, the deployment of autonomous robots in safety-critical applications requires these accelerators to be error-resilient. 
%Soft errors can degrade resilience and lead to catastrophic failures in robots, and blanket application of traditional mitigation techniques such as ECC is ill-suited for these accelerators due to cost and performance overheads.
We propose Collision Exposure Factor (CEF), a {\em novel metric} to assess the failure vulnerability of circuits processing spatial relationships, including motion planning.
CEF is based on the insight that the safety violation probability increases with the surface area of the physical space exposed by a bit-flip.
We evaluate CEF on four MPAs. %motion planning accelerators, %and examine the effect of different design aspects on resilience.
We demonstrate empirically that CEF is correlated with safety violation probability, and that CEF-aware selective error mitigation provides {$12.3\times$, $9.6\times$, and $4.2\times$} lower Failures-In-Time (FIT) rate on average for the same amount of protected memory compared to uniform, bit-position, and access-frequency-aware selection of critical data. 
Furthermore, we show how to employ CEF to enable fault characterization using {$23,000\times$} fewer fault injection (FI) experiments than exhaustive FI, and evaluate our FI approach on different robots and MPAs. We demonstrate that CEF-aware FI can provide insights on vulnerable bits in an MPA while taking the same amount of time as uniform statistical FI. Finally, we use the CEF to formulate guidelines for designing soft-error resilient MPAs.
%%\textcolor{red}{\em [{\bf Tor:} For what benchmark or workload? What about versus statistical fault simulation?]} \blue{{\bf Deval:} added}
%We propose Collision Exposure Factor (CEF), a {\em novel metric} to assess the failure vulnerability of circuits processing spatial relationships, including motion planning. 
%The insight underlying CEF is that the probability of safety violation increases with the amount of physical space exposed by a bit flip. 
%We evaluate CEF on three motion planning accelerators, and examine the effect of different design aspects on resilience.
%Our analysis shows that CEF is positively correlated with safety violation probability and can find the critical portions of the design with {$14000\times$} fewer fault injection (FI) experiments than exhaustive FI. 
%CEF can also be used for selective error mitigation provides average $22\times$, $2.34\times$, and $1.51\times$ lower FIT rate for the same amount of protected on-chip memory as compared to uniform, bit position, and access-frequency-aware selection of critical data. 

\end{abstract}

}

\maketitle

\IEEEdisplaynontitleabstractindextext

% \fi
\IEEEpeerreviewmaketitle

\IEEEraisesectionheading{\section{Introduction}\label{sec:intro}}

Autonomous robots are increasingly used for real-time and safety-critical tasks, including medical care~\cite{medical,moxi}, autonomous driving~\cite{autopilot,uberatg}, and home assistance~\cite{eldercare, smarthomes}.
The rate of autonomous robot deployment is expected to reach {2.2 million} units per year by 2022~\cite{service, autonomousvehicles, industrial}.
As autonomous robots are becoming an integral part of our lives,
it is crucial to ensure that an autonomous robot does not collide with other objects, and thereby harm the safety of its surroundings. 

Motion planning allows an autonomous robot to navigate and reach its end goal safely without collisions. Therefore, motion planning is key to the many tasks performed by autonomous robots, including navigation, object manipulation, footstep planning, and full-body movement. It has been an area of study since the 1970s~\cite{5751929, Kavraki2008},  and is today one of the key research topics in robotics. For example, motion planning constitutes about  $\sim$$10\%$ of the total publications in top robotics conferences~\cite{icra, iros} (ICRA-2020 had 1073 papers, 121 of which were in motion planning). 
Motion planning has also received significant attention from industry, with an increasing number of patents. %~\cite{US20190217857A1,US9229453B1,US10518770B2,US10007269B1}.  
For example, the number of patents granted every year by the United States Patent Office (USPTO) on collision detection and motion planning has increased by $4\times$ from 2015 to 2020~\cite{uspto,googlepatent}.
%%\tor{from which year to which year?}\deval{added}

The computational and real-time demands of motion planning exceed those provided by typical CPUs.
Recently, several approaches have been proposed to accelerate motion planning on different hardware platforms, 
including GPUs~\cite{Bialkowski2011, Gayle}, FPGAs~\cite{Atay2006, Murray2016ori, Shi2018}, and ASICs~\cite{Murray, sorin, Lian2018,daducd}.
Motion Planning Accelerators (MPAs) %~\cite{Murray, sorin} 
have achieved impressive performance gains and are being adopted in industry~\cite{RTR,8662463,siemens}. 
%%\tor{can we add to this list?}\deval{added a two more references }. 

\begin{figure}[t]
     \centering
     \begin{subfigure}[b]{0.45\linewidth}
         \centering
         \includegraphics[width=\textwidth]{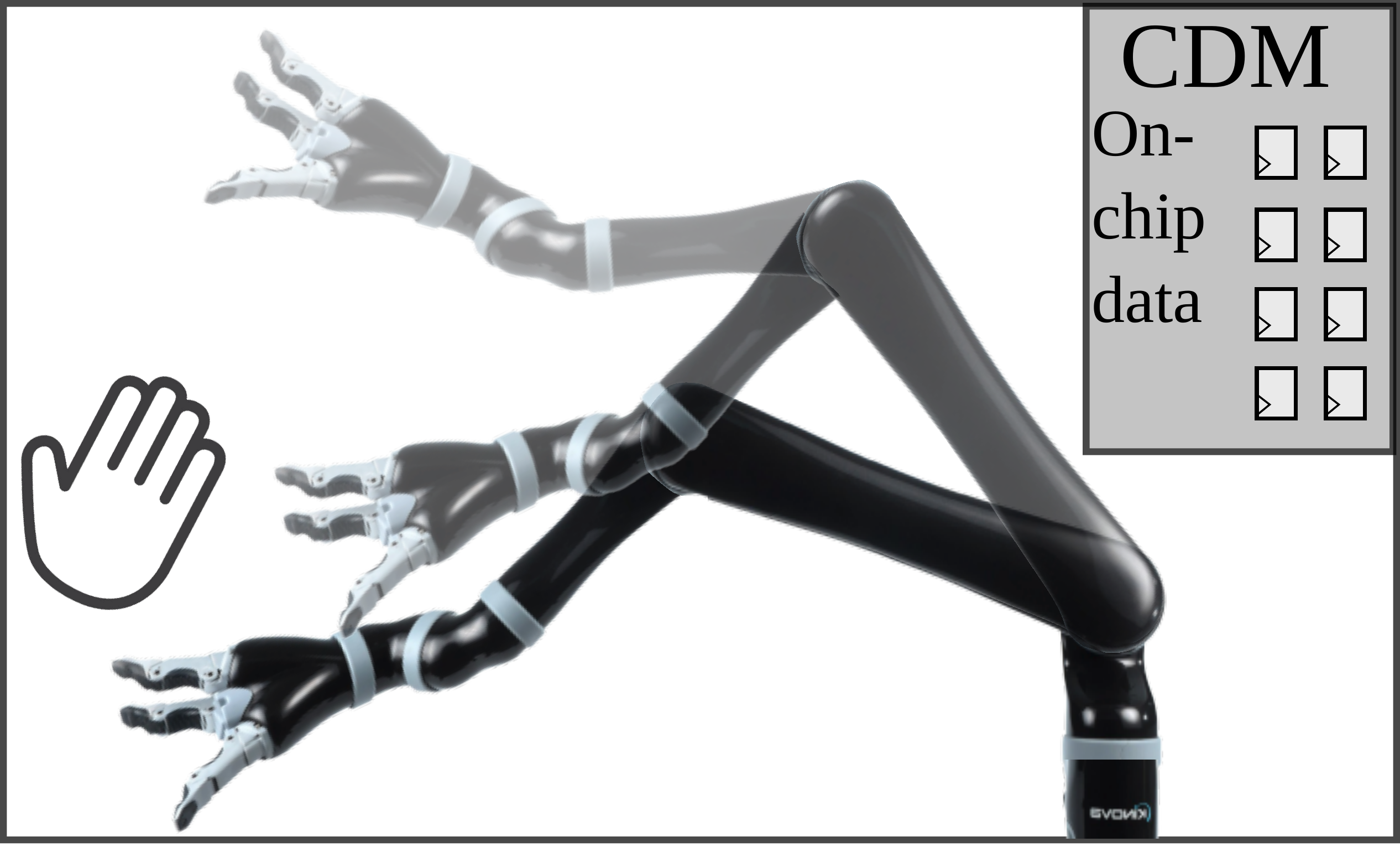}
        \caption{Error-free scenario}
        \label{fig:1a}
     \end{subfigure}
     \hfill
     \begin{subfigure}[b]{0.45\linewidth}
         \centering
         \includegraphics[width=\textwidth]{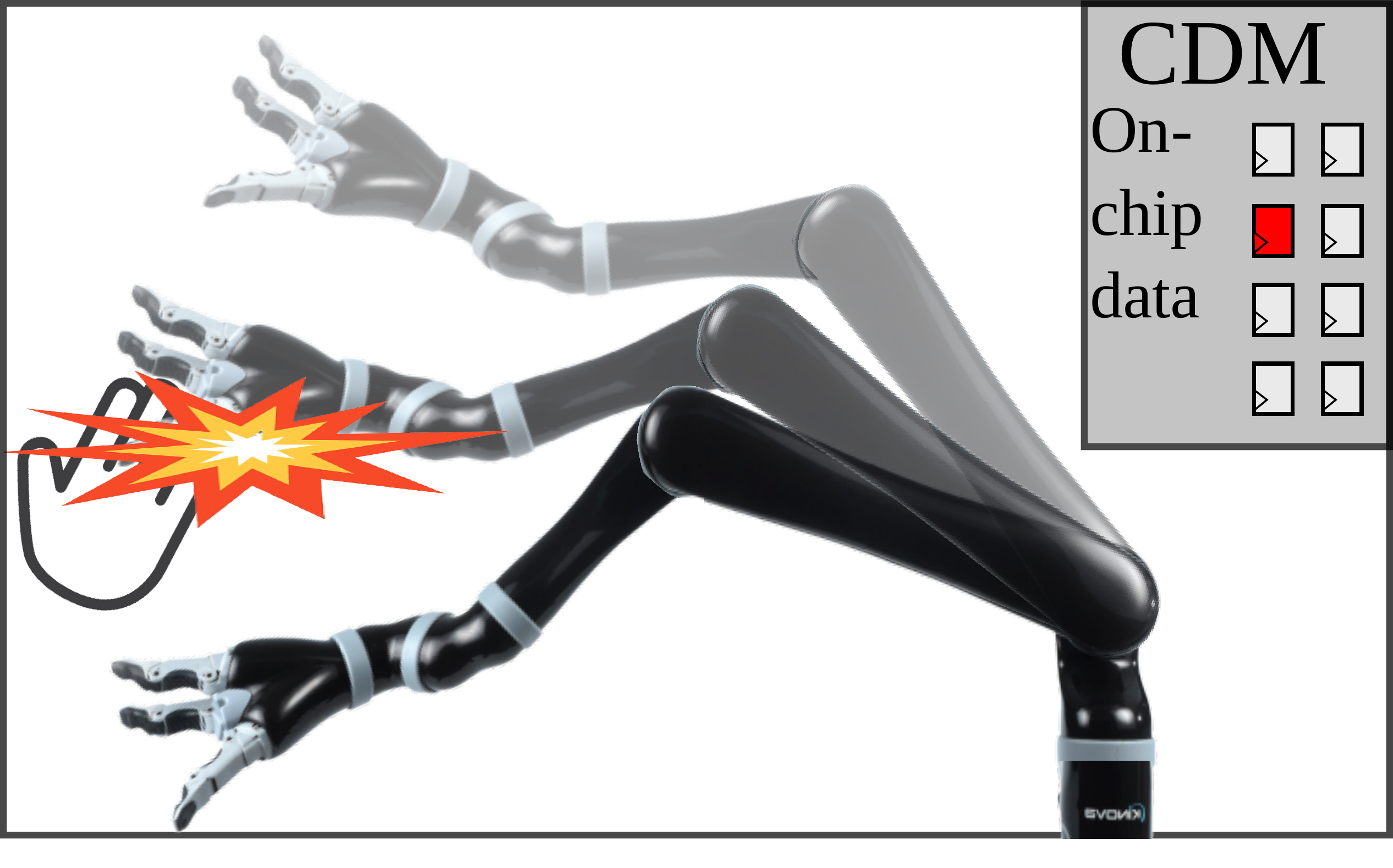}
        \caption{Faulty scenario}
        \label{fig:1b}
     \end{subfigure}
        \caption{The effect of soft errors on safety in autonomous robots.}
        \vspace{-3mm}
        \label{fig:motivation}
        \hfill
\end{figure} 
However, the use of MPAs in robotics applications has significant safety implications. 
For example, the IEC~61508 \cite{61508} provides functional safety standards for {electronic systems} used in applications such as robotics in terms of allowable FIT (Section ~\ref{subsec:SIL}), 
where 1~FIT is one failure per billion hours. %\karthik{Provide citation}
In modern systems, errors induced by particle strikes and radiation, or soft errors, make up the majority of SRAM and register-level faults~\cite{Mitra2005,Sridharan2015}. Hence, soft errors are a major threat to FIT level compliance in MPAs used in robotics applications. 
Furthermore, due to their transient nature, soft errors cannot be eliminated during the design and test phases of a chip, and hence need runtime mitigation. 

There have been many studies on the effects of soft errors on CPUs and GPUs~\cite{Mukherjee2003, Tselonis2016, Li2018, Vallero2018, Hari2017, 8809540, Bo2016}, 
deep learning accelerators~\cite{Li2017, Reagan2018}, and autonomous vehicle systems~\cite{Banerjee2018, Jha2019, Jha2018}. However, their impact on accelerators for robotics has not been studied. 
Application-agnostic blanket error mitigation techniques such as error-correcting codes (ECC) and dual/triple modular redundancy (DMR/TMR) can increase the area, cost, and power by $12\%$-$125\%$ ({Table}~\ref{tab:accel}), and degrade the performance of these hardware accelerators. 
With consumer applications driving growth of robotics, the electronics controlling these systems will become increasingly cost-sensitive~\cite{cost,costr,keynoterel}.
%\tor{Check added sentence. We need to say something like this given many prior reviewers seem to think costs are less important than reliability. Would be best to cite something to back up the sentence I added here.}
%\deval{added reference to some articles.}
Furthermore, any increase in the MPA's power consumption significantly reduces the operation time of a mobile robot with a single charge as MPAs suitable for real-time motion planning contribute to $15\%$-$50\%$ of the total power consumption of robotic systems~\cite{ur5,jaco,puma,al5d}. 
While there has been work on sensor and actuator faults in autonomous robots~\cite{Christensen2008},
there has been no study of the reliability of MPAs in the presence of soft errors. {\em To the best of our knowledge, we are the first paper on characterizing and improving the reliability of MPAs.}

In this paper, we study soft errors in the collision detection module (CDM), which is the largest, most energy-consuming, and safety-critical component in MPAs~\cite{Murray, sorin, Lian2018,Atay2006}.  
We find that CDMs in these accelerators consist of on-chip storage elements to store the information about space that the robot passes through for possible motions. 
These storage elements account for more than {$97$\%} of the sequential elements,  and so we focus on it in this paper.
 
Figure~\ref{fig:motivation} shows the effect of soft errors in the CDM. 
In the error-free scenario (Figure~\ref{fig:1a}), the robot navigates safely to the end goal.
However, in Figure~\ref{fig:1b}, a soft error causes the MPA to misidentify a path taken by the robot as safe, thereby resulting in a collision. 
This mis-identification is due to an error modifying the geometric representation of the space that the robot passes through. 

%%One way to find such soft errors causing safety violations in a robot is via exhaustive fault injection (FI). %\karthik{Added}
Prior work has explored memory and register file designs that allow flexible partition into protected and non-protected regions %low overhead techniques
%\tor{``low overhead techniques'' is vague; can we describe the approach taken in a few more words?  The next sentence does not seem to explain the low overhead part.} 
%\deval{Added}
for incorporating selective error mitigation %and reliability-aware data placement 
in systems using CPUs and GPUs~\cite{selectiveeccisca,7357094,4272980,intelpatent,5661755}. 
%In applying selective error mitigation to an MPA, 
These techniques protect only the most vulnerable data and do so by placing it in protected memory. 
A challenge with applying such an approach to error mitigation is that it requires accurate and fast identification of the most critical data. %bits in the data. % to be stored in memory. 
One approach
%\tor{Is this what these earlier papers propose to do?} \deval{Those papers mainly focus on architecture design that allows flexible partitioning of memory or register files. Works that focus  register file proposes to use register file lifetime to select vulnerable data. Next paragraph convers related work for this. }
is to use exhaustive fault injection (FI) to identify storage bits that exhibit the highest resilience improvement when protected from soft errors. 
%%Exhaustive FI is typically required to gain insights into the bits of data that exhibit the highest resilience improvement when protected from soft errors.
%\karthik{This was controversial in the DSN Paper - we need to explain why we need exhaustive FI rather than statistical FI.} \deval{Toned down the claim}
 %
Unfortunately, exhaustive FI can take up to $24,000$ CPU hours for a typical MPA (Section~\ref{sec:cefawareFI}). 
%, and it varies with the MPA's configuration., requiring robot and application-specific analysis. 
%\tor{24,000 hrs is not a large cost if one is going to mass produce something and this cost is paid only once.  Wasn't there another arguement about how CEF helps provide design insight?} 
%\deval{This cost has to be paid each time the user configures it with a new motion set. I think, some simple robots, which are used for specific tasks, we can get mass-produced robots with pre-configured MPAs. However, for fully autonomous complex robots, the user must be provided some configurability to change the motion set. This paper from ICRA 2018~\cite{46570} talks about combining RL and dynamic motion set generation for robot navigation. There are other works also on dynamically changing the motion set~\cite{Jaillet2004APM, Hauser2009IntegratingTA}.  In such a case, a simple heuristic such as CEF can be used to determine the vulnerability of bits at runtime. I added some text (highlighted) to explain this and emphasize the design insight point. }
%Finally, FI is highly sensitive to the characteristics of each accelerator and workload and needs to be performed for each such combination, further exacerbating its high computational cost.
Such high FI time overhead for error mitigation each time the MPA is reconfigured for a different task or robot introduces practical deployment challenges. 
For fully autonomous robots, the MPA's storage data can be generated or modified at runtime~\cite{Jaillet2004APM, Hauser2009IntegratingTA,46570}, requiring runtime characterization of vulnerable data for selective error mitigation.  
Also, as noted in earlier work~\cite{Mukherjee2003} exhaustive FI is an inefficient way to gain insight during architecture design. 

Prior work on FI techniques for CPUs and GPUs~\cite{Tselonis2016, Li2018, Hari2017, 8809540, Bo2016, Nie2018} has obtained significant reductions in the FI time. 
%\blue{However, these techniques are }
These FI tools and techniques are targeted towards specific languages, Instruction Set Architectures (ISA), or CPU/GPU system simulators and often exploit the microarchitecture or ISA to reduce the FI time and estimate the failure probability (Section~\ref{sec:rw}). 
However, these techniques cannot be directly applied to robotic accelerators that use specialized microarchitectures and instruction sets. 
%\tor{What about work on speeding up FI that is published in architecture conferences?} 
%\deval{There are works on FI tools for CPU and GPU. These works can not be directly applied to MPA as their techniques are specific to CPU or GPU microarchitecture. We talk about this in the related work, and I have also added one line here.  }
%\tor{We should elaborate a bit on why these cannot be applied to MPA here (1-2 sentences) and give a forward reference to related work (assuming we expand on the reasons there).}
Thus, there is a need for techniques that efficiently characterize the effect of soft errors in robotics applications. %causing safety violations. % and  apply selective protection techniques for MPAs.
%\deval{added}

Architectural Vulnerability Factor (AVF) has been widely used to define the vulnerability of a system and can be measured using an analytical method such as Architecturally Correct Execution (ACE) analysis~\cite{Mukherjee2003} or FI~\cite{Leveugle2013}. 
%%\red{Efficient techniques to characterize the vulnerability of different microprocessor structures are Architectural Vulnerability Factor (AVF)~\cite{Mukherjee2003}, and Program Vulnerability Factor (PVF)~\cite{Sridharan2009} analysis.
%%AVF and PVF identify Architecturally Correct Execution (ACE) bits--values that if changed will change the observed output.
Directly applying AVF methodology such as ACE analysis to MPAs requires considering the positions of obstacles in the environment, thus leading to the need to run a large number of simulations to accurately estimate the fraction of time a hardware storage element contains an ACE bit. 

To overcome these challenges, we introduce a novel heuristic, {\em Collision Exposure Factor (CEF)} that depends only upon the accelerator and robot, not on the environment.
%determining a fast approximation of the fraction of runtime a bit is ACE. 
%%CEF gives a measure of the fraction of runtime for which a bit is ACE. 
%Efficient techniques to characterize the vulnerability of different bits for microprocessors are Architectural Vulnerability Factor (AVF) \cite{Mukherjee2003} and Program Vulnerability Factor (PVF) \cite{Sridharan2009}. They use Architecturally Correct Execution (ACE) analysis of bits in processors. Unfortunately, these cannot be applied to MPAs without significant modifications as MPAs lack an Instruction Set Architecture (ISA).
%\tor{After rereading the AVF paper, I'm not convinced the above sentence is correct.}
%Further, not all architecturally visible faults lead to safety violations in motion planning, making ACE analysis overly conservative. 
%\tor{Again, I don't think the above sentence is correct--as presented in the AVF paper, ACE analysis does NOT equate architecturally visible faults with errors.}
%\deval{Modified it}
Other heuristics, such as bit position~\cite{Li2017} and  access-frequency~\cite{765955,Mehrara2008} have been proposed to find critical bits for deep learning accelerators and embedded applications, respectively. 
%\karthik{such as?} \deval{added}
{However, our analysis shows that considerable variation exists in the failure probabilities of bits with the same access-frequency or bit position in MPAs. In contrast, our approach provides a higher reduction in the overall failure probability as our proposed heuristic is more accurate at finding critical bits in the MPAs (Section~\ref{sec:sol}). }
%{However, our analysis shows that these heuristics do not capture the failure probabilities of bits in MPAs accurately and considerable variation exists in the failure probabilities of bits with the same access-frequency or bit-position. In contrast, our approach provides a higher reduction in the overall failure probability as our proposed heuristic is more accurate at finding critical bits in the MPAs (Section~\ref{sec:sol}). }
%\tor{Saying they are not specific to MPAs is unsatisfactory as an explanation in my view. I believe we had deeper insights into this outcome.  Can we provide a brief summary of the underlying reason.} \deval{Added}
%However, we find these provide a lower reduction in failure probabilities than the approach we propose (Section~\ref{sec:sol}).

%We propose \textit{Collision Exposure Factor} (CEF), a novel metric 
%that alleviates the above challenges by estimating
%\tor{not sure what ``representing'' means} \deval{yes ``representing'' was incorrect, ``estimate'' makes sense}
The CEF estimates the architectural vulnerability factor 
%\tor{What is ``vulnerability factor''?  Do we mean AVF?}\deval{Yes, it should be AVF. Just ``vulnerability factor'' is ambiguous. Modified } 
of memory locations storing spatial information. 
The 3D model data of the swept spaces of a robot's possible motions play a key role in real-time collision detection and motion planning. %, and requires a significant amount of on-chip and off-chip memory. 
The main reason for safety violations due to a soft error is a modification of this stored 3D model of the swept space, which 
% that the robot passes through, leading to 
potentially leads to erroneous collision detection (Figure \ref{fig:motivation}). 
{\em To account for this violation, we define the CEF as the surface area of the swept space of the robot exposed to obstacles by a soft error.}

%%\karthik{I added this paragraph to bring out the key insight behind CEF - please check.}
%The main insight behind the CEF is that 
By considering the entire swept space of the robot's motion, the CEF decouples the effects of the robot model and the environment on safety violations.  CEF can be calculated once for a given MPA and robot {\em without} needing to consider obstacles in the environment (which might be highly dynamic). 
The underlying reason this separation is possible is that, in the most widely used approaches for real-time motion planning, a robot has a fixed set of possible short motions
that are precomputed independently of obstacle positions~\cite{leven}.   
During operation a subset of these motions is selected to navigate to a given goal depending upon obstacle positions.
For purposes of fault analysis we decouple these steps by assuming a conservative distribution (e.g., uniform) on where objects will appear.
%This is because most robots (in our study) have a fixed trajectory of motion that is highly repetitive. 
%\karthik{Please check}
%deval{I modified this with reference}
%Hence, CEF decouples the effects of the robot model and the environment on safety violations, which can be exploited for error mitigation.  
%We show empirically that the CEF computed this way is positively correlated with the average failure probability due to a soft error {\em independent of the actual environment.} 
We show empirically that the CEF computed this way {\em independent of the actual environment} is positively correlated with the average failure probability due to a soft error.  
%%CEF can be used as a reliability metric across different MPAs and robots.
%%\tor{Not sure what this last sentence is saying.  Is this trying to say we can use CEF to compare robots or MPAs?} \deval{Yes its placement is not good. I have removed this last sentence. We added this sentence as at some point, a reviewer was confused whether CEF can be used for different robots and MPAs. But I think now we can remove this sentence as we also mention in the abstract that CEF is evaluated for different robots and MPAs. }

Further, we propose a CEF-aware error mitigation technique to selectively protect values with higher CEF in an MPA's on-chip storage. 
%,  enabling resilience for a single MPA across different robots.
Finally, we propose a two-phase FI methodology: Phase~1 FI to find the CEF of all the bits (for a given robot and MPA),
and Phase~2 FI to find the relation between the CEF and failure probability with fault site pruning. % for different workloads. 
{Uniform statistical FI-based characterization of the CDM provides similar speedup over exhaustive FI as CEF-aware FI (Section~\ref{sec:cefawareFI}).  However, it does not find the safety-violation probability of an individual bit nor does it find relative vulnerability of different bits, which is needed for error mitigation.  }
In CEF-aware FI, on the other hand, decoupling the two FI phases allows efficient calculation of the CEF and safety-violation probability for every bit in the CDM.  

%%%\karthik{Removed the numbers here as they're misleading - we discuss speedups later.}  
%(within $2$ CPU hours) for all the bits in the circuit, and FIT rate measurement of all the bits within $4$ CPU hours. 

In summary, we make the following contributions in this paper:
\begin{itemize}
\item Establish the necessary conditions for safety violations (i.e., collisions), and propose Collision Exposure Factor (CEF), a reliability metric for CDM storage elements. 
\item Propose an efficient \emph{CEF-aware error mitigation} technique that selectively protects values with higher CEF, and  compare it to uniform, exhaustive FI, bit position-aware, and access-frequency-aware application of DMR, TMR, and ECC techniques for four different CDM designs. 
\item Propose a two-phase FI methodology using the CEF of storage elements for fault site pruning to reduce FI time by orders of magnitude. % with statistical FI. 
\end{itemize}
%\karthik{Moved this after the contributions}

Our results show that CEF-aware selective error mitigation results in  {$12.3\times$, $9.6\times$, and $4.2\times$} lower FIT rate for the same amount of protected memory compared to uniform, bit position, and access-frequency-aware selection of critical data. 
Further, CEF-aware FI reduces the FI time by $23,000\times$ with minimal loss of accuracy, and finds the FIT rate of the MPA and individual bits. 
Finally, we study the impact of architectural design parameters on resilience and error mitigation overheads, and demonstrate the potential for developing efficient soft error resilient MPA architectures using the CEF metric.
%Our CEF metric enables the estimation of resilience characteristics in the early design phase of MPAs.
\section{Background and Motivation}
\label{sec:bm}

This section briefly summarizes relevant background information on motion planning, general CDM architecture, and safety in robotics. 
Finally, we summarize the limitations of current FI methodologies, and discuss the need for application-specific reliability metric and FI methodology for MPAs. 

\subsection{Motion Planning}
\label{sec:mp}
\begin{figure}[t!]
    \centering
    \begin{subfigure}[b]{0.13\textwidth}
         \centering
         \includegraphics[width=\textwidth]{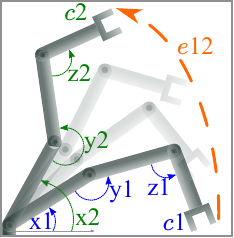}
         \caption{  }
         \label{fig:cspace1}
     \end{subfigure}
     %\hfill
     \begin{subfigure}[b]{0.13\textwidth}
         \centering
         \includegraphics[width=\textwidth]{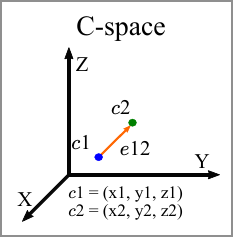}
         \caption{  }
         \label{fig:cspace2}
     \end{subfigure}
     %\hfill
     \begin{subfigure}[b]{0.13\textwidth}
      \centering
      \includegraphics[width=\textwidth]{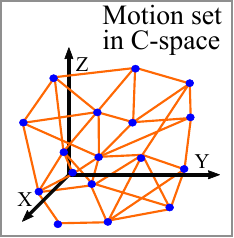}
      \caption{  }
      \label{fig:cspace3}
  \end{subfigure}
    \caption{(a) shows the spatial poses and motion of a 2D robot with three degrees of freedom (x, y, z), (b) represents these poses and motion in the robot's C-space, and (c) shows a motion set in the C-space. }
    \label{fig:cspace}
\end{figure}

The objective of motion planning is to compute a collision-free path for a robot.
Motion planning is performed in the robot's configuration space (C-space), which has the same number of dimensions as the robot's degrees of freedom (DOFs). 
%%Each dimension represents the range of values of a degree of freedom of the robot. 
Figure~\ref{fig:cspace1} shows a planar robot with three DOFs, while Figure~\ref{fig:cspace2} represents its C-space. 
%%Although the robot moves on a plane, its pose is represented by the angle of its three joints $x$, $y$, and $z$ in 3-dimensional C-space. 
Each dimension of the C-space represents the range of angles of a joint of the robot. 
The point $c1$ with coordinates (x1, y1, z1) in the C-space corresponds to the pose $c1$ in Figure~\ref{fig:cspace1}. Similarly, edge $e12$ represents the robot's motion from $c1$ to $c2$.
 % \iffalse

The complexity of motion planning increases exponentially with the robot's DOFs~\cite{Salzman}. 
%Thus, approximate methods such as sampling-based motion planning have been explored and are commonly used~\cite{lavalle,Salzman}. 
%Of these, probabilistic roadmap-based algorithms~\cite{Kavraki} are widely used, and many works have built upon this strategy over the last 20 years. 
Thus, approximate methods such as probabilistic roadmaps~\cite{Kavraki} are widely used~\cite{lavalle,Salzman}  %, and many papers have built upon this strategy 
over the last 20 years. 
% to solve the problem, and are commonly used~\cite{Salzman,lavalle}. 
Many hardware accelerators~\cite{Atay2006, Shi2018, Murray, Murray2016, Lian2018,daducd} have been proposed for these algorithms.
\subsubsection{Probabilistic roadmaps}
\begin{figure}[b!]
  \centering
 % \begin{subfigure}[b]{0.15\textwidth}
 %      \centering
 %      \includegraphics[width=\textwidth]{figures/cspaceprma.pdf}
 %      \caption{  }
 %      \label{fig:prma}
 %  \end{subfigure}
 %  \hfill
   \begin{subfigure}[b]{0.18\textwidth}
       \centering
       \includegraphics[width=\textwidth]{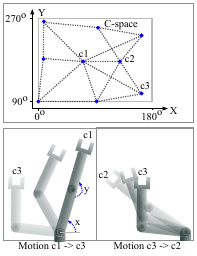}
       \caption{  }
       \label{fig:prmb}
   \end{subfigure}
   %\hfill
   \begin{subfigure}[b]{0.27\textwidth}
       \centering
       \includegraphics[width=\textwidth]{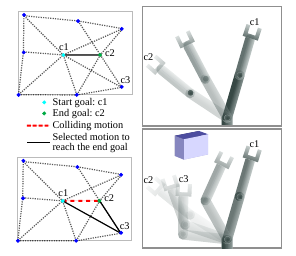}
       \caption{ }
       \label{fig:prmc}
   \end{subfigure}
  \caption{(a) shows the C-space and the motion set of a planar robot with two DOFs. The bottom figures give examples of two motions (linearly interpolated path between two poses) highlighted in the motion set. (b) The  top figure shows the path between c1 and c2 in the absence of any obstacle. Whereas the bottom figure shows the path in the presence of an obstacle. }
  \label{fig:prm}
\end{figure}
Probabilistic motion planning consists of two phases. 
In the preprocessing stage, a graph, called a \emph{motion set}, is constructed in the robot's C-space. Figure~\ref{fig:cspace3} and Figure~\ref{fig:prmb} give examples of motion sets for a robot with three and two DOFs, respectively. 
The nodes in the graph correspond to the robot's spatial poses, and an edge between close-by nodes represent motion generated by a local planner (e.g., linear interpolation between two poses) from one pose to another. 
Motion set consists of collision-free poses and motions of the robot for a given environmental scenario. 
%%A simple local planner is used to generate the motion between close-by nodes connected by edges~\cite{Kavraki}, e.g., a linearly interpolated motion between two poses, as shown in Figure~\ref{fig:prmb}. 
%%The motion set (graph) is modified to be collision-free in the query stage. 
%%A path search algorithm is used to find a feasible path between any two poses. This path can consist of more than one edge/motion. 
%%Such an approach results in multiple possible paths between two poses, which increases the probability of finding a feasible path in a cluttered environment as some of the edges/motions must be avoided due to possible collisions.  
In the query phase, a path search algorithm (e.g., Dijkstra's algorithm) is used to find a path between given start and end poses using the precomputed motion set.  
A path consists of one or more motions from the motion set. 
The probabilistic roadmap is a multi-query method, i.e., the same motion set is used to perform motion planning for multiple combinations of start and goal poses. 

Leven and Hutchinson~\cite{leven} proposed a real-time motion planning approach based on probabilistic roadmaps for a dynamic environment and is also used by MPAs~\cite{Shi2018, Murray, Murray2016, Lian2018,daducd,daduseries}. 
In this approach, the motion set is generated for an obstacle-free environment. 
At runtime, collision detection is performed to find collision-free motions in this set for a given placement of obstacles. 
The path search module considers collision-free motions to find a feasible path between the start and end pose in the resulting \textit{collision-free} motion set graph. 
For example, in Figure~\ref{fig:prmc} (bottom), an obstacle in the robot's environment makes some of the motions in the motion set flagged as ``colliding motion'', which are avoided by the path search module to find a path between c1 and c2. 
%For example, as shown in Figure~\ref{fig:prmc}, the path search module avoids colliding motions and gives a collision-free path between c1 and c2. 

\begin{figure}[t!]
  \centering
   \begin{subfigure}[b]{0.12\textwidth}
       \centering
       \includegraphics[width=\textwidth]{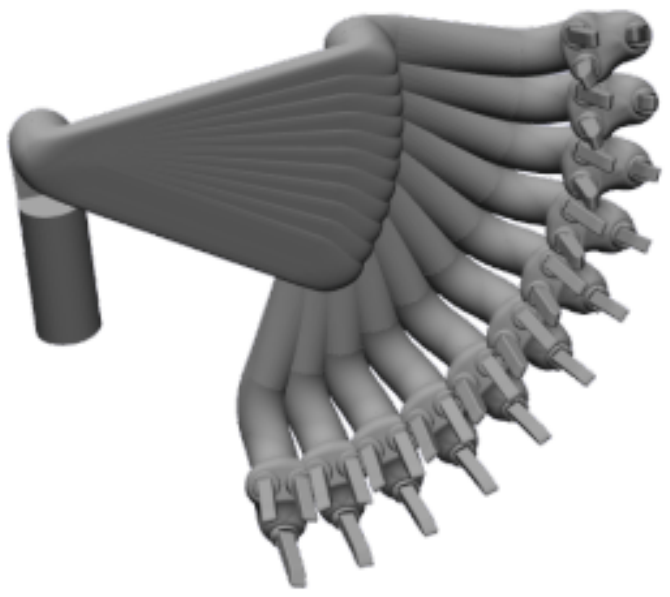}
       \caption{ }
       \label{fig:cdm1}
   \end{subfigure}
   \hfill
   \begin{subfigure}[b]{0.12\textwidth}
       \centering
       \includegraphics[width=\textwidth]{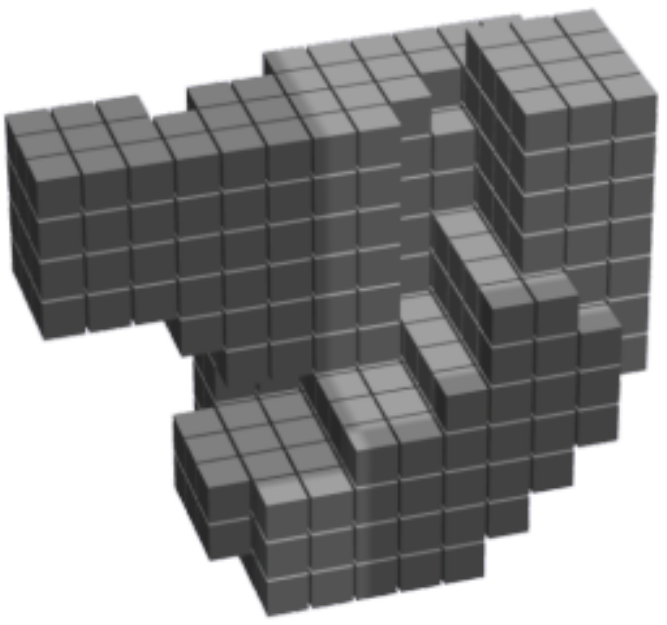}
       \caption{ }
       \label{fig:cdm3}
   \end{subfigure}
   \hfill
   \begin{subfigure}[b]{0.13\textwidth}
       \centering
       \includegraphics[width=\textwidth]{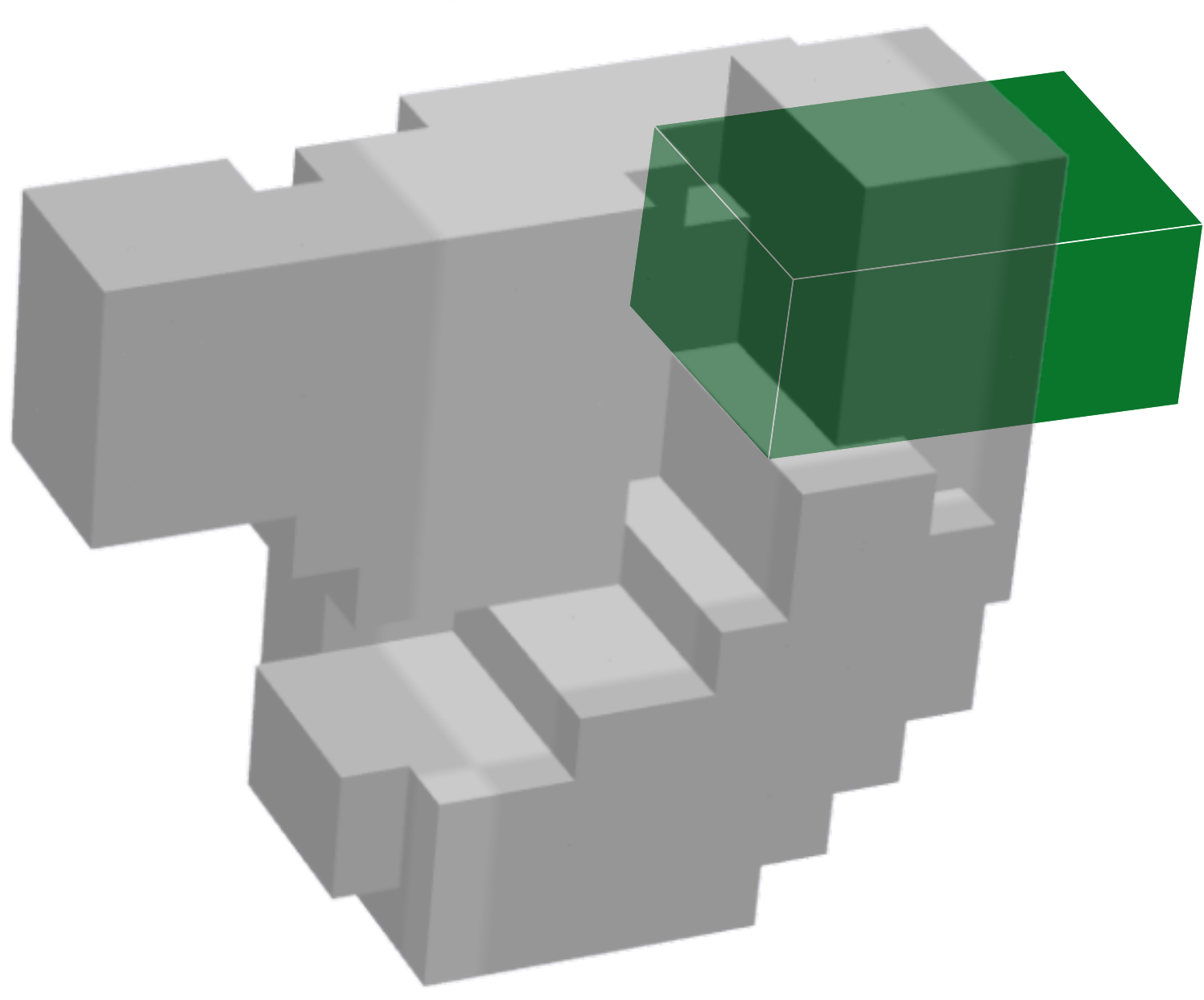}
       \caption{ }
       \label{fig:cdm4}
   \end{subfigure}
  \caption{(a) shows the swept space of a robot motion, (b) represents the voxelization of the swept space (explained in Section~\ref{sec:accelerator}),  and (c) shows an obstacle that overlaps with swept space of this motion. Collision detection process identifies  this motion to be in collision. }
  \label{fig:collision_detection}
\end{figure}

The information about the swept space of each motion in the motion set is stored in the memory to facilitate real-time collision detection. 
% to facilitate real-time collision detection. 
Swept space of a motion is the part of space occupied by a robot as the given motion is followed. 
%%The key advantage of the given approach is that the same motion set can be used for multiple queries in a dynamic environment and enables real-time motion planning. 
For example, as shown in Figure~\ref{fig:cdm1} and Figure~\ref{fig:cdm3}, a motion's  swept space  is discretized, and its 3D model is stored. 
%This step facilitates collision detection to find the feasibility of motions in real-time. 
%%Collision detection identifies motions in the motion set that will cause collisions with obstacles. 
As shown in Figure~\ref{fig:cdm4}, the motion's swept space is checked for overlap with obstacles for collision detection at runtime. 
%%\textbf{4. Path search (runtime): }
%Such an approach results in multiple possible paths between two poses, which increases the probability of finding a feasible path in a cluttered environment as some of the edges/motions must be avoided due to possible collisions.
%%The probability of finding a collision-free path between two poses in a cluttered environment increases with the number of poses and motions in the motion set~\cite{lavalle}. 
%In more recent approaches 
For fully autonomous robots, depending upon the environment, the motion set is modified ~\cite{Jaillet2004APM, Hauser2009IntegratingTA,46570} and swept spaces of new motions are generated at runtime. This update is performed outside of the critical path of motion planning.  
%%Collision detection and path search are both performed in real-time.
The collision detection step is the most time- and energy-consuming in motion planning and takes up to 99\% of the total runtime on a CPU~\cite{Bialkowski2011}. 
As collision detection must be performed in real-time with low latency to ensure safety in an environment with dynamic obstacles, MPAs typically dedicate more than 85\% of their total area to accelerate collision detection~\cite{Murray, Murray2016, Lian2018}, making the CDM the most vulnerable component in an MPA. 
Unfortunately, an erroneous collision detection can potentially lead to a collision between the robot and an obstacle in its surroundings, making collision detection safety-critical. 
%%Further, the CDM occupies most of the area in an MPA~\cite{Murray2016,Murray,Lian2018}, making it the most vulnerable component in an MPA. 
Thus, we focus on the CDM and perform fault characterization of CDMs of four MPAs, which we refer to as A1~\cite{Murray2016ori}, A2~\cite{Murray}, A3~\cite{Lian2018}, and A4~\cite{daducd} (Section~\ref{sec:bench}) throughout this paper.
\subsection{Collision Detection Modules (CDM)}
\label{sec:accelerator}
Many approaches have been proposed to accelerate collision detection on different hardware platforms. 
The architecture proposed by Murray et al.~\cite{Murray2016ori} uses specialized combinational circuits for a given motion set, but it is not reconfigurable to different motion sets at runtime. 
In contrast, GPU-based collision detection acceleration~\cite{Bialkowski2011, Gayle} provides high flexibility, but it is not energy-efficient. 
Configurable collision detection hardware accelerators~\cite{Atay2006, Shi2018, Murray, Murray2016, Lian2018,daducd} provide a balance between flexibility and performance/energy-efficiency. 
%%Configurable CDMs consist of highly optimized collision detection logic and storage structures, and the on-chip storage can be configured for different motion sets at runtime. 
We focus on such configurable CDM architectures. 

Figure~\ref{fig:cdm} illustrates the architecture of a generic configurable CDM. 
The CDM consists of collision detection circuits (CDCs). It exploits inter-motion parallelism to check multiple motions for collision in parallel. 
Each CDC consists of on-chip memory to store a motion's swept space, and exploits inter-query and intra-query data reuse. 
The on-chip storage can be configured for different motions at runtime.
Based on our synthesis experiments, the on-chip memory to store swept spaces constitutes more than $97\%$ of the sequential elements in CDMs. 

The key design consideration of CDMs is the geometric representation used to store swept spaces in the memory. Several representations for storing the 3D model have been proposed for motion planning, including polygonal meshes~\cite{Amanatides}, bounding box hierarchies~\cite{vandenBergen,Gottschalk}, voxels~\cite{30724}, and octrees~\cite{JACKINS1980249}. These approaches differ in storage requirements, representation accuracy, and computational complexity. 
%% Added this to set the height of cdc1,2, and 3 properly. 
\newdimen\imageheight
\settoheight{\imageheight}{%
\includegraphics[width=0.10\textwidth]{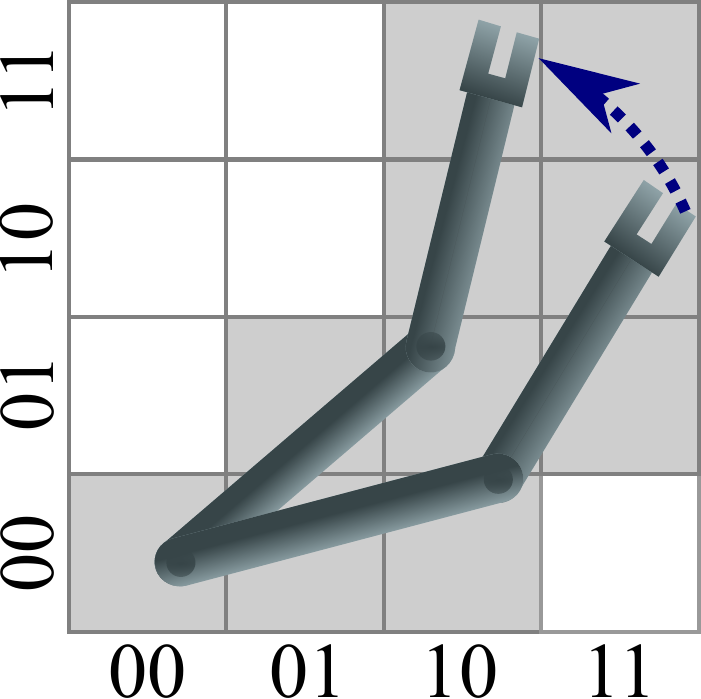}}
\newdimen\octheight
\settoheight{\octheight}{%
\includegraphics[width=0.30\textwidth]{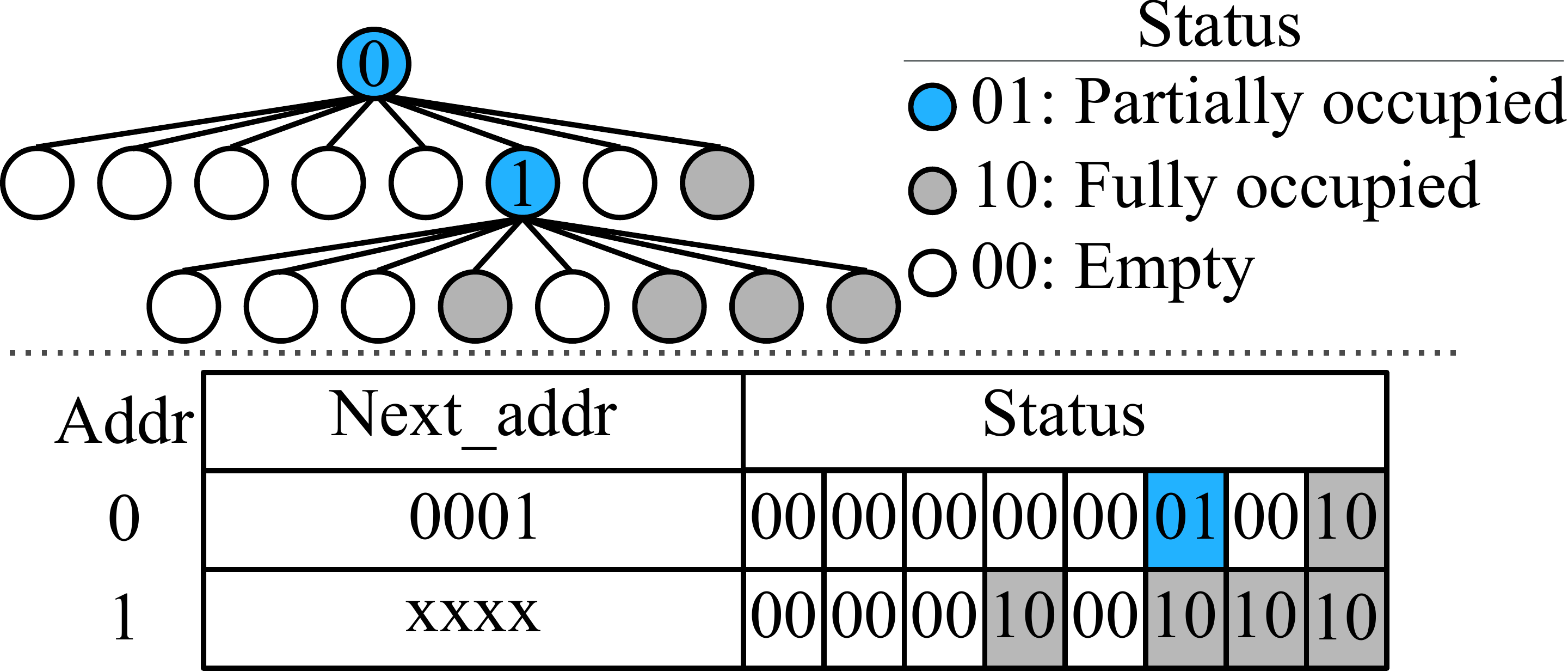}}
%\showthe\imageheight - this will print the height in pt, so you can add height manually if needed
\begin{figure}[]
         \centering
         \includegraphics[width=0.4\textwidth]{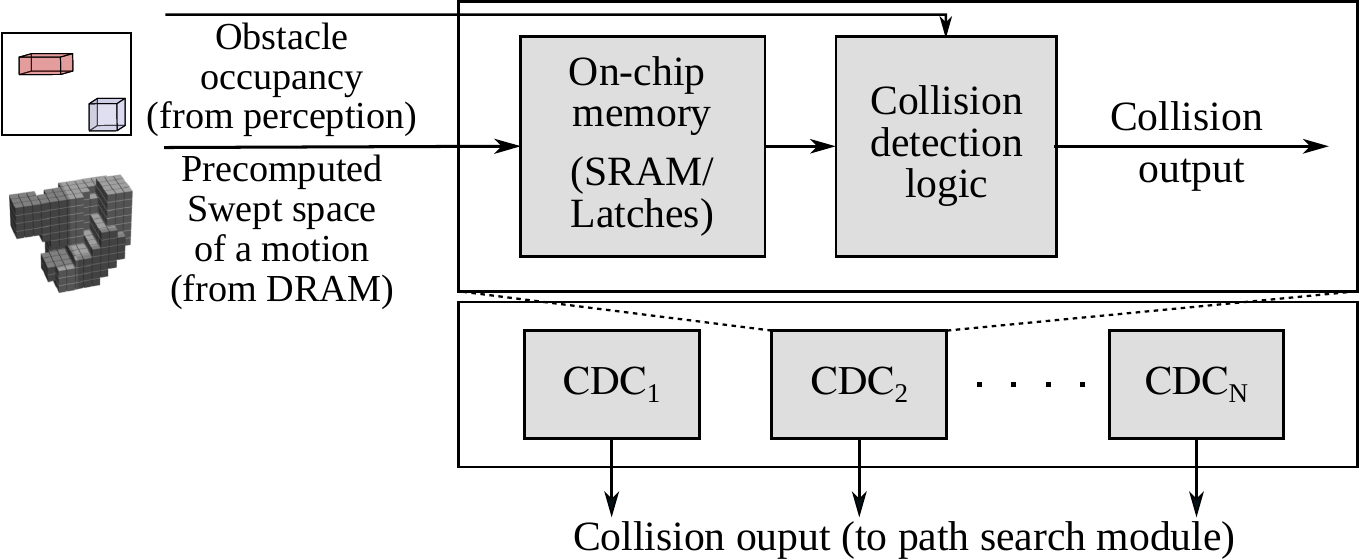}
        \caption{Architecture of a Collision Detection Module (CDM). }
        \label{fig:cdm}
%\end{figure}
\vspace{5mm}
%\begin{figure}[t]
     \centering
     \begin{subfigure}[t]{0.13\textwidth}
         \centering
         \includegraphics[height=\imageheight]{figures/cdc1.pdf}
        \caption{Swept space}
        \label{fig:cdc1}
     \end{subfigure}
     %\hfill
     \begin{subfigure}[t]{0.17\textwidth}
         \centering
         \includegraphics[height=\imageheight]{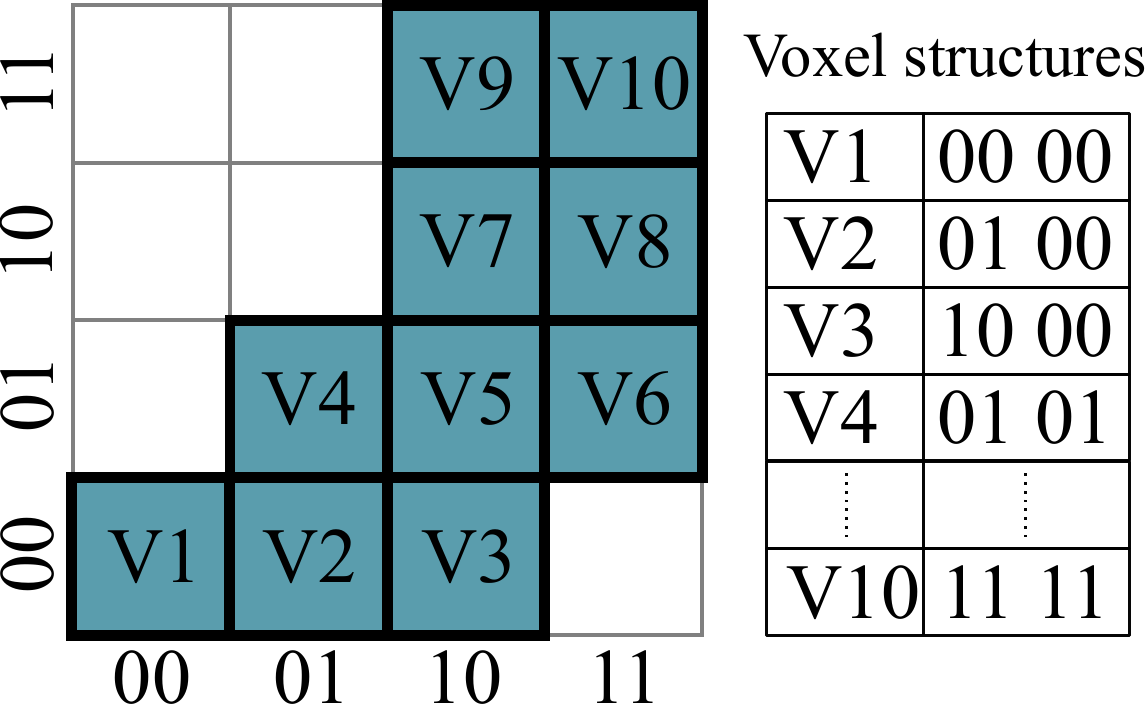}
        \caption{Voxel (A1)}
        \label{fig:cdc2}
     \end{subfigure}
     %\hfill
      \begin{subfigure}[t]{0.17\textwidth}
         \centering
         \includegraphics[height=\imageheight]{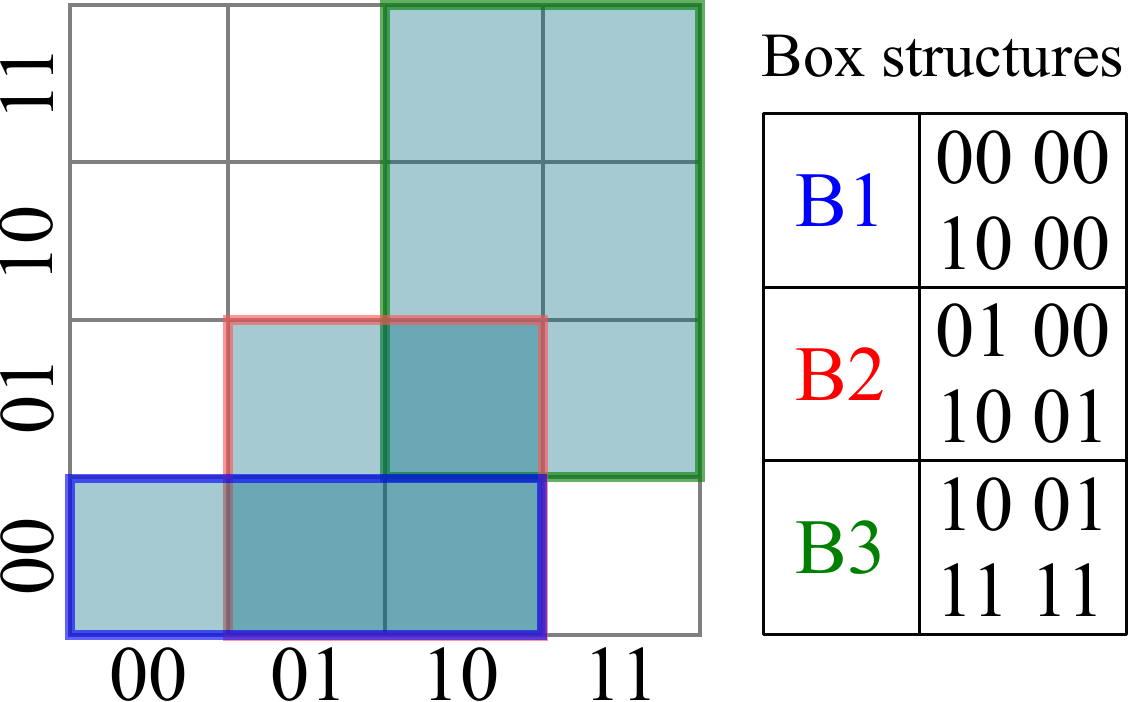}
        \caption{Box (A2)}
        \label{fig:cdc3}
     \end{subfigure}
     \hfill
        \caption{Voxelization and voxel/box based representation example in 2D. (a) Represents a robot and its motion in 2D. Shaded voxels represent the voxelized swept space. (b) Swept space is stored using coordinates (x,y) of individual voxels. (c) Contiguous voxels combine into boxes and stored using coordinates of diagonal voxels; darker regions represent overlapping of multiple boxes.}
        \label{fig:cdc}
        \vspace{5mm}
%\end{figure}
%\begin{figure}[t]
     \centering
     %\begin{subfigure}[t]{0.08\textwidth}
     %    \centering
     %    \includegraphics[height=0.7\octheight]{figures/octree_1.pdf}
     %   \caption{Swept space}
     %   \label{fig:octree1}
     %\end{subfigure}
     %\hfill
     \begin{subfigure}[t]{0.15\textwidth}
         \centering
         \includegraphics[height=\octheight]{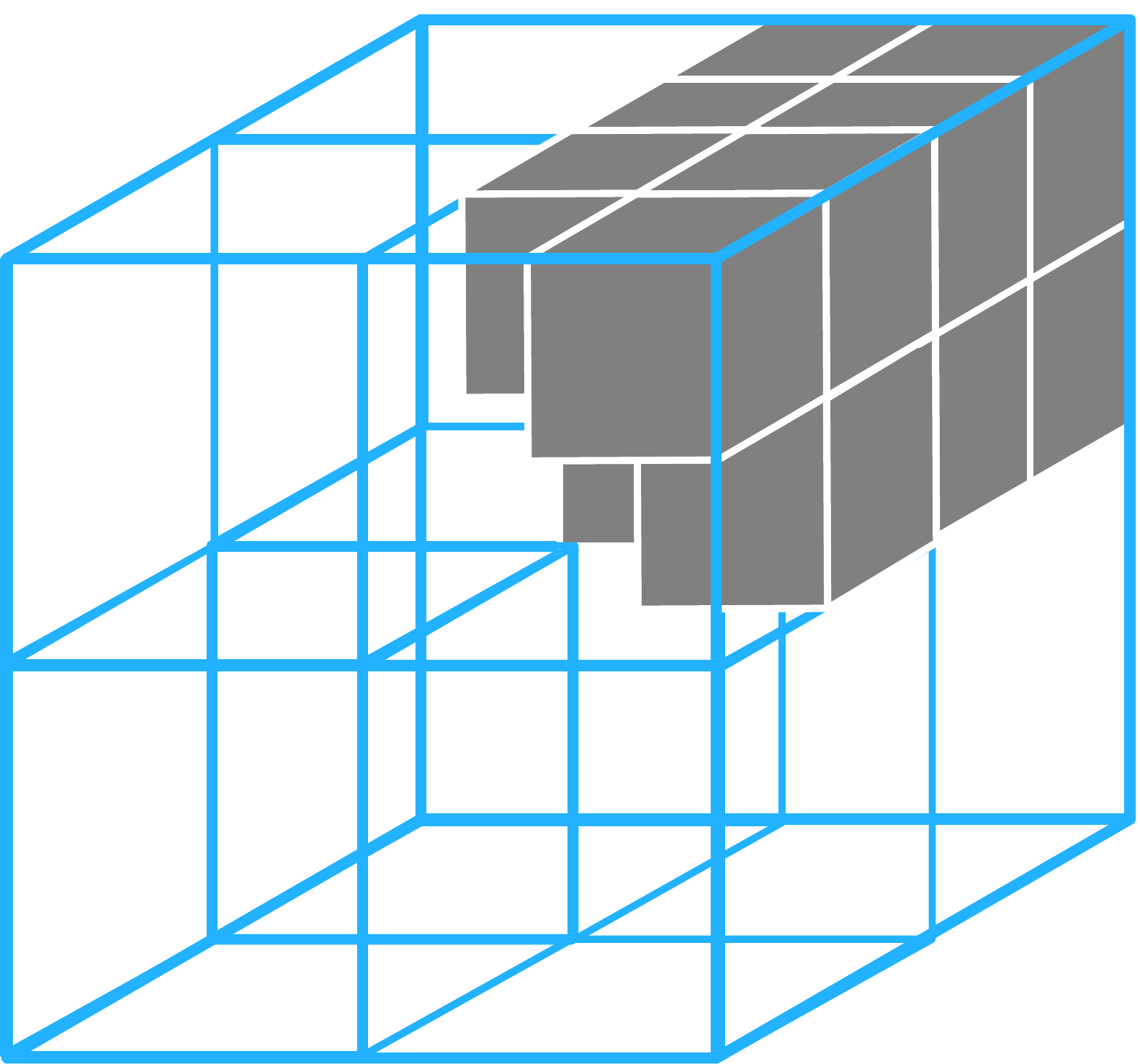}
        \caption{Spatial division of the swept space}
        \label{fig:octree2}
     \end{subfigure}
     \hfill
      \begin{subfigure}[t]{0.30\textwidth}
         \centering
         \includegraphics[height=\octheight]{figures/octree_3.pdf}
        \caption{Octree structure (A3)}
        \label{fig:octree3}
     \end{subfigure}
        \caption{Octree representation (figure reproduced from~\cite{Lian2018}).}
        \label{fig:octree}
\end{figure}
%Because grid-based representations are less compute- and memory-intensive, 
Note that the proposed metric CEF is applicable regardless of the geometric representation used. 
All CDM hardware accelerators use either grid-based (Figure~\ref{fig:cdc2}, Figure~\ref{fig:cdc3}) or octree-based representations (Figure~\ref{fig:octree}) (explained in Section~\ref{subsec:cef}) as these representations are less compute- and memory-intensive. %These representations are further explained in Section~\ref{subsec:cef}. 
For both representations, the swept space of a motion is discretized into fixed-size cubes known as {\em voxels}.  
Voxelized swept space is then stored using a set of structures specific to the representation used. 
At runtime, a perception sensor module senses obstacles occupancy information, converts it to voxels, and sends it to the CDCs. 
%The CDCs check which motions are in a collision with the obstacles, and send the output to the path search module. 
The CDCs perform collision detection for stored motions and send the output to the path search module. 

\subsection{Functional Safety in Robotics}\label{subsec:SIL}
\begin{table}[b]
  \begin{center}
  \scriptsize
  \setlength\tabcolsep{3pt}
  \caption{Comparison of complete and selective ECC. The die cost calculation is based on the equation provided in~\cite{hp} (Chapter 1.6). The wafer cost, yield, and impurity factors are for 16nm technology node~\cite{wafer,yield,8203849}.}
  \label{tab:compcost}
  \begin{tabular}{ccC{1.5cm}C{1.5cm}C{1.5cm}}
  \hline
  & No ECC & Full ECC & Selective ECC (SIL~2) & Selective ECC (SIL~3) \\ \hline
  Total area (mm$^2$) & 450 & 502.5 & 454.5 & 469.8 \\ \hline
  Cost/die (\$) & 59.9 & 70.0 & 60.8 & 63.6 \\ \hline
  \end{tabular}
\end{center}
  \end{table}
%\mbox{FIT} = \sum_{c \in \mbox{components}} S_{c} \times \mbox{SDC}_{c} \times N \times \mbox{FIT}_{\mbox{\small Raw}}
Safety is a crucial consideration in robotics. 
Hence, the failure rate of circuits used in robotics applications, including MPAs, is an important factor. 
The FIT rate of a circuit consisting of multiple components can be computed using Equation~\eqref{eq:3}~\cite{Mukherjee2003, Li2017}. 
\begin{equation} 
  \mbox{FIT} = \sum_{i \in \mbox{components}} \mbox{S}_{i} \times \mbox{SDC}_{i} \times \mbox{FIT}_{\mbox{\small Raw}}
  \label{eq:3}
\end{equation}
$\text{FIT}_{\mbox{\small Raw}}$ is the raw FIT rate defined in FIT/Mb units and depends upon multiple factors including technology node, ambient conditions, and elevation~\cite{Marc}. 
$\text{S}_{i}$ is the number (in Mb) of sequential elements/latches in component $i$. 
$\text{SDC}_{i}$ is the probability that a fault in component $i$ affects the output of the application. 

IEC~61508~\cite{61508} defines an international safety standard for safety-critical electronic systems. 
This standard is based on the risks of failure and defines four Safety Integrity Levels (SIL).
Each SIL expresses the upper bound on the FIT rate.
SIL~1 is the least stringent, while SIL~4 is the most stringent. The maximum allowable FIT rate decreases by three orders of magnitude from SIL~1 ($10,000$) to SIL~4 ($10$). 
Note that the IEC~61508 standard considers the entire electronic system, not only the MPA. %, and hence this conservative.

One approach to make circuits soft error-resilient for certifiable safety is to use hardware error mitigation techniques on storage structures, which incurs high  cost/power/performance overheads. 
Autonomous vehicles and robotics industries typically have shallow profit margins. 
For example, the profit margin per unit is under $\$1000$ for several automobile industries~\cite{profitmargin}.  
Electronic systems contribute up to $40\%$ to the total cost of a car~\cite{electronicscar,electronicswp} (at the time of writing). 
Hence cost-effective solutions to make MPAs more reliable are imperative~\cite{keynoterel}. 
The overheads for complete protection of storage structures from soft errors increase with their size. 
In such a case, the protection may be sacrificed entirely if the area/power overheads are not under budget. 
In comparison, selective protection is flexible and provides certifiable safety with less overhead than complete protection of storage structures by protecting only the most vulnerable data. 

Table~\ref{tab:compcost} compares the die area and cost for complete and selective ECC for the A3~\cite{Lian2018}. 
As shown in Table~\ref{tab:compFI}, selective ECC can reduce the cost by $10\%$ for SIL~3 ($~1\%$ increase in the profit margin). % for certifiable-safety. 
While the table compares only the die cost, an increase in the die area has a cumulative effect on the total cost of an electronic system, amplifying the need for selective error mitigation. 
\subsection{Fault Injection (FI)}
\label{subsec:FI}
%Fault injection (FI) is a method for characterizing a component or system's behavior under the presence of faults. % removed for DSN
In a circuit, a soft error can occur at any location and time. 
Assuming a single-bit error model, where only one bit is affected by a soft error in the component, exhaustive fault characterization typically requires A $\times$ B FI runs. In this equation, A is the number of fault sites in space, and B is the number of fault sites in time.
A is determined by the number of bits, and B is determined by the application's total execution cycles and the number of possible inputs to the application.
Unfortunately, this requires very high numbers of FI runs. 
In MPAs, different combinations of obstacle positions (input to collision detection) add to the number of FI runs, making the space even larger. 

Accelerators' fault characterization is typically carried out by statistical FI experiments~\cite{sfigpu,Hari2017,Leveugle2013,hpcsfi,Li2017,tdscconv}.
%%%\tor{Is there a reference to back up the above claim?} \deval{I added some references that use SFI or mention ``most papers use SFI''. }
%Statistical FI  performs random sampling in the fault space, and can determine the overall FIT rate with high confidence with just a few hundred trials~\cite{Leveugle2013}. 
%Statistical FI  performs random sampling in the fault space and can determine the overall FIT rate with good confidence with a significant reduction in the number of FI runs compared to exhaustive FI~\cite{Leveugle2013}. 
Statistical FI  performs random sampling in the fault space and allows to tune the number of FI experiments as per the required confidence of the estimated FIT rate for a system~\cite{Leveugle2013}. 
%It has been shown that statistical FI can determine the overall FIT rate with good confidence with just a few hundred trials for some systems~\cite{Leveugle2013}. 
%%%\tor{The prior sentence seems inaccurate/misleading as whether  ``a few hundred trials'' suffices must depend upon the system.} \deval{yes it depends upon the system, I rephrased it. }
Often, statistical FI is performed per component to find the safety-critical components in a circuit. 
%However, it requires a strategic grouping of bits to find safety-critical components in a circuit. 
However, the CDM component occupies more than $85\%$ of the area in an MPA and mainly consists of sizeable on-chip storage structures to store swept spaces. Hence, selective error mitigation of CDM requires a strategic approach to determine the safety-critical data in the on-chip storage. 
Further, the number of FI runs needed for statistical FI increases with a decrease in the probability of a bit error resulting in an output error, i.e., the failure probability of a component~\cite{Leveugle2013}. 
%%\tor{First mention of the term ``error probability''.  What is our definition of this term?} \deval{rephrased it. }
%Also, the number of FI runs for statistical FI increases as the error probability
Typical CDMs have a low overall failure probability ($\sim0.3\%$) which increases the number of FI runs required when using uniform random statistical FI (Section~\ref{sec:cefawareFI}).
%\karthik{Should we also mention that statistical FI needs a lot of runs as the SDC rates are quite low?}
%\deval{I added this. So statistical FI and CEF-aware FI require the same number of FI runs, but CEF-aware FI performs more detailed fault characterization}
Therefore, it would be beneficial to develop an application-aware FI strategy in which the sites for FI are strategically selected while keeping the number of FI runs required low. 

\section{Reliability Metric for Motion Planning} 
\label{sec:cef}
\begin{figure*}[t]
  \centering
       \begin{subfigure}[t]{0.11\textwidth}
      \centering
      \includegraphics[width=\textwidth]{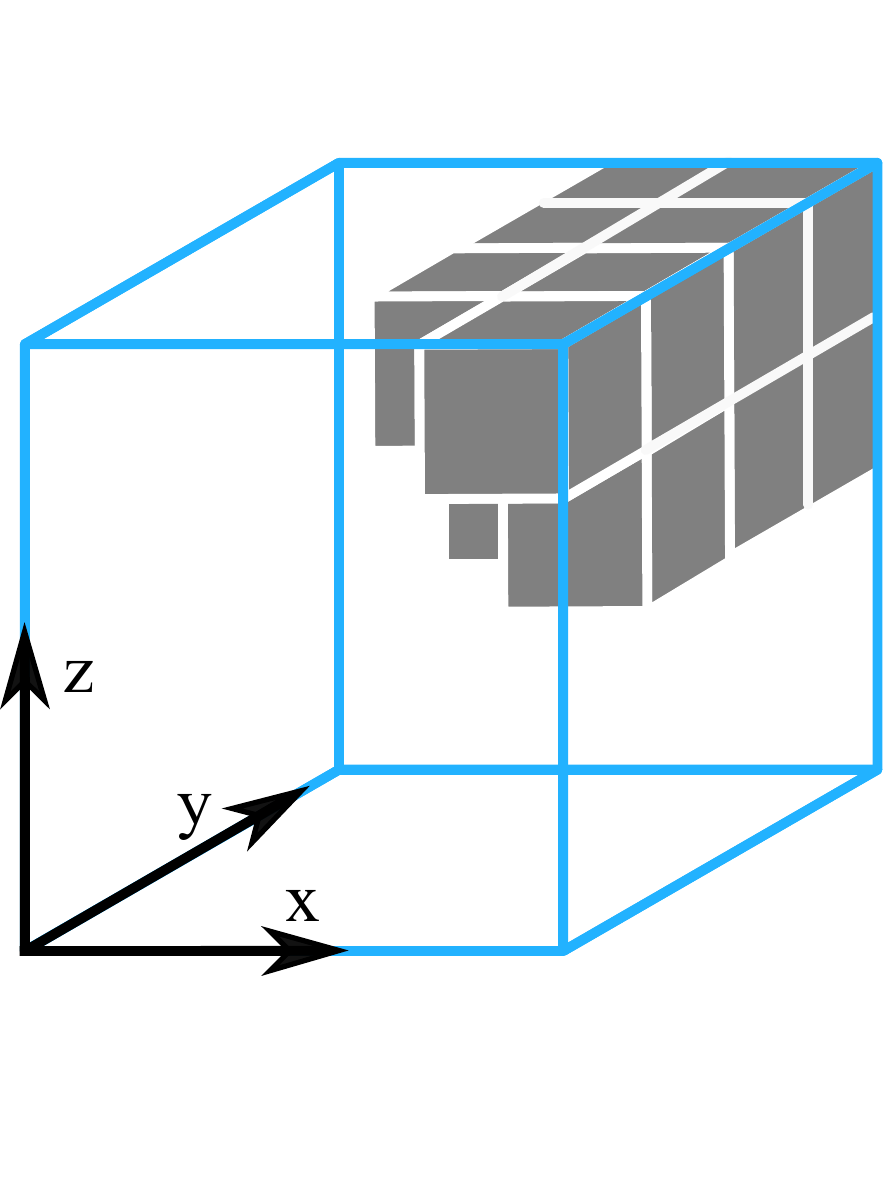}
      \caption{Swept space}
    \label{fig:critical_base}
  \end{subfigure}
  \hfill
  \begin{subfigure}[t]{0.12\textwidth}
      \centering
      \includegraphics[width=\textwidth]{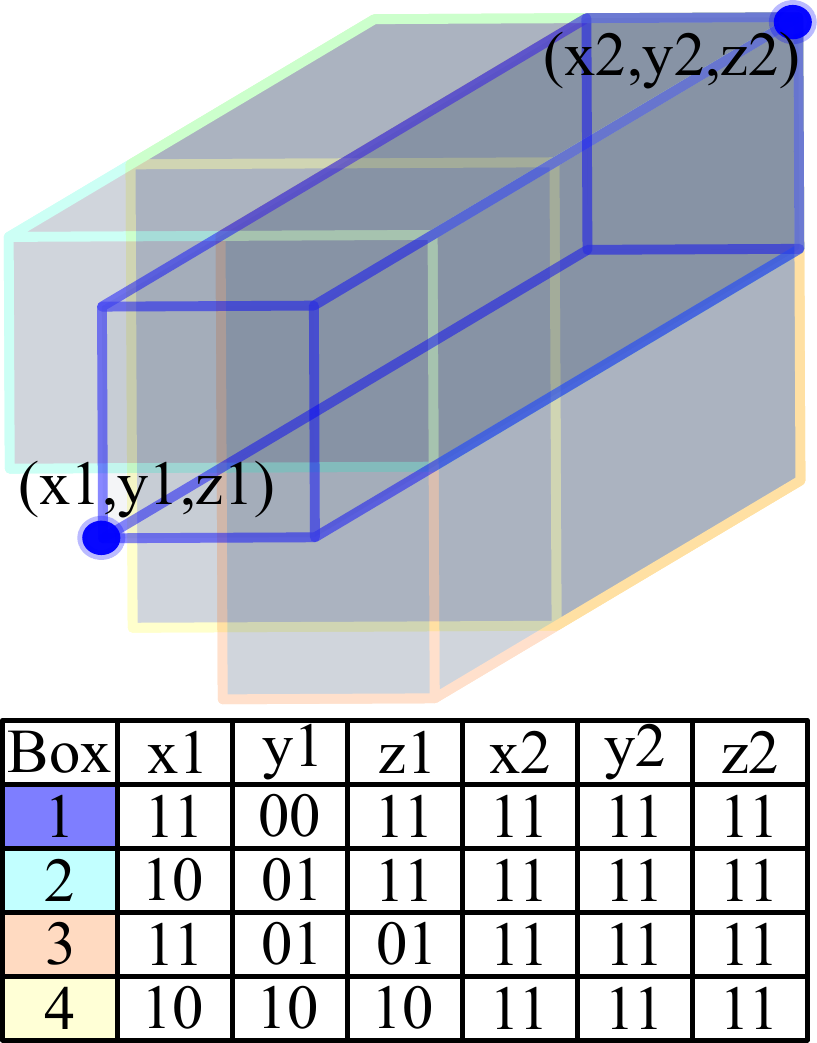}
      \caption{No fault}
    \label{fig:no_fault}
  \end{subfigure}
  \hfill
   \begin{subfigure}[t]{0.12\textwidth}
      \centering
      \includegraphics[width=\textwidth]{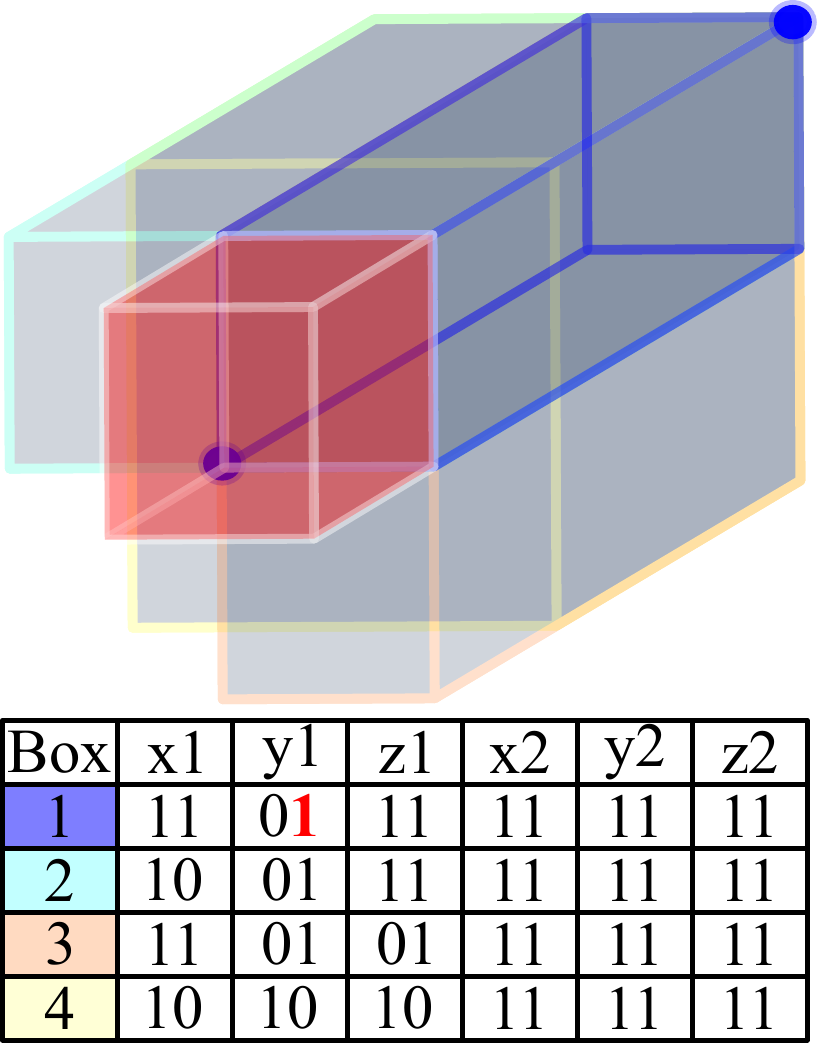}
      \caption{Critical voxel (CEF = $5$)}
     \label{fig:critical_voxel}
  \end{subfigure}
       \hfill
   \begin{subfigure}[t]{0.12\textwidth}
      \centering
      \includegraphics[width=\textwidth]{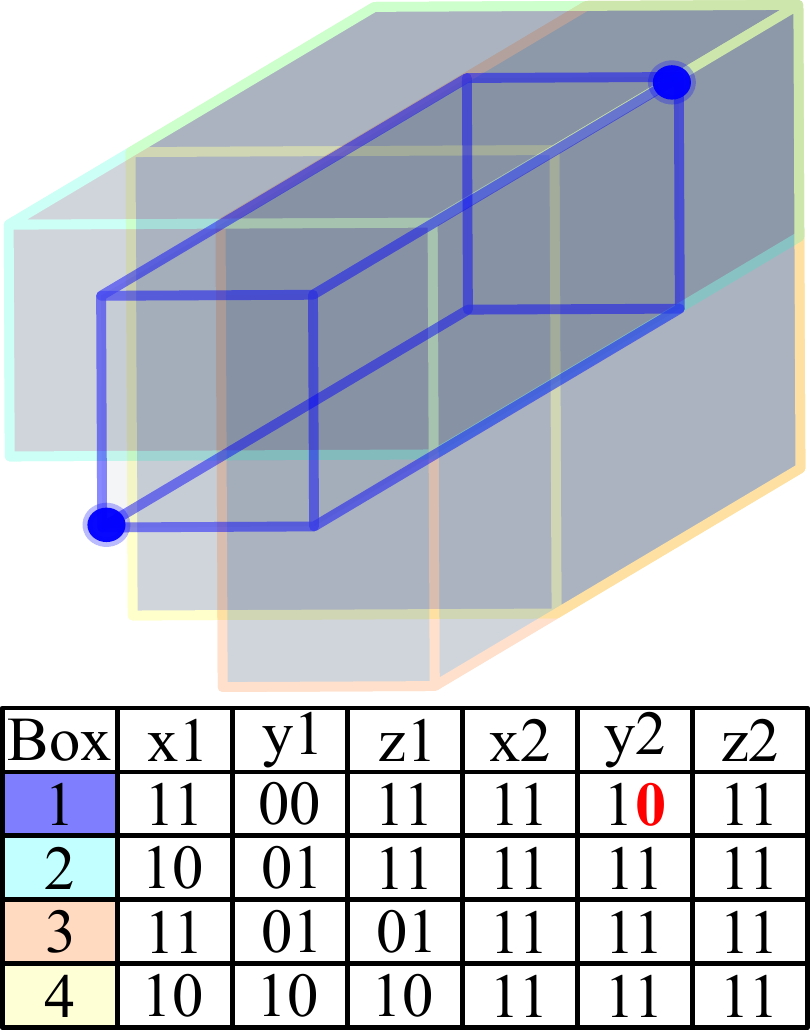}
      \caption{Fault masked (CEF = $0$)}
     \label{fig:critical_no_error}
  \end{subfigure}      
  \hfill
  \begin{subfigure}[t]{0.16\textwidth}
      \centering
      \includegraphics[width=\textwidth]{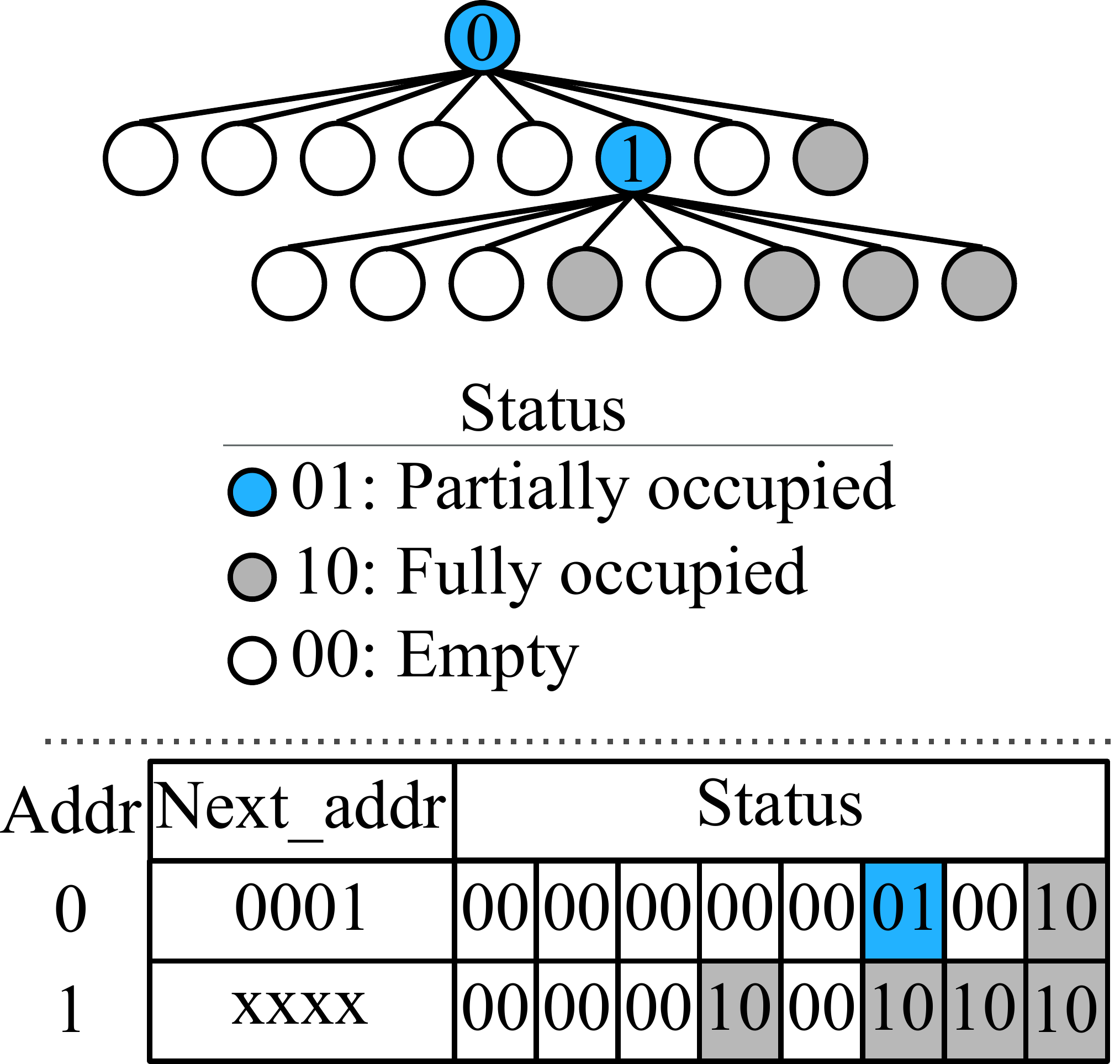}
      \caption{No fault}
    \label{fig:critical_baseA3}
  \end{subfigure}
  \hfill
  \begin{subfigure}[t]{0.145\textwidth}
      \centering
      \includegraphics[width=\textwidth]{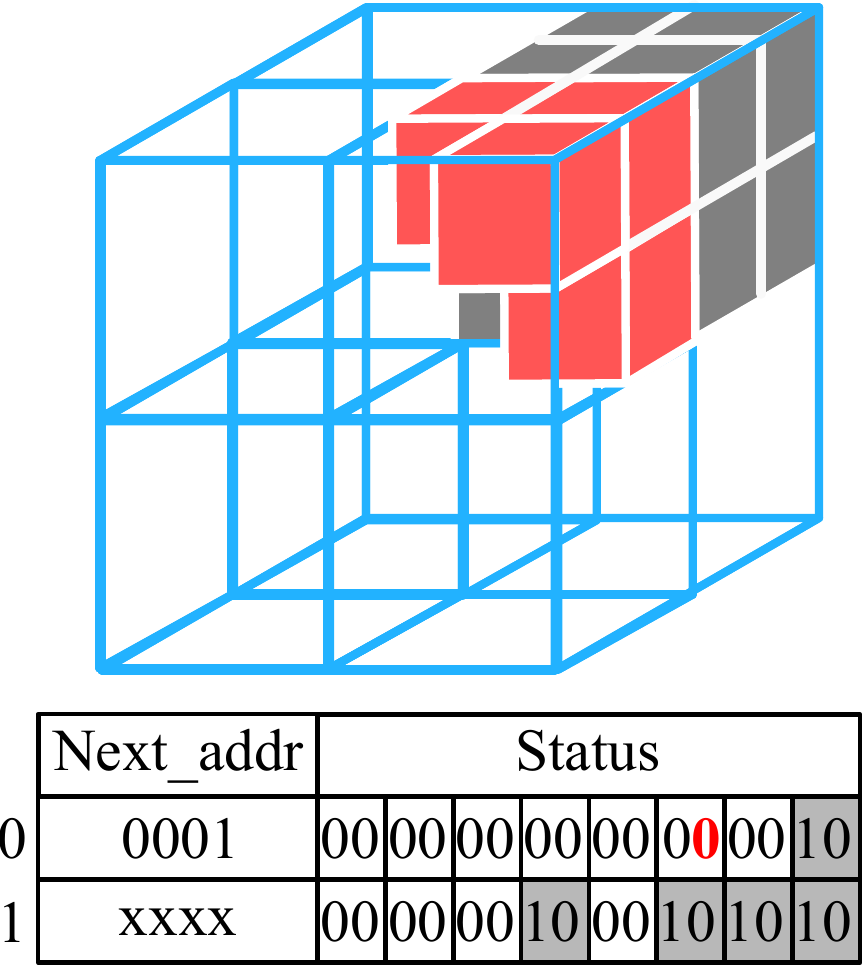}
      \caption{Critical voxel (CEF = $16$)}
    \label{fig:critical_voxelA3}
  \end{subfigure}
  \hfill
  \begin{subfigure}[t]{0.145\textwidth}
      \centering
      \includegraphics[width=\textwidth]{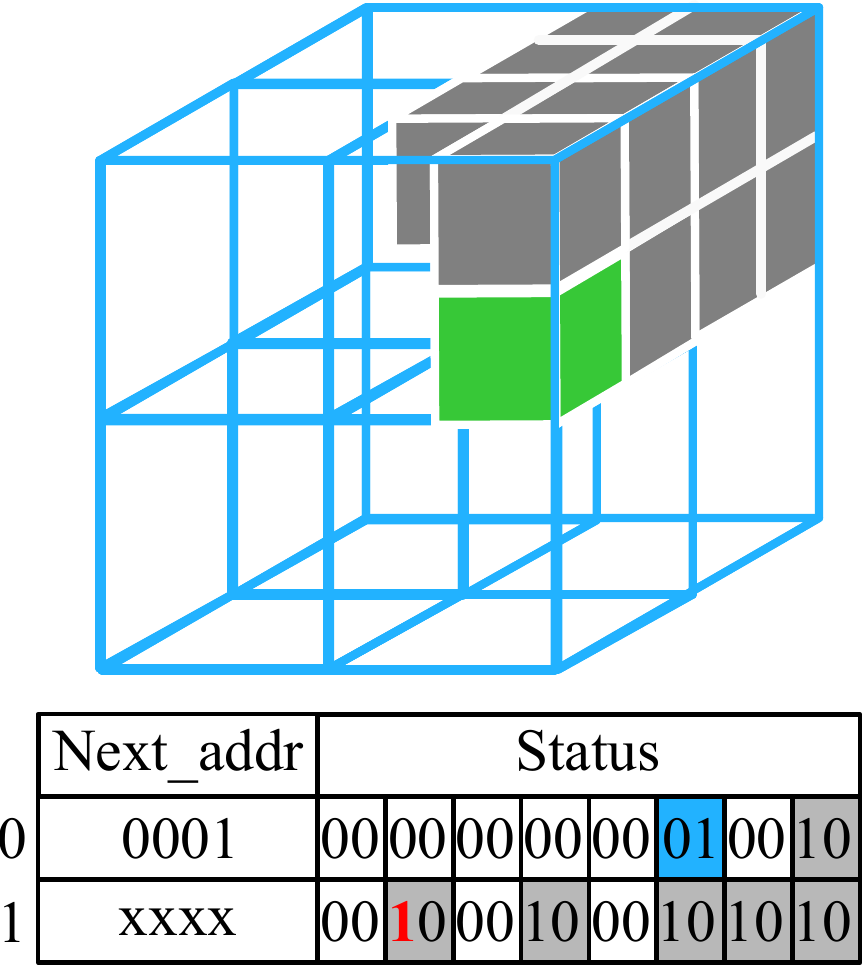}
      \caption{False-positive (CEF = $0$)}
    \label{fig:critical_false_positiveA3}
  \end{subfigure}
 
     \caption{Analyzing impact of bit flips in a CDM. (a) represents the swept space of a motion stored in CDM, (b)-(d) represent box-based CDM (A2), and (e)-(g) represent octree-based CDM (A3). CEF values are normalized to the surface area of one face of a voxel. } 
    \label{fig:error}
\end{figure*}
In this section we start by considering how erroneous collision-detection outcomes can lead to safety-critical events, 
then describe \textit{Collision Exposure Factor} (CEF) in detail,
and finally discuss how to apply CEF to building resilient MPAs.
%%\subsection{Silent Data Corruption for Collision Detection (SDC-C)}
%%\label{subsec:sdc}

Specialized accelerators, such as a CDM, obtain efficiency by replacing long sequences of software instructions with specialized hardware.
Such accelerators may perform computations under the supervision of a command processor via an ISA interface (e.g., Google's TPU~\cite{tpu}), 
and/or may start computation triggered by an event such as the arrival of a new frame of data in a buffer~\cite{eyeriss}. 
%\tor{Can we find an example of the latter to cite?}
%\deval{added deep learning inference engine example}
%\tor{I wrote this but re-reading it now I suppose a fault could propagate from the accelerator to the command processor.  Reviewers might get stuck on that point.  What was the original concern reviewers had about crashing?   Maybe we can just say while a fault could progress from accelerator to command processor our focus in this work is on the accelerator not the command processor (or something like that).   Could perhaps say that whether a fault propagates from an accelerator to command processor depends upon specifics of the interface.}
%\blue{
While an error originating in the accelerator could potentially propagate to the command processor and thereby result in erroneous operation (e.g., hang or system software crash)
%a hang, the system can be restarted. However, 
this paper focuses on Silent Data Corruption (SDC) within the CDM that can result in the robot colliding with an object in its environment. 
% which we refer to as SDC that may result in a Collision, or SDC-C.
%safety violation and requires protection.  
%}
%\deval{DSN 2020 reviewers had asked whether we measure only SDC related to a collision or also other SDC where a robot would crash or hang. I think we can safely say that we are focusing on SDC related to the collision as they require protection and constitute safety violations. It is possible that an SDC propagates to the command processor and results in robot hang, the system can be restarted.  }
%Hence an error in the datapath of such an accelerator does not generally result in a crash or hang of the command processor. 
%While an accelerator could be designed to signal an exception (e.g., a machine-check exception due to an ECC failure) to the command processor, we focus on SDC in this work 
%as SDCs can not be protected using low cost detectors and requires~\cite{Venkatagiri2016}. 
%\blue{as detectable errors do not require protection and can be found using low cost detectors~\cite{Venkatagiri2016}. }
%\blue{as low cost detectors can be used to find detectable errors~\cite{Venkatagiri2016}, whereas silent data corruptions need to be protected using error-mitigation techniques}
%\tor{because... ?}\deval{added}
 
Given a motion set and the current positions of obstacles collision detection finds motions that may lead to collisions.
%% XXX: Have we not pointed out the above already?
We define an \sdc{} as an event involving a false-negative result when performing collision detection for a proposed motion and an object
(i.e., a colliding motion is misidentified to be collision-free even though sensors detected an object the proposed motion would collide with if chosen during path search). 
Since, in general, an obstacle might appear anywhere in the environment, 
a naive (but expensive) approach to estimate \sdc{} probability is to simulate many sample environments, each with randomly placed obstacles. 
Below we consider a more efficient approach suitable, for example, during CDM architecture design.

\subsection{Collision Exposure Factor (CEF)}
\label{subsec:cef}
To analyze the errors that can occur in a CDM circuit and their effect on \sdc{} probability, we focus on bit changes that can lead to false-negative collision detection. 
Specifically, we consider the impact of a change in a single bit used to represent a portion of a motion's swept space. 
As mentioned in Section~\ref{sec:accelerator}, swept spaces of the motions are stored in CDM memory and used to find possible collisions with the obstacles. 
%%This on-chip storage accounts for more than $97\%$ of the sequential elements in the CDM circuits we study. 
Each bit in the on-chip storage helps specify the bounds of some motion's swept space. % depending upon the geometric representation.  
We define the \emph{critical space} of a bit as the region excluded from the swept space if that bit changes value due to a fault. 
This region represents a part of the swept space that can potentially lead to a false-negative collision detection. 
We then define the {\em collision exposure factor} (CEF) of a bit as the surface area of that bit's critical space exposed to obstacles. 
If the geometric representation uses voxelized swept spaces, the CEF can be normalized to the surface area of one face of a voxel.

For a given erroneous bit, it is more likely that an obstacle occupies its critical space and results in an \sdc{ }as the exposed surface area of this critical space increases. 
Hence, intuitively, the \sdc{ }probability of a bit is proportional to its CEF value (Section~\ref{sec:part2}). 
%\new{Add: why surface is better than volume}
Note that the CEF of a bit is {\em independent of the position of the obstacles in the environment}. 
The CEF captures the probability of obstacles appearing in critical space and decouples the effect of a soft-error and exact position of obstacles on \sdc{ }probability.   
%The CEF decouples the effect of a soft error and the exact position of obstacles on \sdc{ }probability by considering the probability of obstacles appearing in critical space. 
While the CEF definition assumes a uniform distribution of obstacles, it can also be extended to nonuniform distributions. 
For example, the CEF value can be scaled by the estimated probability of obstacles occupying the critical space for a nonuniform distribution of obstacles in the environment. 

Figure~\ref{fig:error} illustrates the critical space of specific bits due to single-bit errors for different geometric representations. 
Figure~\ref{fig:critical_base} reproduces Figure~\ref{fig:octree2}, and represents the swept space of a motion. % stored in the CDM. 
Geometric representation methods convert the swept space to a set of \emph{structures}, such as voxels (used in A1 described in Section~\ref{sec:bench}), boxes (used in A2), or octree nodes (used in A3 and A4). 
Each of these structures is encoded into bits and stored in the on-chip storage of the CDM. For example, a voxel structure is stored using its coordinates, a box structure is stored using the coordinates of the diagonal voxels, and an octree node structure is stored using a custom data structure described below.
A bit flip caused by a soft error would modify the space represented by the structure in different ways depending on the location of the bit in the structure and the geometric representation. 

Figure~\ref{fig:no_fault} shows a box-based representation of the swept space, where four boxes are required to cover the swept space. 
Since space is divided into four voxels in $x$, $y$, and $z$ directions, a total of six ($\text{log}_24 \times 3$) bits are used to store a coordinate. 
The highlighted box is represented with coordinates $(x1,y1,z1)$ and $(x2,y2,z2)$ of the diagonal voxels, before any error is introduced. 
In Figure~\ref{fig:critical_voxel}, flipping a specific bit causes the value of $y1$ to increase by one voxel. 
As the voxel highlighted in red is now excluded from the box, and no other box covers the voxel, it becomes part of the bit's critical space. 
%If an obstacle occupies this critical space, the motion should be detected as a potential collision. However, a soft error in this bit results in the exclusion of the critical space from swept space; hence the collision detection circuit fails to detect this collision, leading to an \sdc. 
A soft error in this bit results in the exclusion of the critical space from swept space; hence the collision detection circuit fails to detect a collision if an obstacle occupies this critical space, leading to an \sdc. 
In contrast, in Figure~\ref{fig:critical_no_error}, a bit flip causes $y2$ to decrease, but %no collision would occur as other boxes cover the voxels exposed. Hence, 
this bit flip is masked and does not impact the CDM's output as other boxes cover the voxels exposed. 
%%\red{This soft error does not impact architecturally visible state and does not affect the outcome. }

Figure~\ref{fig:critical_baseA3} shows the octree representation of the same swept space. 
In this representation, the root node and all partially occupied nodes are stored in memory using a tree data structure.
Each node in the tree contains two fields: ``{\tt status}'' and ``{\tt next\_addr}''. 
The {\tt status} field contains an entry for each octant within the node indicating whether that octant is empty, partially or fully occupied.
Only partially occupied octants are further divided.
The {\tt next\_addr} field contains the start address of an array holding the resulting children nodes. 
For example, in Figure~\ref{fig:critical_baseA3}, the node stored in memory at address~0 contains only one partially occupied octant, and the node containing information about it is stored at address $1$. % as next\_addr is $1$ for structure~0. 
In Figure~\ref{fig:critical_voxelA3}, a bit flip in the node at address~0 modifies the status of a particular octant to be ``fully empty'', thereby adding all the voxels in that octant to the critical space.
In Figure~\ref{fig:critical_false_positiveA3}, a bit flip in the node at address~1 modifies the status of an octant to ``fully occupied'', resulting in a false-positive rather than a false-negative. In this case, motions that would not lead to a collision may be disallowed. However, this false-positive outcome does not result in potential for a collision (an \sdc{}). Therefore, this voxel (highlighted in green) is not a part of the critical space.

\begin{algorithm}
  \caption{CEF measurement (Phase~1 FI)}\label{alg:cef}
  \textbf{Input:} Motion\_set, Swept\_data, Swept\_voxels, CDM;\\
  \textbf{Output:} bit\_info     = \{bitID: (CEF, Critical\_space)\};\\
  \begin{algorithmic}[1]
  \For{\text{Motion} $\in$ \text{Motion\_set}}\label{alg1:ms}
    \State bits = Swept\_data(Motion) \label{alg1:ss}
    \State voxels = Swept\_voxels(Motion)\label{alg1:4}
    \For{b $\in$ bits}\label{alg1:5}
        \State Critical\_space = $\varnothing$ \label{alg1:6}
        \State Collision\_vector = ${\text{FI}}$(CDM, bits, voxels, b)\label{alg1:FI}
        \For{(\text{v}, \text{collision}) $\in$ (\text{voxels}, \text{Collision\_vector})}\label{alg1:sv}
          %\State collision = ${\text{FI}}$(CDM, bits, v, b)\label{alg1:FI}
          \If{ $\neg$ collision } \label{alg1:if}
            \State Critical\_space = Critical\_space $\cup$ v \label{alg1:append}
          \EndIf
        \EndFor \label{alg1:endv}
        \State CEF = Calculate$_{\text{CEF}}$ (Critical\_space, voxels)\label{alg1:cefcal}
        \State bit\_info[b]= (CEF, Critical\_space) \label{alg1:7}
    \EndFor \label{alg1:endbits}
  \EndFor
  \end{algorithmic}
\end{algorithm}
For a given motion set of the robot and accelerator, the CEF of all the bits can be calculated using Algorithm~\ref{alg:cef}. 
The algorithm works by considering each possible motion in turn (Line~\ref{alg1:ms}).
For a given motion, {\tt Swept\_data} returns the storage bits used to represent its swept space (Line~\ref{alg1:ss}), and {\tt Swept\_voxel} returns the voxels in the swept space of that motion (Line~\ref{alg1:4}). 
The loop between Line~\ref{alg1:5} and \ref{alg1:endbits} considers each bit in the {\tt Swept\_data}. % of the motion.
%For each such bit, the CEF is computed by performing a fault injection for each voxel in the motion's swept space on Line~\ref{alg1:FI}.
The CDM takes precomputed swept space (i.e., {\tt bits}) and obstacle occupancy voxels %{\tt v} 
as inputs and performs collision detection (Figure~\ref{fig:cdm}).  
The storage elements of the CDM are initialized with {\tt bits}. 
To find the critical space an FI run is performed setting {\tt voxels} as the obstacle occupancy voxels input to the CDM (Line~\ref{alg1:FI}). 
Specifically, a fault is injected into a low-level (e.g., RTL or microarchitectural simulator) model of the CDM at bit {\tt b} (Line~\ref{alg1:FI}). 
The CDM outputs a {\tt Collision\_vector} containing collision detection output for each voxel in {\tt voxels}. 
To find the critical space, each bit of the {\tt Collision\_vector} is checked (Lines~\ref{alg1:sv}-\ref{alg1:endv}). 
For an error-free run, the CDM detects collision for all the voxels in the swept space (i.e., {\tt voxels}). 
%For each such bit, the CEF is computed by performing a fault injection for each voxel in the motion's swept space on Line~\ref{alg1:FI}.
However, for an FI run, a voxel {\tt v} is added to the {\tt Critical\_space} if the CDM does not detect a collision (Line~\ref{alg1:if} and \ref{alg1:append}). 
{\tt CEF} is then calculated by measuring the exposed surface area of the {\tt Critical\_space} (Line~\ref{alg1:cefcal}). 
%The {\tt Critical\_space} and {\tt CEF} are stored for use in Phase~2 (Line~\ref{alg1:7}).
Note that the CEF is obtained without the need to consider a potentially unbounded set of environments and object placements. 
{Thus the number of FI runs to measure CEF using Algorithm~\ref{alg:cef} is orders of magnitude lower than exhaustive FI in which for each bit, multiple FI runs need to be performed with a large number of environment scenarios. }

As presented above, Algorithm~\ref{alg:cef} assumes an RTL or architecture model for the CDM.  
We note that the resulting reliance upon fault injection to determine CEF could be avoided provided an analytical model is available to compute the critical space on Lines~\ref{alg1:6} to \ref{alg1:cefcal}. 
Such a model could be used to analyze CEF prior to the development of RTL or architecture simulators. 

Algorithm~\ref{alg:cef} is proposed for CDMs that use a voxelized representation of spatial data and output collision decisions (A1-A4). 
The majority of the collision detection acceleration approaches for robotics use voxelized representation as it consumes less memory and computational resources compared to polygonal mesh-based representation~\cite{Ericson}. 
%, and the proposed Algorithm~\ref{alg:cef} can be used. 
However, the concept behind CEF is applicable regardless of the underlying design parameters, and Algorithm~\ref{alg:cef} to measure the CEF needs can be modified for a different CDM. 
For example, for a triangle meshes-based representation of swept space, Line~\ref{alg1:6} to Line~\ref{alg1:cefcal} can be replaced by a geometry-based calculation of the exposed surface area. 

CEF is useful for CDMs that dedicate a significant area to storage structures for spatial data. Earlier work on FPGA-based accelerators~\cite{Murray} uses combinational logic to encode spatial data. Combinational logic is known to be less prone to soft errors compared to sequential elements~\cite{5173251}. However, the definition of CEF applies to erroneous combinational logic gates if needed.

\subsection{CEF-aware Error Mitigation}
\label{subsec:cefem}
For a fixed budget of area/power overhead, selective protection of the most vulnerable bits provides the optimal reduction in the FIT rate. 
Selective error mitigation in storage structures can be implemented by reliability-aware data placement to partially protected memory. 
However, this requires efficient ways to identify vulnerable data as the data can be generated/modified at runtime. 
%However, this requires fast and efficient detection of vulnerable data. 
%%Such techniques have been used for CPUs and GPUs~\cite{MehraraTACO, 1466436} by adding architecture or compiler supper for selective protection of data. 
%Existing techniques  efficient memory design~\cite{selectiveeccisca,7357094} and register file design~\cite{4272980,intelpatent} for selective protection. 
%, which is exploited by reliability-aware data placement. 
%%Similarly, other works have compiler and architecture support for reliability-aware selective placement to reduce error mitigation cost~\cite{MehraraTACO, 1466436} for CPU applications. 
We find that the CEF of a bit gives a measure of its impact on the \sdc{ }probability of the CDM (Section~\ref{sec:part2}). 
Therefore, we can use information about the CEF of each bit in an input design to selectively apply error mitigation techniques such as ECC, DMR/TMR, and strike suppression.
%redundant node latch hardening and  DMR/TMR. 
This reduces the cost of providing resilience compared to blanket protection of all the bits. 
%and is much more efficient than other selective protection techniques~\cite{Chen2005,Mehrara2008} (Section~\ref{sec:sol}).

%Selective mitigation is applied at the programming stage of the accelerator for a specific robot and motion set by CEF-aware data placement. 
Selective mitigation is applied by CEF-aware data placement to the on-chip memory of the accelerator. % by CEF-aware data placement. 
%The user is given an MPA with a certain amount of protected on-chip storage in the CDM. 
%In a configurable MPA, the user configures the CDM's on-chip storage with swept spaces of motions from a motion set customized to the robot and its tasks. 
Different structures such as voxel (A1), box (A2), and octree node (A3) are stored in the on-chip storage of the CDM depending upon the geometric representation method used. 
CEF-aware error mitigation is performed by placing structures with higher CEF in the protected storage regions. 
% for a given MPA with a certain amount of protected on-chip storage in the CDM. 
%%%Further, CEF-aware data placement is integrated with the MPA configuration step. 
In our evaluation, the sum of CEFs over all the bits in a structure is used to sort the structures in descending order of CEF for selective error mitigation. 
Because CEF of a bit is proportional to the \sdc{ }probability, CEF-aware data placement results in a higher reduction of the overall FIT rate for a given fraction of protected storage compared to other heuristics (Section~\ref{sec:sol}). 
%The CEF measurement is done offline using Algorithm~\ref{alg:cef} after motion set generation. 
The CEF measurement is done offline or outside of the critical path using Algorithm~\ref{alg:cef} after motion set generation. 
%This step is done offline and does not affect the performance of the accelerator. 

Selective error mitigation can also be used to guide the accelerator design process. 
At the accelerator design stage, the designer can perform CEF-aware FI (Section~\ref{subsec:methodcefawareFI}) for a set of target robots, motion sets, and the expected number of obstacles in the environment to find the distribution of CEF, \sdc{ }probability of bits, and the FIT rate of the accelerator without any error mitigation. 
Based on this information, a designer can determine the required fraction of protected memory for a given FIT rate requirement or achievable FIT rate for a given area/power budget for error mitigation. 
%\tor{This has been explained earlier in the intro but not the same words.}\deval{We need Phase 1 FI to find the exposed swept space and CEF of a bit, and Phase 2 FI to find the FIT rate. To determine how much reduction is required in the FIT rate for a given safety level, we need to determine the FIT rate of unprotected design. I have added text}
%Based on this information, a designer can judiciously apply error mitigation to on-chip storage elements for a given FIT rate requirement and area/power budget. 
As we show in Section~\ref{subsec:lh}, selective error mitigation reduces the required fraction of protected memory for a given FIT rate requirement compared to the complete protection of memory. 
\section{CEF-aware Reliability Characterization} 
\label{sec:porposed_solution}
\begin{figure}[t]
  \centering
  \includegraphics[width=0.47\textwidth]{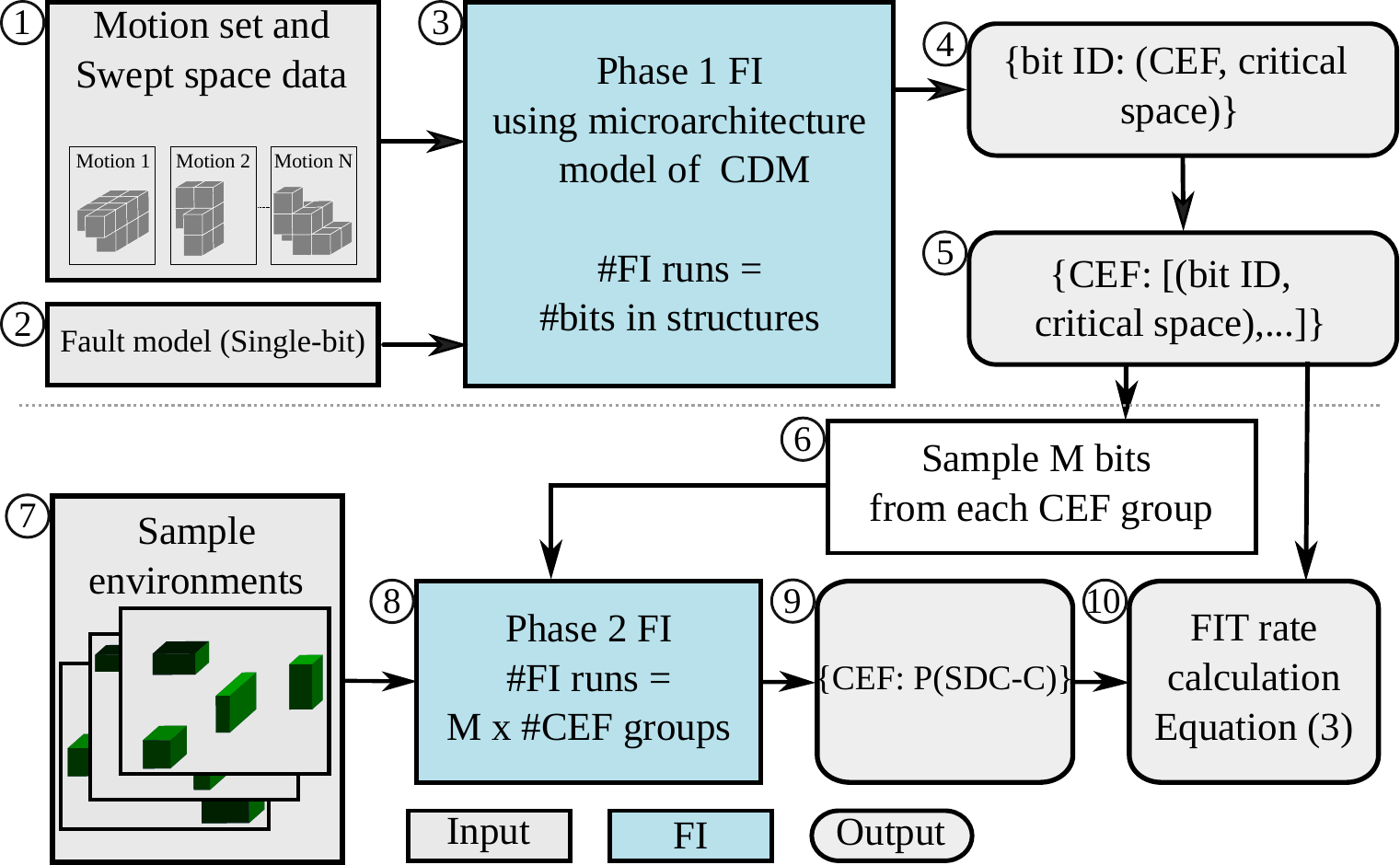}
  \caption{CEF-aware FI} 
  \label{fig:fimethod}
\end{figure} 
\sdc{ }probability and the FIT rate measurements are important to find the error mitigation requirements for certifiable safety. 
Exhaustive FI to find the \sdc{ }probability of {all the} bits takes a long time (Section~\ref{subsec:FI}). %\deval{all the or individual?}
In this section, we demonstrate how to use the CEF to enable a hierarchical fault analysis methodology that reduces the number of FI runs to measure \sdc{ } probability. 
\subsection{CEF-aware FI}
\label{subsec:methodcefawareFI} 
\begin{table*}[t!]
  \setlength\tabcolsep{3.2pt} 
  \caption{Comparison of different fault injection (FI) approaches.}
  \label{tab:compFI}
  \begin{tabular}{l|C{2.3cm}|C{3.4cm}|C{2.8cm}|C{2.8cm}|}
  \cline{2-5}
   & \multirow{2}{*}{Exhaustive FI} & \multirow{2}{*}{Uniform random statistical FI} & \multicolumn{2}{c|}{CEF-aware FI} \\ \cline{4-5} 
   &  &  & Phase~1 (CEF~measurement) & Phase~1 + Phase~2 (SDC-C measurement) \\ \hline
  \multicolumn{1}{|l|}{Description} & Performs FI on all combinations of bits and environment scenarios & Samples a few combinations of bits and environment scenarios to perform FI   & Performs environment-agnostic FI to measure CEF of all the bits (Algorithm ~\ref{alg:cef})  & Samples combinations of environment scenarios and bits and from each CEF group to perform FI  \\ \hline
  \multicolumn{1}{|l|}{Number of FI runs} & $10^{10} $ & $7 \times 10^6$ & $10^6$ & $7 \times 10^6$ \\ \hline
  \multicolumn{1}{|l|}{FI time} & $24,000$ hours & 2-4 hours & 1-2 hours & 2-4 hours \\ \hline
  \multicolumn{1}{|l|}{Measures overall FIT rate} & \cmark & \cmark & \xmark & \cmark (~2.5\% error) \\ \hline
  \multicolumn{1}{|l|}{\begin{tabular}[c]{@{}l@{}}Finds vulnerable bits for selective error  mitigation\end{tabular}} & \cmark & \xmark & \cmark (Approx.) & \cmark (Approx.) \\ \hline
  \multicolumn{1}{|l|}{Measures SDC probability of different bits} & \cmark & \xmark & \xmark & \cmark (Approx.) \\ \hline
  \end{tabular}
  \end{table*}
%%%%As mentioned (Section~\ref{subsec:FI}), exhaustive FI to find the \sdc{ }probability of {all the} bits takes a long time. %\deval{all the or individual?}
%Further, for a given MPA, the \sdc{ }probability of bits depends upon the motion set and the average number/size of obstacles in the environment. 
%Therefore, whenever there is a change in any of these factors, the FI needs to be redone from scratch. 
%Thus, it is essential to reduce the time taken for FI.
We propose a two-phase CEF-aware FI (shown in Figure~\ref{fig:fimethod} and described below), a technique to speed up FI for MPAs. 
%CEF-aware FI consists of two phases. 
%CEF-aware FI consists of two phases, as it divides the flow of error from the bit flip to erroneous collision detection into two parts.  
\subsubsection*{Phase 1: CEF measurement (Environment independent)}
\label{subsec:p1}
%\circled{1}
For a given motion set~\circled{1} and fault model~\circled{2}, Phase~1 performs microarchitecture- or RTL-level FI~\circled{3} to find the CEF and critical space of each bit in the swept space data. 
%For a given motion set~\circled{1} and fault model~\circled{2}, Phase~1 FI finds the CEF and critical space of each bit in the motion set swept space data using microarchitectural or RTL model of the CDM~\circled{3}. 
Algorithm~\ref{alg:cef} is used for this phase. 
%Phase~1 obtains the critical space and CEF for each bit in the CDM for a given motion set (\circled{1}) and the fault model (\circled{2}). 
%Phase~1 obtains the CEF for each bit in the CDM.  %While there may be more efficient ways to obtain this measure,
%%Note that the FI run does not require any information about the environment and obstacle as critical space is independent of these parameters.
%\tor{what about statistical FI?}\deval{removed that line }
%%\blue{Whereas, Phase~1 obtains the CEF of a bit. } 
The number of FI runs for this phase is limited to the number of bits in swept space data of all the motions in the robot's motion set. 
The CEF and critical space of all bits are stored to be used in the next phase~\circled{4}. 
Note that environmental information, such as the position of obstacles or the robot, is not needed for this phase, as CEF does not depend upon the environment. 
\subsubsection*{Phase 2: SDC-C probability measurement}
\label{subsec:p2}
A bit flip might lead to an SDC-C depending upon the position of obstacles as shown in Section~\ref{sec:cef}. 
Since for a dynamic environment, obstacles can appear anywhere in the space, a large number of FI runs with random environmental scenarios are required to measure the \sdc{ }probability of a bit with statistical significance. 
%A large number of random environmental scenarios are required for statistical significance of measured SDC-C probability. 
%Since the position of obstacles is dynamically changing; we need to perform FI with several environmental scenarios to measure the probability of a bit-flip resulting in SDC-C. 
In our experiments, the representative environment scenarios are generated using apriori information about the environment, such as the distribution and average number/size of the obstacles. % to find the average SDC. 

In Phase~2, we use the CEF information gathered in Phase~1 to speed up the fault analysis. 
As mentioned, there is a strong (positive) correlation between the CEF and \sdc{ }probability (Section~\ref{sec:part2}).  
Thus, we can speed up the FI experiments many-fold by performing FI for only a subset of bits with a given CEF value to measure the approximate \sdc{ }probability for all bits with the same CEF. 
%In Phase~2, we leverage this observation to speed up the FI experiments many-fold.  
The CEF information gathered in Phase~1 is used to group bits with equal CEF values together \circled{5}. 
Then, for each CEF value, $M$ bits are selected at random \circled{6}. 
Finally, using multiple sample environment scenarios~\circled{7} and FI simulations~\circled{8}, the \sdc{ }probability for each CEF value is measured \circled{9}. 
%%Multiple runs with a set of representative sample environmental scenarios \circled{7} is then used to measure the average \sdc{ }probability of a bit. 
Note that $M$ is a tunable parameter in the above heuristic. 
We use the analytical model in Leveugle et al.~\cite{Leveugle2013} to determine the value of $M$ to measure \sdc{ }probability with the required confidence level and error margin. 
%%As the value of $M$ increases, the measured \sdc{ }probability approaches its true value, and hence accuracy increases. 
%For exhaustive FI, where $M$ is equal to the number of bits in a CEF group, Equation~\eqref{eq:1} gives the true \sdc{ }probability value. 
The CEF-aware sampling, while being faster than exhaustive FI, may introduce inaccuracies due to the approximations it makes. 
However, the accuracy can be increased by increasing the value of $M$ - we evaluate this trade-off in Section~\ref{sec:cefawareFI}. 
%%%Thus, the CEF can be used to estimate the contribution of the different bits stored in the CDM to the overall FIT independent of the environment scenario.
%%In Phase~2, we use the CEF and critical space information gathered in Phase~1 to measure the \sdc{ }probability, i.e., probability of a soft error resulting in a collision (safety-violation).} % using the results of phase 1. 
%As we show in section~\ref{sec:part2}, \sdc probability of CDM depends upon the average density of obstacles in the sample environments. 
%%The \sdc{ }analysis process uses critical space information gathered in Phase~1 to find t. 
%%Thus, phase~1 and phase~2 together measure the impact of soft errors on the collision detection output.  

 To speed up the FI simulation, we further exploit the fault propagation in the studied CDMs.  
All the accelerators studied in this work use a geometric representation that divides the space into voxels. 
For a given motion and environment scenario, a non-empty subset of swept space and obstacle occupancy voxels signify potential collision. 
Thus the effect of a bit flip can be captured by storing the erroneous swept space voxels of the corresponding motion, instead of performing slow microarchitectural and RTL simulations for multiple environmental scenarios.  Simple set operations can be used to find the \sdc{ }probability. 

For a given bit flip, an FI simulation is used to find the erroneous swept space (i.e., set of voxels) using Algorithm~\ref{alg:cef}. 
Line~\ref{alg1:4} of the Algorithm~\ref{alg:cef} is modified to use the set of all environment voxels as {\tt voxels}. 
Similarly, Lines~\ref{alg1:if} and \ref{alg1:append} are modified to measure the erroneous swept space (i.e., colliding voxels). 
%%Erreneous swept space can be measured by setting full environment voxels as the obstacle occupancy voxels and perform FI. 
For each environment scenario, the subsets of obstacles occupancy voxels and erroneous and error-free swept space are measured. If the erroneous subset is empty, but the error-free subset is non-empty, the bit flip will result in an \sdc. 
%%Thus, instead of performing microarchitectural or RTL-level FI for multiple environmental scenarios, the effect of bit-flip can be captured using a single FI run. We use Python-based implementation to measure \sdc{ }probability.  
%%Thus, Phase~1 FI information can be reused in this phase to measure \sdc{ }probability, eliminating the need for an FI run for each environmental scenario. 
The same strategy is used for exhaustive, statistical, and CEF-aware FI. 
While we focus on FI to measure CEF and \sdc{ }probability,  we believe that an analytical model can be used to replace the multiple runs and find the relation between critical space and \sdc{ }probability. We defer this to future work. 
%%\deval{I believe that it is possible to build an analytical model to find the SDC probability for given critical space and obstacle distribution. I added a sentence for that, but I am not sure if this is the correct location. Should we mention that here or in conclusion?}

Equation~\ref{eq:1} is used to calculate the \sdc{ }probability of a CDM. Here $\text{bits}_{\text{CDM}}$ is the set of all bits stored in the CDM's on-chip memory.   
%$\text{Pr}_{\text{edge}}$ 
CEF($x$) is the CEF value of bit $x$, and $\text{P(SDC-C}_{\text{CEF}(x)})$ is the \sdc{ }probability for the CEF value $\text{CEF}(x)$.
%We calculate the $\text{SDC}_{\text{CDM}}$ for a robot as:
\begin{align} \label{eq:1}
  \text{P}(\text{\sdc}_{\text{CDM}}) \approx \frac{1}{|\text{bits}_{\text{CDM}}|} \times \left[  \sum_{x \in \text{bits}_{\text{CDM}}} \text{P}(\text{\sdc}_{\text{CEF}(x)}) \right] 
\end{align}
%\begin{align} \label{eq:1}
%  \text{P}(\text{\sdc}_{\text{CDM}}) \approx \frac{1}{|\text{bits}_{\text{CDM}}|} \times \left[  \sum_{x = \text{Min}(\text{CEF}_{\text{CDM}})}^{\text{Max}(\text{CEF}_{\text{CDM}})} \text{P}(\text{\sdc}_{x}) \times  |\text{CEF[bits]}==x|\right] 
%\end{align}
% pr_edge

Table~\ref{tab:compFI} compares the different phases of CEF-aware FI with exhaustive and uniform random statistical FI approaches for CDM fault characterization and error mitigation. 
We also use CEF and CEF-aware FI to analyze and compare the effects of microarchitectural design parameters of the CDM,  and to derive the principles of {\em resilience-aware} MPA design (Section~\ref{sec:arch}). 
%%, similar to prior work on performance and energy efficiency of MPAs~\cite{ Murray, sorin, Lian2018}. 

\subsection{FIT Rate Calculation}
\label{subsec:fitN}
%%%%The failure-in-time (FIT) rate of a circuit consisting of multiple components can be computed using Equation~\eqref{eq:3}~\cite{Mukherjee2003, Li2017}. 
%%%%\begin{equation} 
%%%%  \mbox{FIT} = \sum_{i \in \mbox{components}} \mbox{S}_{i} \times \mbox{SDC}_{i} \times \mbox{FIT}_{\mbox{\small Raw}}
%%%%  \label{eq:3}
%%%%\end{equation}
%%%%$\text{FIT}_{\mbox{\small Raw}}$ is the raw FIT rate defined in FIT/Mb units and depends upon multiple factors including technology node, ambient conditions, and elevation~\cite{Marc}. 
%%%%$\text{S}_{i}$ is the number (in Mb) of sequential elements/latches in component $i$. 
%%%%$\text{SDC}_{i}$ is the probability that a fault in component $i$ affects the output of the application. 
{We derived P($\text{SDC-C}_{\text{CDM}}$) in the previous section (CEF-aware FI). To calculate the FIT rate of a CDM, we modify Equation~\eqref{eq:3} as:}
\begin{equation} 
  \text{FIT}_{\text{CDM}} =  |\text{bits}_{\text{CDM}}| \times \text{P}(\text{SDC-C}_{\text{CDM}}) \times \text{FIT}_{\text{\small Raw}} \times N
  \label{eq:2}
\end{equation}
P($ \text{SDC-C}_{\text{CDM}}$) is the probability that a soft error results in an \sdc{ }(i.e., number of failures per soft error). 
Since $\text{FIT}_{\text{\small Raw}}$ gives the number of soft errors per billion hours per Mbit, the term ``P($ \text{SDC-C}_{\text{CDM}}) \times \text{FIT}_{\text{\small Raw}} $''  gives the number of failures per billion hours per Mbit due to soft errors.
Multiplying this term with the total number of bits gives the resultant FIT rate for the entire CDM. 

Equation~\eqref{eq:3} is based on the assumption that a soft error only affects a single execution of the application, and storage elements are reloaded between executions. 
%This is typically so for CPUs and GPUs.  
However, in the MPAs we study, the on-chip data is typically reused across multiple executions of the same application ~\cite{Han2016, sorin, Murray, daducd} to reduce the DRAM-bandwidth requirement and data movement. 
%%The DRAM-bandwidth requirement for real-time motion planning can be very high without on-chip inter-query data reuse. For example, as mentioned by ~\red{Lian2018}, a naive implementation without on-chip data resue requires 35GBps bandwidth, which is on the higher end for an edge device. ~\red{daducd} proposed processing in memory architecture to tackle this issue.  
%%In MPAs, the swept spaces of motions are stored in the on-chip buffers in the CDM.
%%If the CDM has sufficient storage space, the loaded data is used for multiple collision queries as the swept space information is not modified unless there is a change in the motion set. 
%Similar strategy is used in many deep learning accelerators as well to improve energy efficiency and performance~\ref{ref}. 
In such a case, a bit flip due to a soft error will persist in the buffer and affect multiple executions,  until the buffers are reloaded. 
Therefore, to calculate the effective FIT rate for MPAs with data reuse, we modify Equation~\ref{eq:3}  and add the term $N$ in Equation~\ref{eq:2}.
$N$ is the expected number of executions of the application before the bits are reloaded.
Thus, ``P($\text{SDC-C}_{\text{CDM}}) \times N $'' is the number of failures per soft error, and Equation~\ref{eq:2} yields the FIT rate for accelerators with %inter-query 
data reuse.
\section{Experimental Methodology}
\label{sec:method}
\subsection{Experimental Setup}
\label{subsec:setup}
\begin{table}[b]
    \scriptsize
    \setlength\tabcolsep{2.2pt} 
    \caption{Robots used for fault characterization.}
    \label{tab:robots}
    \begin{tabular}{|C{4.1cm}|C{1.2cm}|C{1.7cm}|C{1.2cm}|}
    \hline
    Robot & Degrees of freedom & Reach in one direction &Mechanical power (W)\\ \hline
    {\color[HTML]{000000} Kinova Jaco2~\cite{jaco}} & 7 & 90 cm & 25\\ \hline
    {\color[HTML]{000000} Programmable Universal Machine for Assembly (PUMA) 761~\cite{puma}} & 6  & 150 cm & 30\\ \hline
    AL5D~\cite{al5d}& 4 & 27 cm & 12\\ \hline
    \end{tabular}
    \end{table}
Table~\ref{tab:robots} summarizes the robots used in our experiments. 
%We use three robots summarized in Table~\ref{tab:robots} in our experiments.
%: (1) Kinova Jaco2, (2) Programmable Universal Machine for Assembly (PUMA) 761, and (3) AL5D. 
% robotic arms. 
These robotic arms are representative of widely used industrial robots~\cite{jaconews,pumanews}, and are also included in larger humanoid robots~\cite{jacowheele}. 
Because we did not have access to the real robots, we use the Klampt~\cite{klampt} software simulator to simulate the robot's movements and the environment- this has also been used in prior work~\cite{7139603,sorin,Murray2016ori}. 

The environment size for a robot is determined by its reach, and the environment is discretized into a grid of $32 \times 32 \times 32$ voxels. 
For each robot, we generate a motion set with $16,384$ poses and $32,768$ motions, using Leven and Hutchinson's strategy~\cite{leven}, similar to prior work~\cite{Murray2016ori}. 
We use uniform motion sets in all our experiments, i.e., poses are distributed uniformly in the C-space. 
%%%%However, we also evaluate CEF on a nonuniform motion set in Section~\ref{sec:nonuni}, which can be used either if the robot needs to perform specific tasks that require more samples of motions in some part of the space %\karthik{This is not clear, but I guess you'll explain it later}\deval{added the next line}
%%%%, or if there are fixed obstacles in the space~\cite{Murray2016ori}. For example, for a robotic arm that picks and places objects in an assembly line, information about the workspace can be used to sample more motions in the more frequently visited region of the space (e.g., placement location)~\cite{1307191}. 
\iffalse
\begin{figure}[t]
    \centering
    \includegraphics[width=0.3\textwidth]{figures/bm4.pdf}
    \caption{Example of an environment for Jaco2 Arm with 4 obstacles.}
    \label{fig:bm4}
\end{figure}
\fi

To measure the \sdc{ }probability, we perform FI with a set of $10,000$ random environment scenarios.  %with randomly placed obstacles. 
Each sample environment contains $3-12$ cuboid-shaped randomly placed obstacles, and the length/height/width of each obstacle is $5-20$cm, which is consistent with other work on motion planning and collision detection~\cite{Murray2016ori,Lian2018}. %\karthik{How did you come up with these numbers ?}\deval{I added references.} 
%%Each environment scenario is generated by the random placement of obstacles in the environment. 
We generate four sets of environments to study the effect of the number of obstacles on the  \sdc{ }probability (Section~\ref{sec:part2}). 
We use the label D{\em $x$} ($x \in [1,2,3,4]$) to represent a set of $10,000$ environment scenarios, 
where obstacles occupy an average $x\%$ volume of the environment. 
The average number of obstacles increases with the value $x$. 
%While the CEF definition assumes a uniform distribution of obstacles; it can also be extended to nonuniform distributions. 
%For example, currently, we define the CEF as the exposed surface area of the critical space of a bit. The CEF value can be scaled by the estimated average probability of obstacles occupying the critical space for the nonuniform distribution of obstacles in the environment. 
%\karthik{Can we say how?}\deval{added}

\subsection{Collision Detection Modules (CDM)} % accelerator
\label{sec:bench}
%\subsection{Synthesis of Storage Elements}
%\label{sec:syn}
\begin{table}[t]
    \scriptsize
    \setlength\tabcolsep{2.6pt} 
    \caption{Accelerators studied. We list power and area for each accelerator from the paper and report suitable error mitigation and accelerator area overhead to protect on-chip memory. We use information gathered from our synthesis of RTL models about the fraction of CDM area and power consumed by the on-chip memory  to calculate the overall CDM area and power overheads. }
    \begin{tabular}{|L{1.4cm}|C{1.0cm}|C{0.5cm}|C{0.7cm}|C{0.6cm}|C{0.6cm}|C{1.2cm}|C{1.2cm}|}
    \hline
    Accelerator  & Representation  & Data reuse & \#CDCs &Power (W) & Area (mm$^2$) &  ECC area/power overhead & TMR area/power overhead\\ \hline
        A1 \cite{Murray, sorin}  & Voxel & Yes    &32,768 &  N/A & N/A & 900/720\%& \textbf{125/100}\%\\ \hline
        A2 \cite{Murray, sorin}    & Box & Yes & 32,768  &  35 & 400 & 900/720\%& \textbf{125/100}\%\\ \hline
        A3 \cite{Lian2018} & Octree    & No  & 128 &  0.47 & 1.7 & \textbf{12/9}\%& 100/75\%\\ \hline
        $\text{A3}_{\text{scaled}}$  \cite{Lian2018}   & Octree   & Yes & 32,768 & 121 & 450 & \textbf{12/9}\%& 100/75\%\\ \hline
        A4 \cite{daducd} & Flattened Octree    & Yes  & 32,768 & 20  & -  & \textbf{12/12}\%& 200/200\%\\ \hline
    \end{tabular}
    \label{tab:accel}
    \end{table}
Table~\ref{tab:accel} summarizes the four accelerators studied in this work and overll CDM area/power overheads for complete protection of on-chip storage using ECC and TMR error mitigation techniques. {\em These constitute the only published work on ASIC-based programmable, real-time accelerators for motion planning and collision detection, to the best of our knowledge. } 

We had to build microarchitectural simulators and RTL models of these CDMs ourselves as there was no existing simulator.  
%(we will publicly release the simulator, RTL models, and FI framework if accepted). 
We synthesized our RTL models using the Synopsys Design Compiler~\cite{synopsys} and the OpenRAM Memory Compiler~\cite{openram} to estimate the area and power of storage elements in CDMs at 45nm technology (FreePDK45 design library ~\cite{freepdk}). Because we are interested in the relative area and power consumption of different storage elements and combinational circuits, %\karthik{can we provide some citation for this?} 
the technology node's choice should not significantly impact the results. 
%\deval{I tried to look for some papers that study technology scaling vs area for combinational and sequential circuits but could not find it. Now I will try to look for some architecture papers that have made the same assumption}
%As mentioned in Section~\ref{sec:porposed_solution}, we study four CDMs~\cite{Murray2016ori,Murray,Lian2018,daducd}. 

\textbf{\underline{A1 (Base accelerator):}}
This architecture was proposed by Murray et al.~\cite{Murray}, and is based on the earlier proposed accelerator for FPGAs \cite{Murray2016ori}. 
A motion's swept space is stored in registers using the 3D Cartesian coordinates of each voxel in the swept space. 
Figure~\ref{fig:cdc1} and \ref{fig:cdc2} show how a swept space is converted to voxels. % and stored in A1.
The collision detection circuit compares the obstacle occupancy voxels with each voxel in the swept space to find if the motion is in collision with obstacles. 

\textbf{\underline{A2 (Spatial locality-aware accelerator):}}
This architecture, proposed by Murray et al.~\cite{Murray} and Sorin et al.~\cite{sorin}, is an optimization of A1.
There is a significant degree of spatial locality in the voxels in a swept space; hence, contiguous voxels are merged into a larger box. 
A box can be represented by the coordinates of two diagonal voxels. 
Figure~\ref{fig:cdc3} gives an example.

%%In both A1 and A2, swept spaces of all the motions in the motion set are stores in on-chip registers, and the on-chip data is reused across multiple collision queries. 
In A1 and A2, all on-chip registers are read in parallel by the collision detection logic, requiring a separate ECC decoder circuit for each register. Hence, the ECC area/power overheads are significantly high for A1 and A2 in Table~\ref{tab:accel}. 
%%A1 and A2 are designed with $32,768$ CDCs and can be used with motion sets consisting of less than or equal to $32,768$ motions. 
%the number of CDCs is the same as the number of motions in the motion set. Therefore, the swept space data is loaded in the on-chip storage elements once and is reused across multiple collision detection queries. 

\textbf{\underline{A3 (Octree-based accelerator):}}
This architecture was proposed by Lian et al. (2018)~\cite{Lian2018}, and uses the octree structure to store the motion's swept space (explained in Section~\ref{sec:cef}). %Octrees are widely used to represent 3D space and facilitate constant-time queries for collision detection \cite{JACKINS1980249}.
%%Figure~\ref{fig:octree} illustrates a voxelized swept space stored using an octree. Each entry in Figure~\ref{fig:octree3} represents a node in the octree. 
%%Each level of the tree divides the space into octants recursively. 
%%The depth of the tree determines the space representation's resolution; for example, a 5-level tree represents $8^{5}$ voxels in 3D space. 
Collision detection is performed by traversing the tree to find if obstacle occupancy voxels overlap with the swept space. % and requires a maximum of 5 read accesses to the table for a 5-level tree.

The proposed design of A3 uses $128$ CDCs, where motions in the motion set are processed for collision detection in batches. Hence, there is no inter-query on-chip data reuse, which results in significant DRAM bandwidth requirement~\cite{daducd}.  
%%It is designed with limited on-chip memory to store the swept spaces of only $128$ motions, and the on-chip memory data is written for each batch of motions within a single query. Hence, there is no on-chip data reuse across queries in A3. 
For comparison, we also study a scaled-up version of A3, called A3$_{\text{scaled}}$, where the number of CDCs is equal to A1 and A2 ($32,768$), and on-chip data is reused across multiple collision queries, reducing DRAM memory traffic.  
%%Further, the scaled-up version of A3 is more suitable for real-time motion planning with larger motion sets for robots that need to navigate in a cluttered environment or have more degrees of freedom.  
While the \sdc{ }probabilities for both A3 and A3$_{\text{scaled}}$ are equal, their FIT rates are different due to differences in the sizes of their components and the value of $N$ (Equation~\ref{eq:2}).

\textbf{\underline{A4 (Flattened octree-based accelerator):}}
This architecture was proposed by Yang et al. (2020)~\cite{daducd}, and proposes processing-in-memory for collision detection with a flattened octree-based representation of the swept space. %Octrees are widely used to represent 3D space 
In the flattened octree-based representation, multiple levels of the trees can be flattened in a single level. For example, if all the levels of a 5-level tree are flattened, the resultant tree consists of a single root node with $32,768$ children, where each child node is $1$ or $0$ specifying occupancy of a single node. 
Such representation consumes more storage but facilitates efficient processing-in-memory, reducing data-movement overhead significantly.  

\subsection{Fault Injection (FI)}
\label{subsec:methodFI}
%%We perform FI on the three MPAs listed in Table~\ref{tab:accel}. 

We use a Python-based implementation of microarchitectural simulators for FI, as RTL-level FI was very slow. 
We use Dell EMC R440 CPU nodes. 
However, to validate the FI accuracy of our microarchitectural simulators, we performed RTL-level FI with the Cadence Incisive Functional Safety Simulator~\cite{cadence}). 
{\em We found that FI using our microarchitectural simulator exhibited $100\%$ correlation with the RTL-level FI. }

We represent soft errors as single-bit flips in hardware registers, which is consistent with most other papers studying the effects of soft errors \cite{Ayatolahi2013, 8809532, 8416519,  Bo2016}. 
While we focus on single-bit flips, Equation~\eqref{eq:1} can be extended to accommodate a multi-bit fault model. 
We defer this to future work. 

For CEF-aware FI, we determine the value of $M$ (number of samples) per CEF group required to measure the \sdc{ }probability with $95\%$ confidence level and $2.5\%$ error margin~\cite{Leveugle2013}. 
The \sdc{ }probability and population size of the highest CEF group are used to determine the value of $M$, as these bits contribute the most to the overall \sdc{ }probability. 
In our experiments, the value of $M$ is $2\times10^{6}$ for A1 and A4, $2\times10^{5}$ for A2, and $15\times10^{3}$ for A3.

\subsection{FIT Rate Calculation}

To calculate the FIT rate of the accelerators, we use Equation~\eqref{eq:2}, where $N$ is the expected number of executions of the application before the bits are reloaded in on-chip storage, and its value depends on the accelerator's architecture and deployment. 
Reloading of data can be overlapped with the collision detection to hide the latency of DRAM accesses. 
%%One real-world example is where collision queries are performed at 60 Hz frequency~\cite{Jha2019, DRL} to detect the dynamic obstacles, and the robot continuously operates for one minute. 
We set the value of $N = 3600$ for the accelerators that exploit inter-query data reuse (A1, A2, A3$_{\text{scaled}}$, and A4), as the power overhead of DRAM accesses is within $1\%$ for refreshing data after $3600$ collision detection queries.
Similarly, $N = 1$ is used for the accelerator that does not reuse data across queries (A3). 
We use $\text{FIT}_{\text{Raw}}$ $= 20.49$ FIT/Mb (for 16nm CMOS~\cite{Li2017}) in Equation~\eqref{eq:2}. 
We choose a 16nm technology node, which is used by all the accelerators studied. Note that while the choice of N and $\text{FIT}_{\text{Raw}}$ affect the absolute values of the FIT rate, they affect neither our fault characterization nor CEF-aware error mitigation. 
%\karthik{Can we say why 16nm is reasonable.}
%\deval{added}

\section{Results}
\label{ref:result}
In this section we present our results for fault characterization (Section~\ref{sec:eval}) and error mitigation (Section~\ref{sec:sol}) using CEF.
\subsection{Fault Characterization}
\label{sec:eval}
\subsubsection{CEF-aware FI}
\label{sec:cefawareFI}

\begin{figure}[]
     \centering
     \begin{subfigure}[b]{0.22\textwidth}
         \centering
         \includegraphics[width=0.9\textwidth]{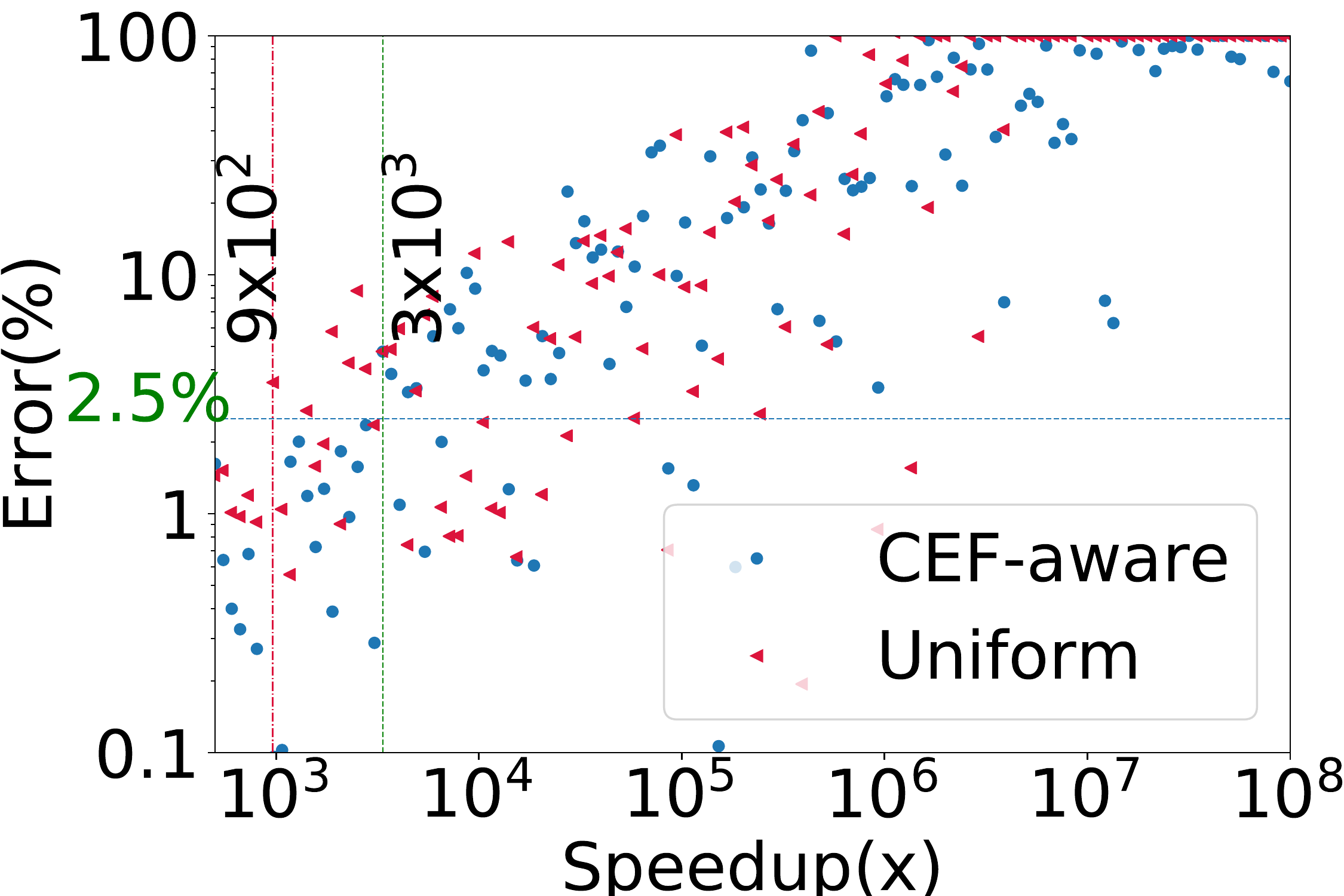}
         \caption{A1 (Voxel-based)}
         \label{fig:a1-ceffi}
     \end{subfigure}
     \begin{subfigure}[b]{0.22\textwidth}
         \centering
         \includegraphics[width=0.9\textwidth]{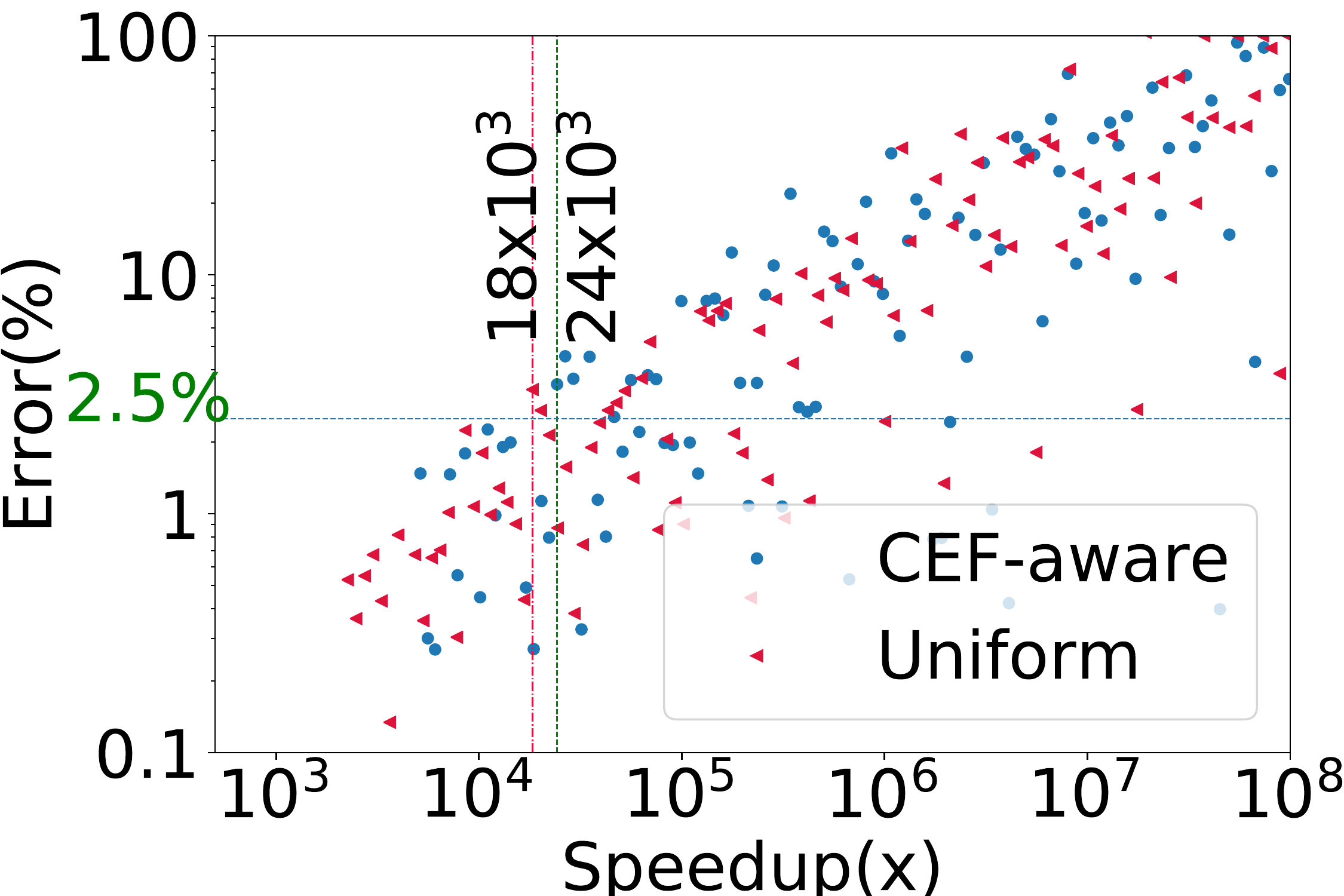}
         \caption{A2 (Box-based)}
         \label{fig:a2-ceffi}
     \end{subfigure}
     \begin{subfigure}[b]{0.22\textwidth}
         \centering
         \includegraphics[width=0.9\textwidth]{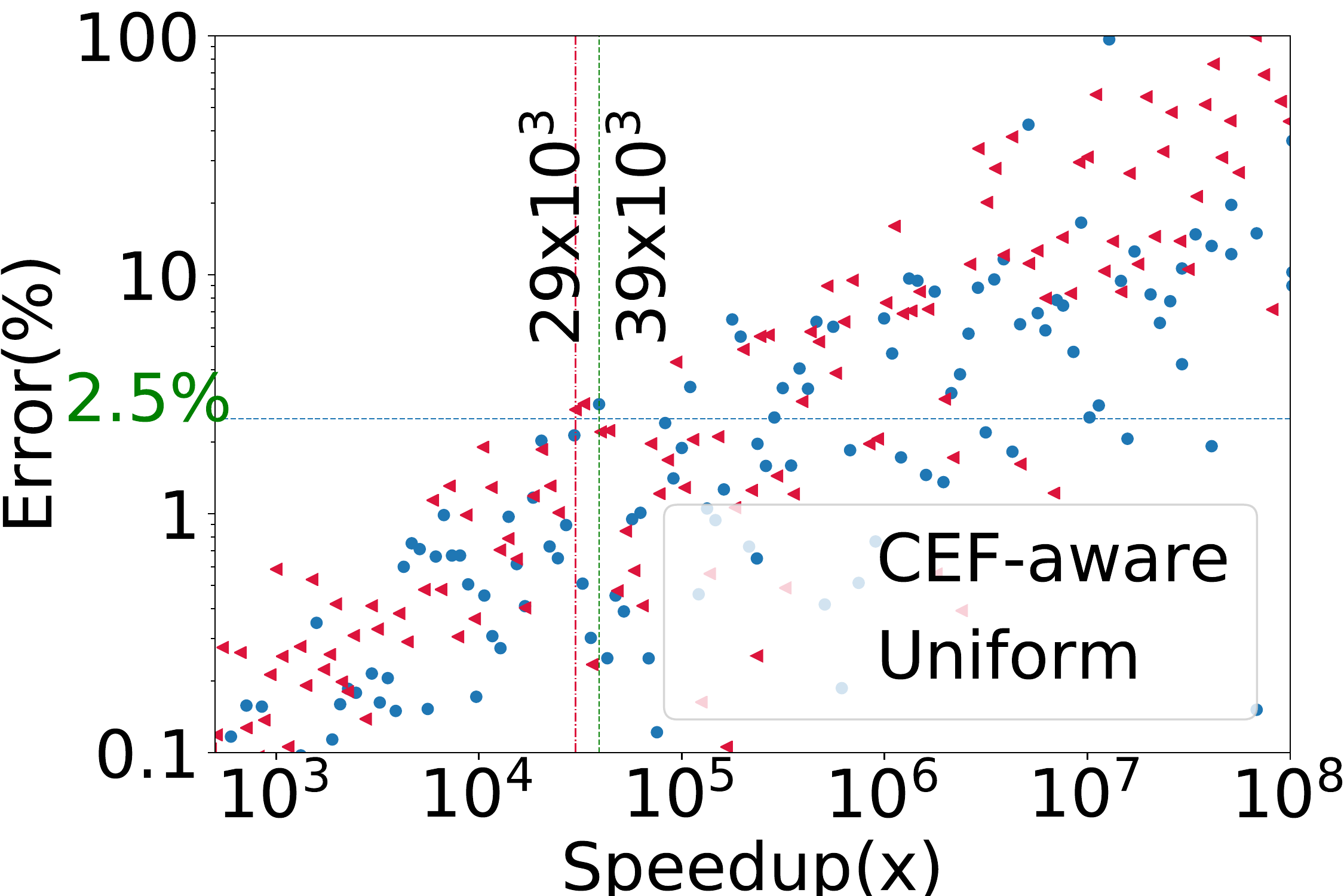}
         \caption{A3 (Octree-based)}
         \label{fig:a3-ceffi}
     \end{subfigure}
     \begin{subfigure}[b]{0.22\textwidth}
        \centering
        \includegraphics[width=0.9\textwidth]{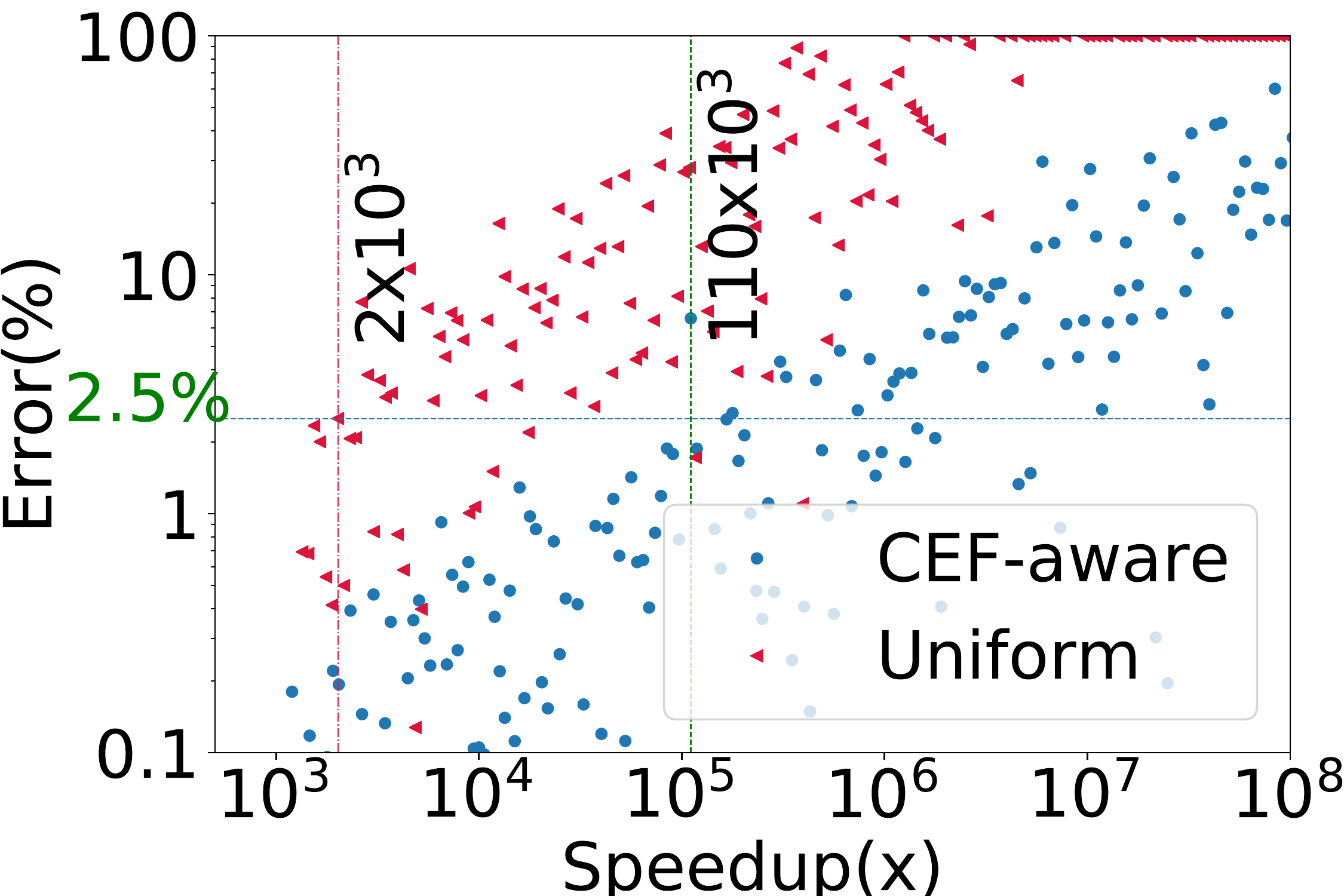}
        \caption{A4 (Flattened octree-based)}
        \label{fig:a3-ceffi}
    \end{subfigure}
        \caption{Error versus speedup for different FI approaches.}
        \label{fig:ceffi}
\end{figure}

In Section \ref{sec:porposed_solution}, we propose two-phase CEF-aware FI to reduce the number of FI runs. 
In CEF-aware FI, instead of performing FI for all the bits for multiple environment scenarios, we sample a subset from a group of bits with the same CEF value. 
This sampling introduces inaccuracies in the measured \sdc{ }probability and FIT rates compared to exhaustive FI. %, which performs FI on all the bits. 
Figure~\ref{fig:ceffi} shows the speedup versus error of the calculated FIT for CEF-aware FI and uniform statistical FI. 
The speedup is the ratio of the number of exhaustive FI runs to that of CEF-aware FI or uniform statistical FI. 

As can be seen from the figure, on average, CEF-aware FI achieves $23,000\times$ speedup over exhaustive FI (geometric mean) with $2.5\%$ error margin. The vertical line represents speedup for error less than $2.5\%$. The error margin can be reduced further at the cost of more FI trials (Section~\ref{sec:porposed_solution}). 
Uniform statistical FI exhibits a similar speedup as CEF-aware FI over exhaustive FI.  
Note however that uniform statistical FI cannot be used to find vulnerable bits for selective error mitigation (Table~\ref{tab:compFI}).
%CEF-aware FI results in $\sim30\%$ speedup compared to uniform statistical FI for the same error margin. 
The speedup of CEF-aware FI over exhaustive FI is due to two reasons: (1) a significant fraction of bits have CEF equal to $0$, and hence have very low or $0$ \sdc{ }probability. CEF-aware FI ignores these bits for FI, (2) Only a few bits have high CEF and \sdc{ }probability, and hence  contribute the most to overall \sdc{ }probability. CEF-aware FI segregates such bits and requires fewer samples to measure \sdc{}. 

Phase~1 of CEF-aware FI consumes $1$, $1$, $2$, and $1$ CPU hours for A1, A2, A3, and A4, respectively. 
Phase~2 of CEF-aware FI consumes less than $2$ CPU hours for all accelerators. In contrast, as per our experiments, exhaustive FI takes $24,000$, $18,000$, $22,000$, and $20,000$ CPU hours for A1, A2, A3, and A4, respectively.
%\karthik{Is this an estimate or did we measure it?}\deval{we measured, added that}
\subsubsection{Evaluation of reliability metric}
\label{sec:part2}
As discussed in Section \ref{sec:porposed_solution}, we propose the CEF as a reliability metric for the MPA's bits.
%%To be useful, the \sdc{ }probability for the workloads should be positively correlated with the CEF metric. 
We study the relationship between the CEF and the \sdc{ }probability to determine if they are indeed positively correlated to demonstrate the validity of CEF metric. 

\begin{figure}[]
     \centering
     \begin{subfigure}[b]{0.157\textwidth}
         \centering
         \includegraphics[width=\textwidth]{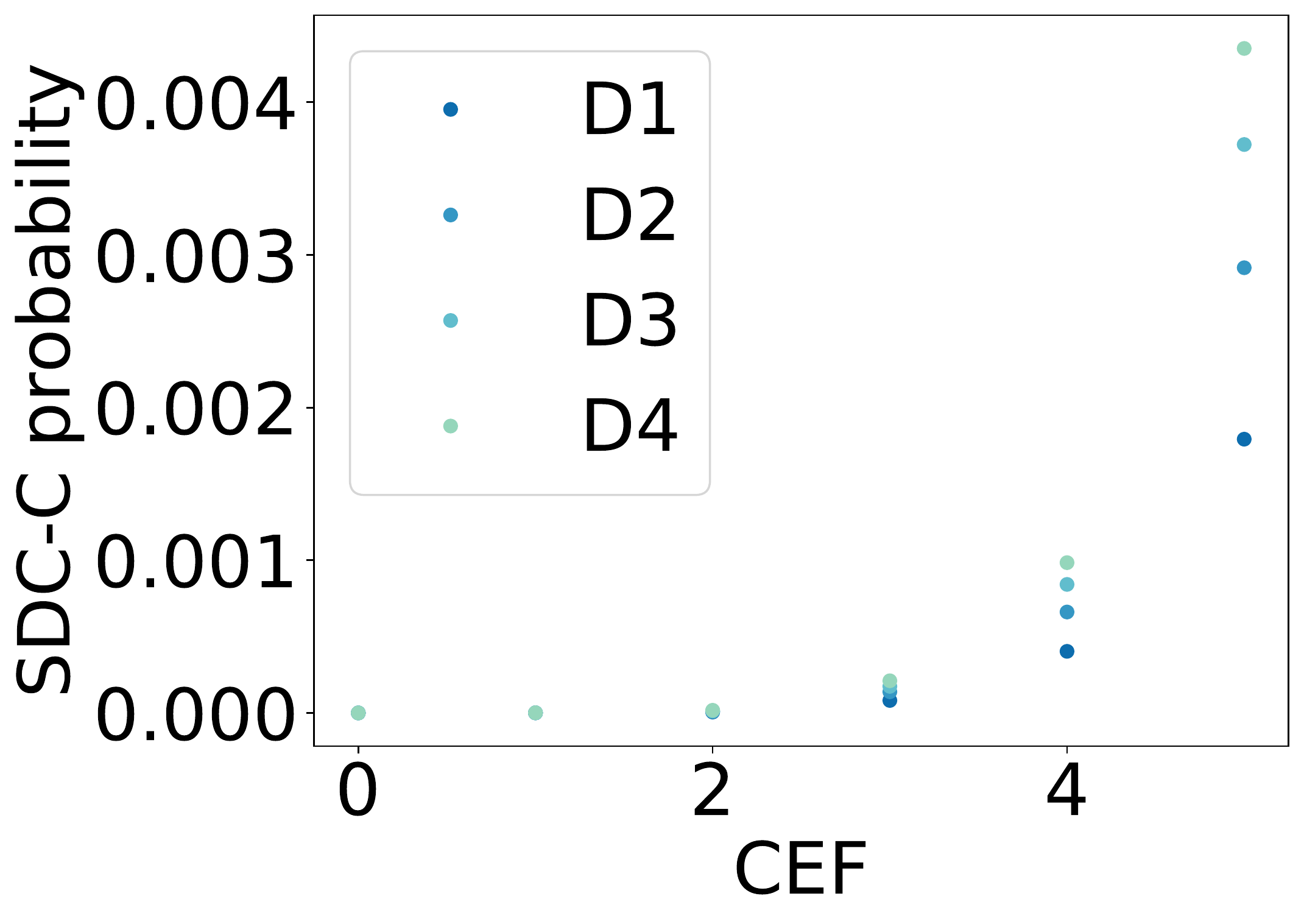}
        \caption{Jaco2:A1}
        \label{fig:bm_box}
     \end{subfigure}
     \hfill
     \begin{subfigure}[b]{0.157\textwidth}
        \centering
        \includegraphics[width=\textwidth]{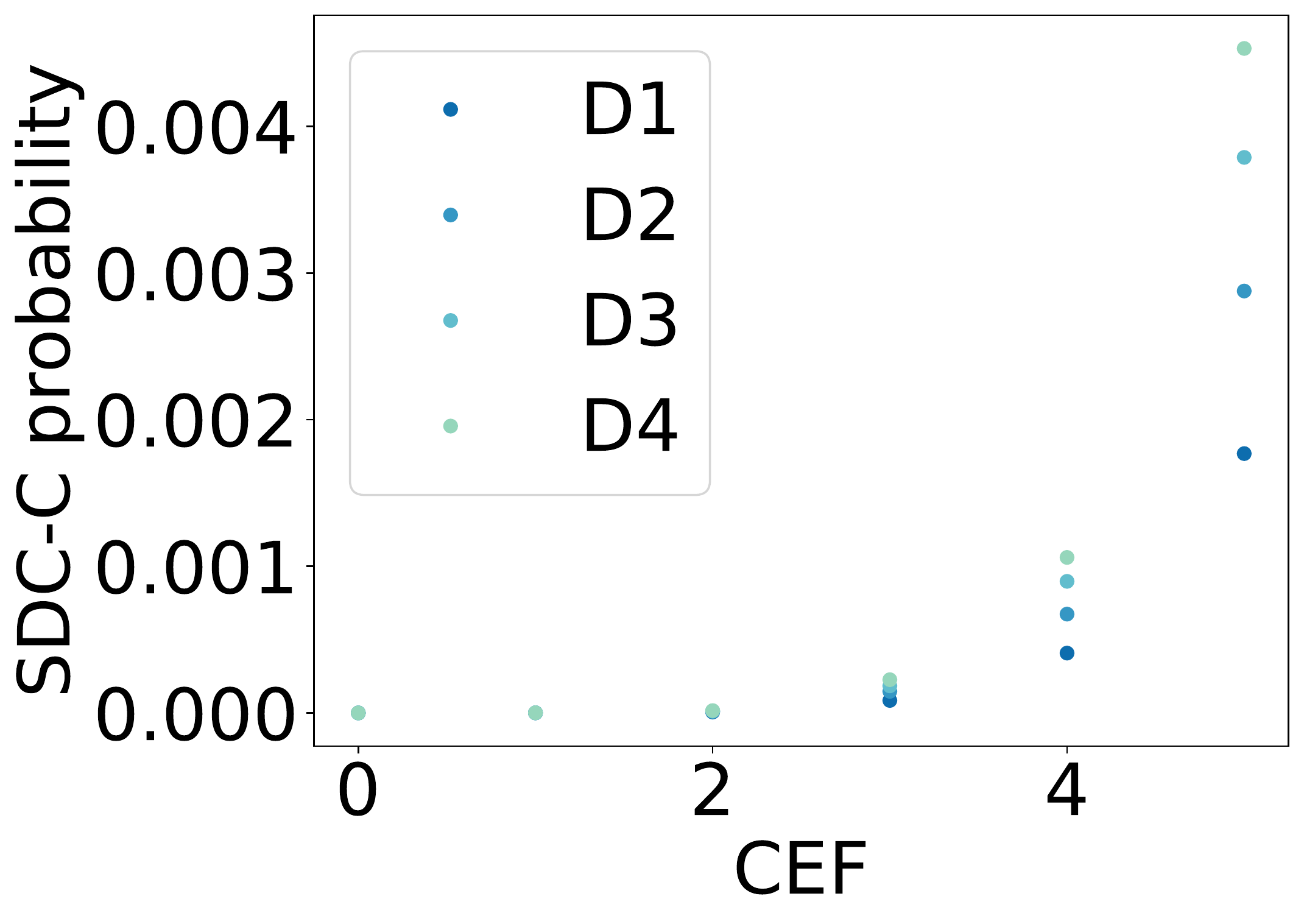}
       \caption{AL5D:A1}
       \label{fig:bm_box}
    \end{subfigure}
    \hfill
     \begin{subfigure}[b]{0.157\textwidth}
        \centering
        \includegraphics[width=\textwidth]{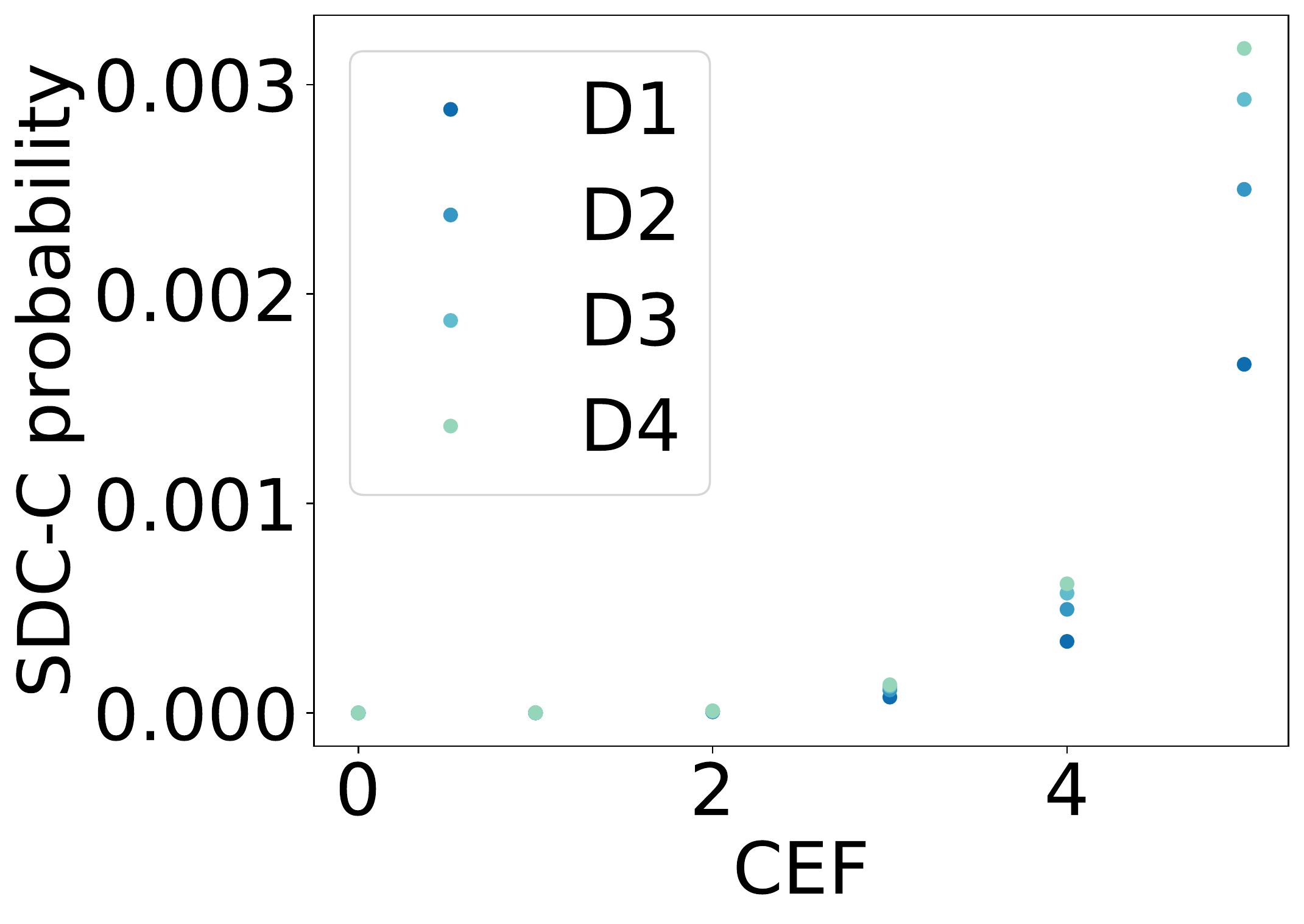}
       \caption{{Puma761:A1}}
       \label{fig:bm_box}
    \end{subfigure}
    \hfill
     \begin{subfigure}[b]{0.157\textwidth}
         \centering
        \includegraphics[width=\textwidth]{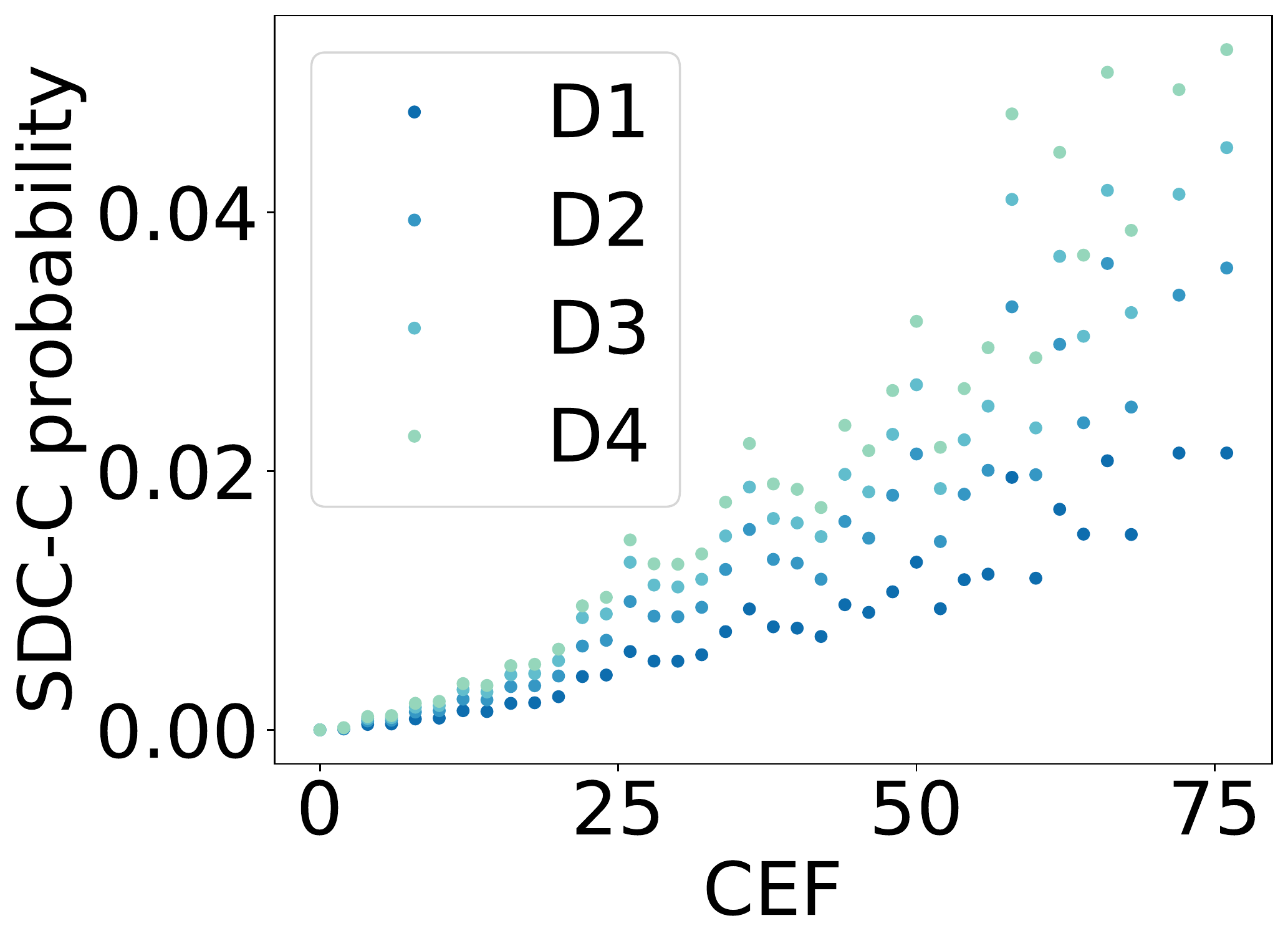}
        \caption{Jaco2:A2}
        \label{fig:bm_octree}
     \end{subfigure}
     \hfill
     \begin{subfigure}[b]{0.157\textwidth}
        \centering
       \includegraphics[width=\textwidth]{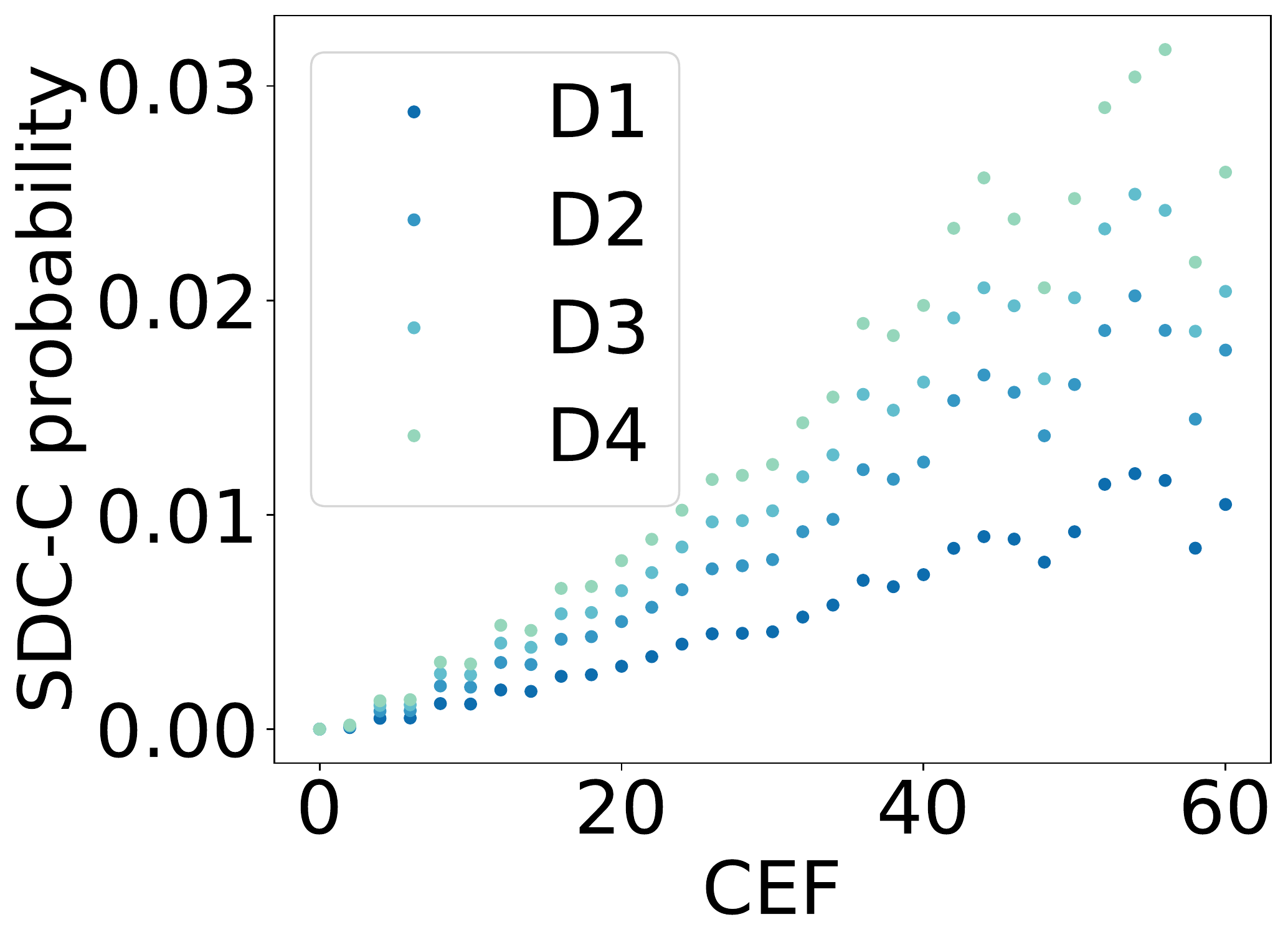}
       \caption{AL5D:A2}
       \label{fig:bm_octree}
    \end{subfigure}
    \hfill
    \begin{subfigure}[b]{0.157\textwidth}
        \centering
       \includegraphics[width=\textwidth]{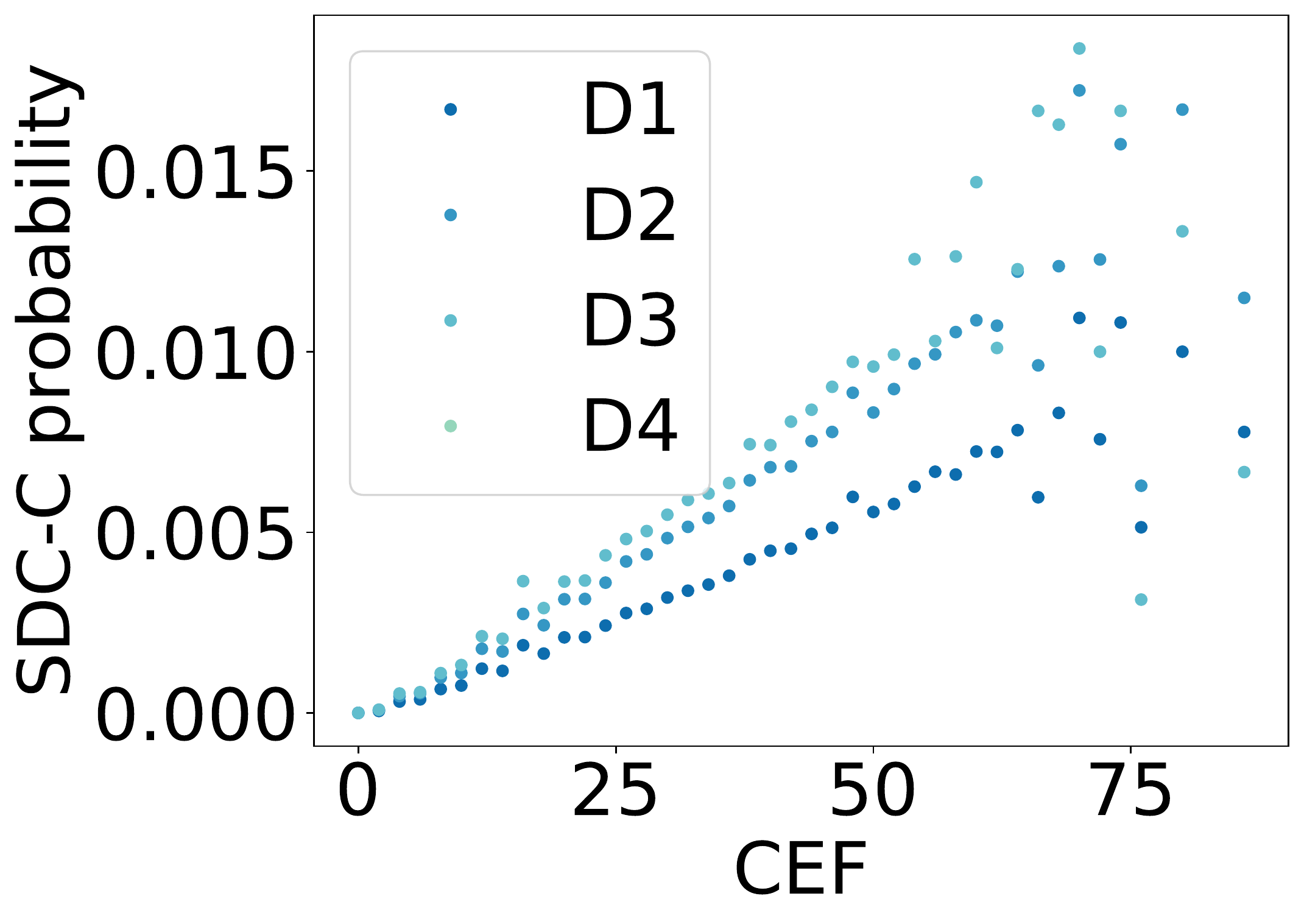}
       \caption{{Puma761:A2}}
       \label{fig:bm_octree}
    \end{subfigure}
    \hfill
     \begin{subfigure}[b]{0.157\textwidth}
         \centering
        \includegraphics[width=\textwidth]{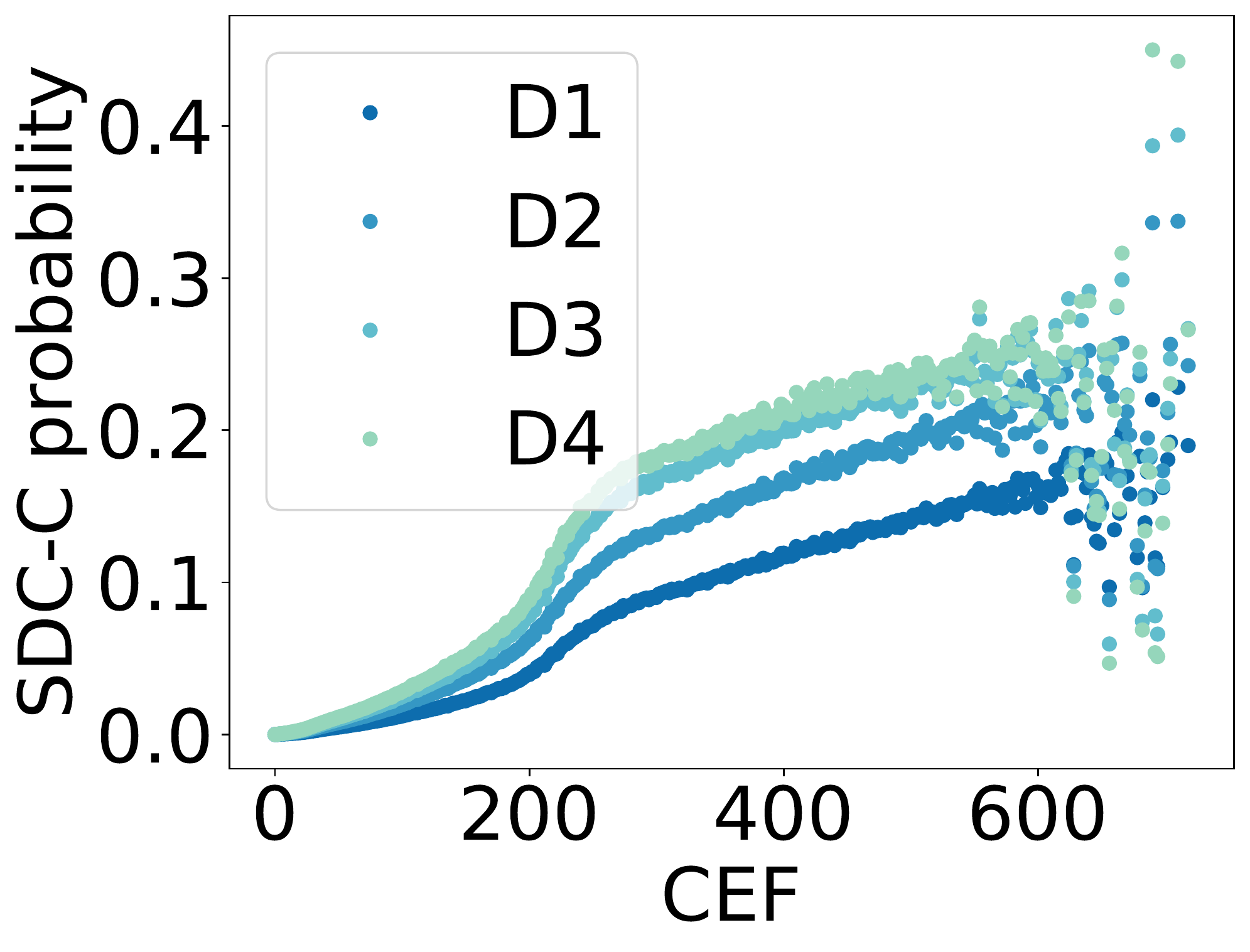}
        \caption{Jaco2:A3}
        \label{fig:bm_octree}
     \end{subfigure}
     \begin{subfigure}[b]{0.157\textwidth}
         \centering
        \includegraphics[width=\textwidth]{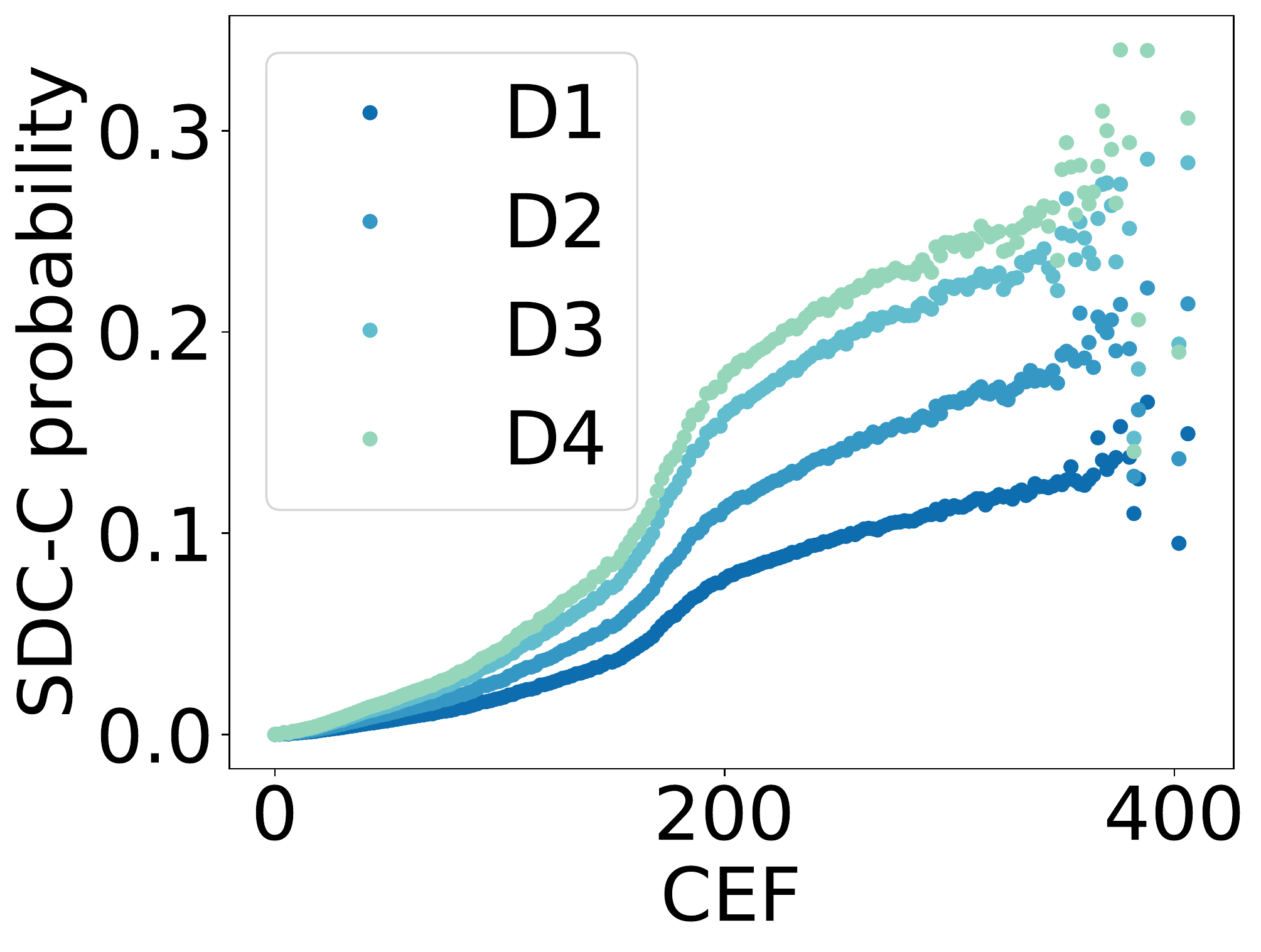}
        \caption{AL5D:A3}
        \label{fig:bm_octree}
     \end{subfigure}
    \begin{subfigure}[b]{0.157\textwidth}
        \centering
       \includegraphics[width=\textwidth]{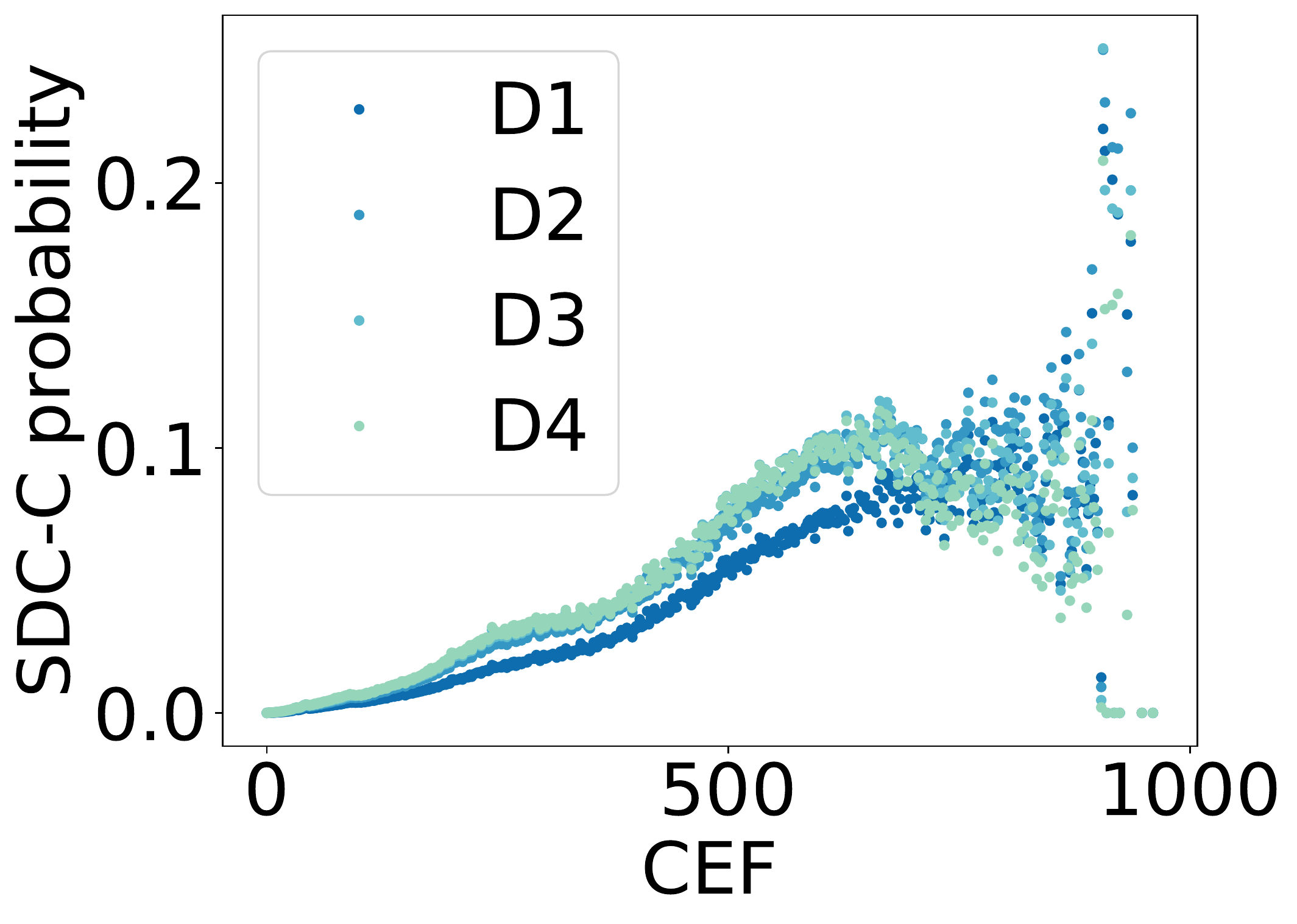}
       \caption{{Puma761:A3}}
       \label{fig:bm_octree}
    \end{subfigure}
    \begin{subfigure}[b]{0.157\textwidth}
        \centering
        \includegraphics[width=\textwidth]{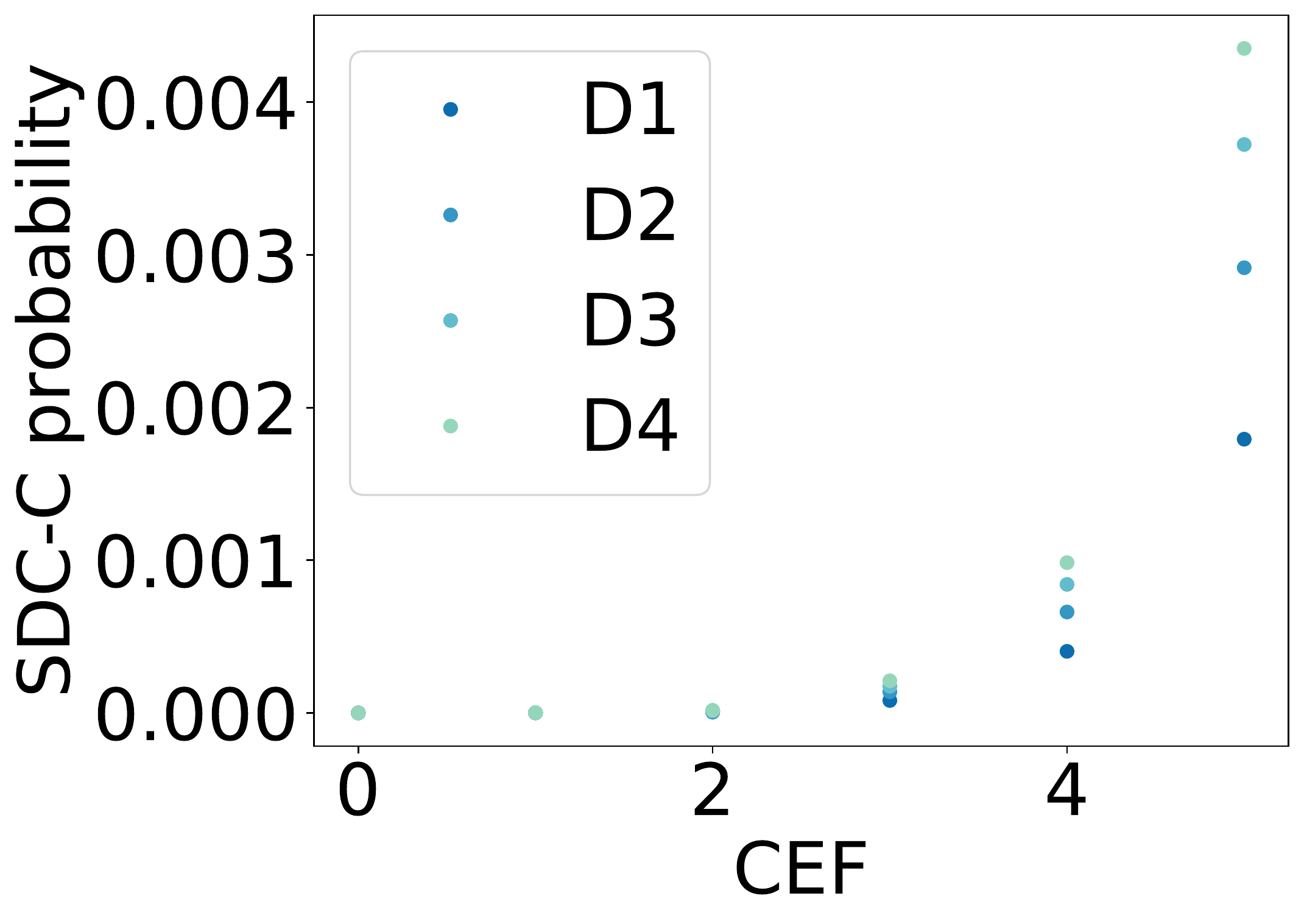}
       \caption{Jaco:A4}
       \label{fig:bm_box}
    \end{subfigure}
    \hfill
    \begin{subfigure}[b]{0.157\textwidth}
        \centering
       \includegraphics[width=\textwidth]{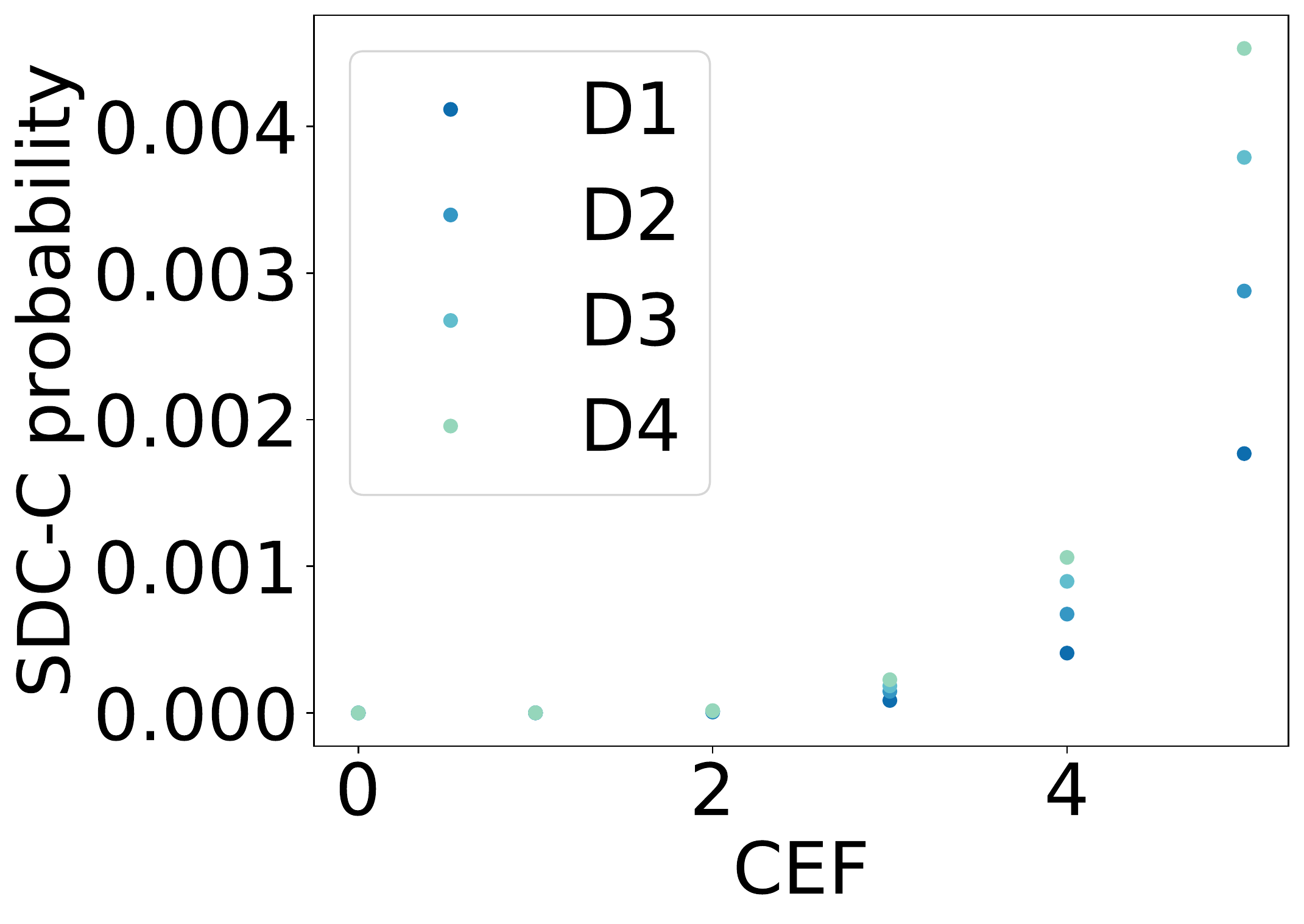}
       \caption{AL5D:A4}
       \label{fig:bm_octree}
    \end{subfigure}
    \hfill
    \begin{subfigure}[b]{0.157\textwidth}
        \centering
       \includegraphics[width=\textwidth]{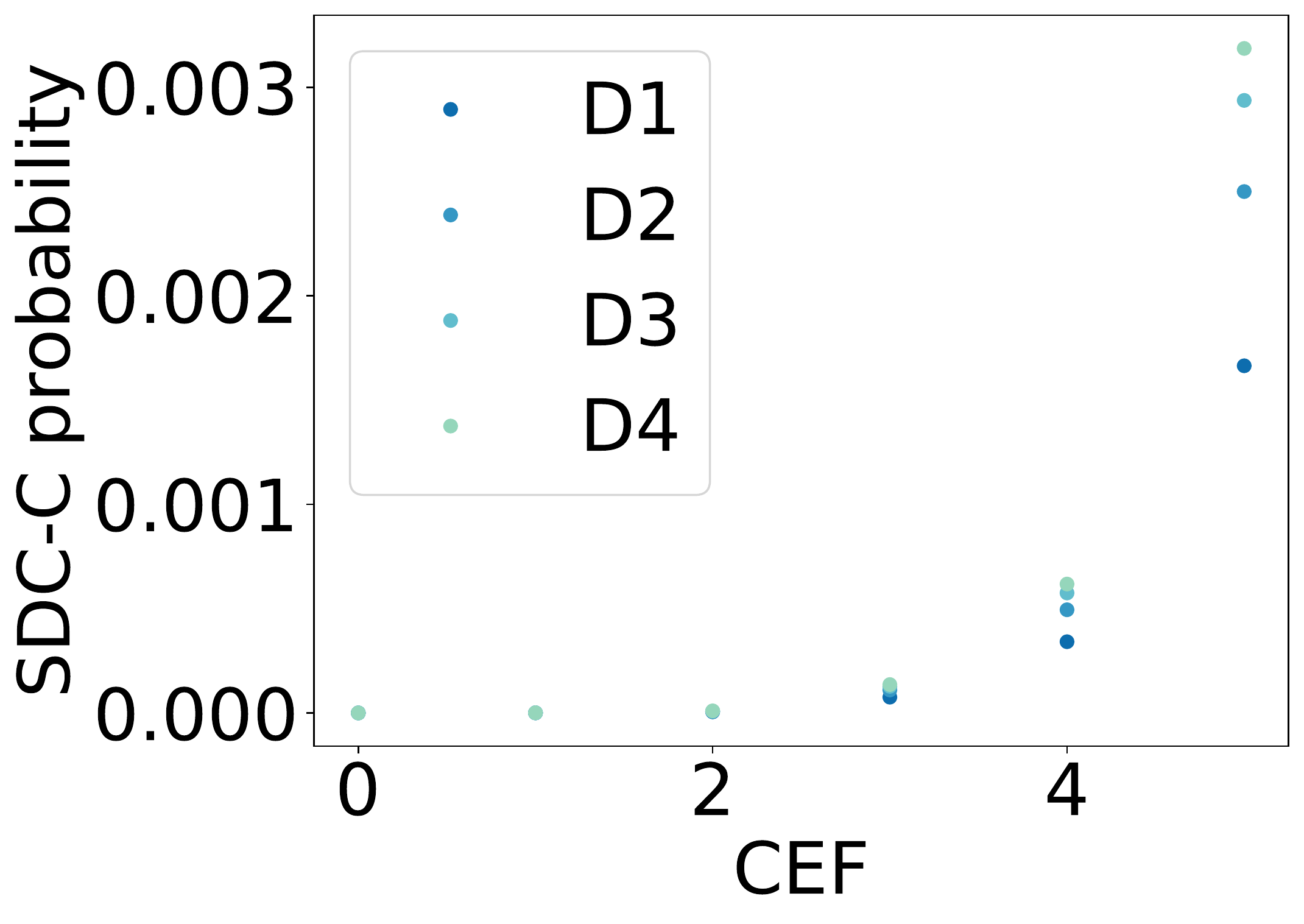}
       \caption{{Puma761:A4}}
       \label{fig:bm_octree}
    \end{subfigure}
        \caption{\sdc{ }probability versus CEF of accelerator A1, A2, A3, and A4 for Jaco2, AL5D, and Puma761 robot on benchmarks D$1$-D$4$.}
        \label{fig:faultAL5D}
\end{figure}

We use the benchmarks [D$1$, D$2$, D$3$, D$4$] (Section~\ref{subsec:setup}) to study the effect of obstacle occupancy density on the \sdc{ }probability. 
Figure~\ref{fig:faultAL5D} shows the \sdc{ }probability for different accelerators, robots, and benchmarks. 
For all the benchmarks, the \sdc{ }probability increases as the CEF increases. 
Note that as the obstacle occupancy density increases, the \sdc{ }probability also increases, 
as there are higher chances that the soft error in a CDM will result in a collision. 
\emph{The \sdc{ }probability is thus strongly positively correlated with CEF values across  benchmarks, which signifies that the CEF can be used as a reliability metric for MPAs.}
\subsubsection{Characterization of CEF}
\label{sec:part1}
\begin{figure}[]
     \centering
     \begin{subfigure}[b]{0.22\textwidth}
         \centering
         \includegraphics[width=0.9\textwidth]{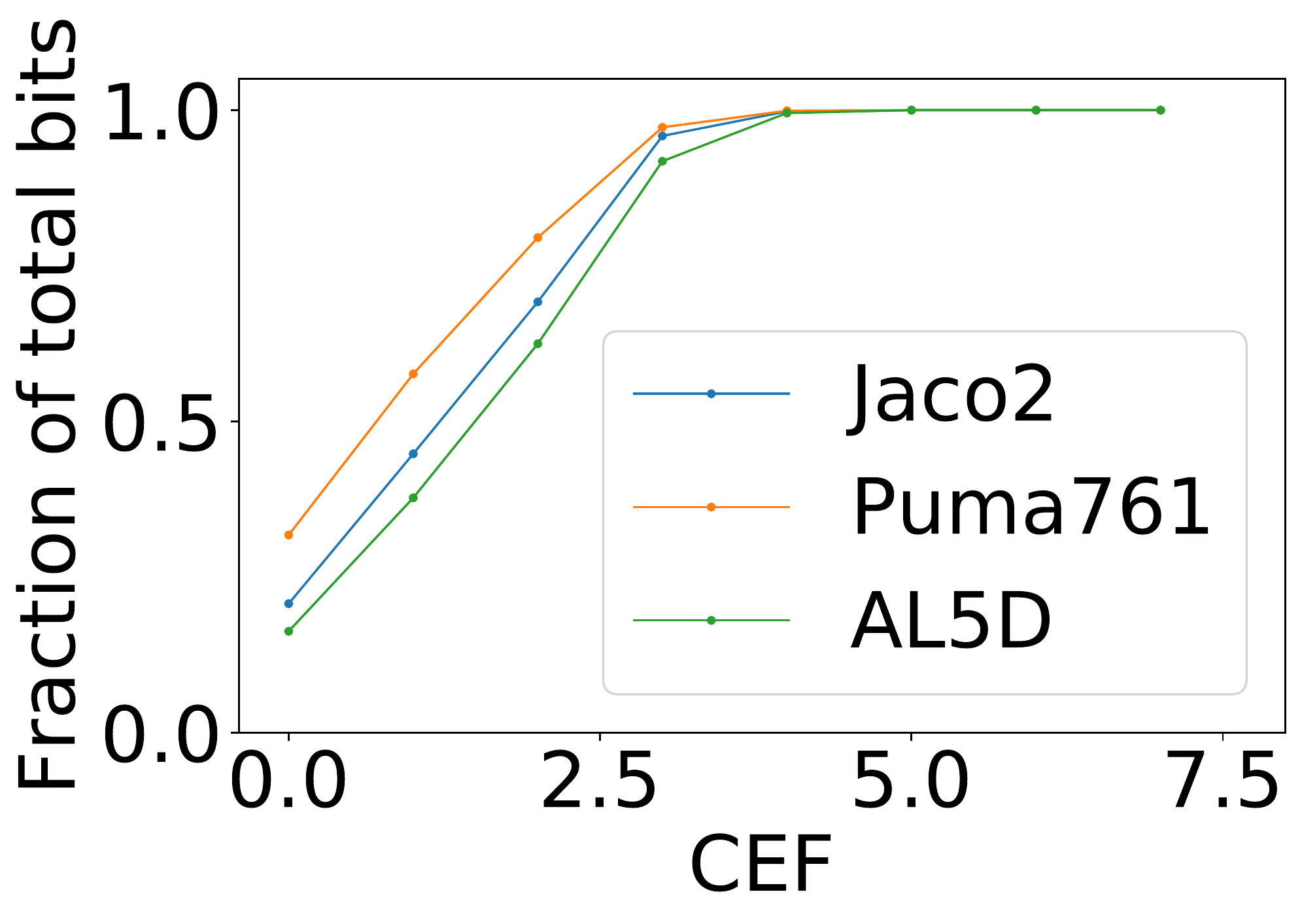}
         \caption{A1 (Voxel-based)}
         \label{fig:a1-cdf}
     \end{subfigure}
     \begin{subfigure}[b]{0.22\textwidth}
         \centering
         \includegraphics[width=0.9\textwidth]{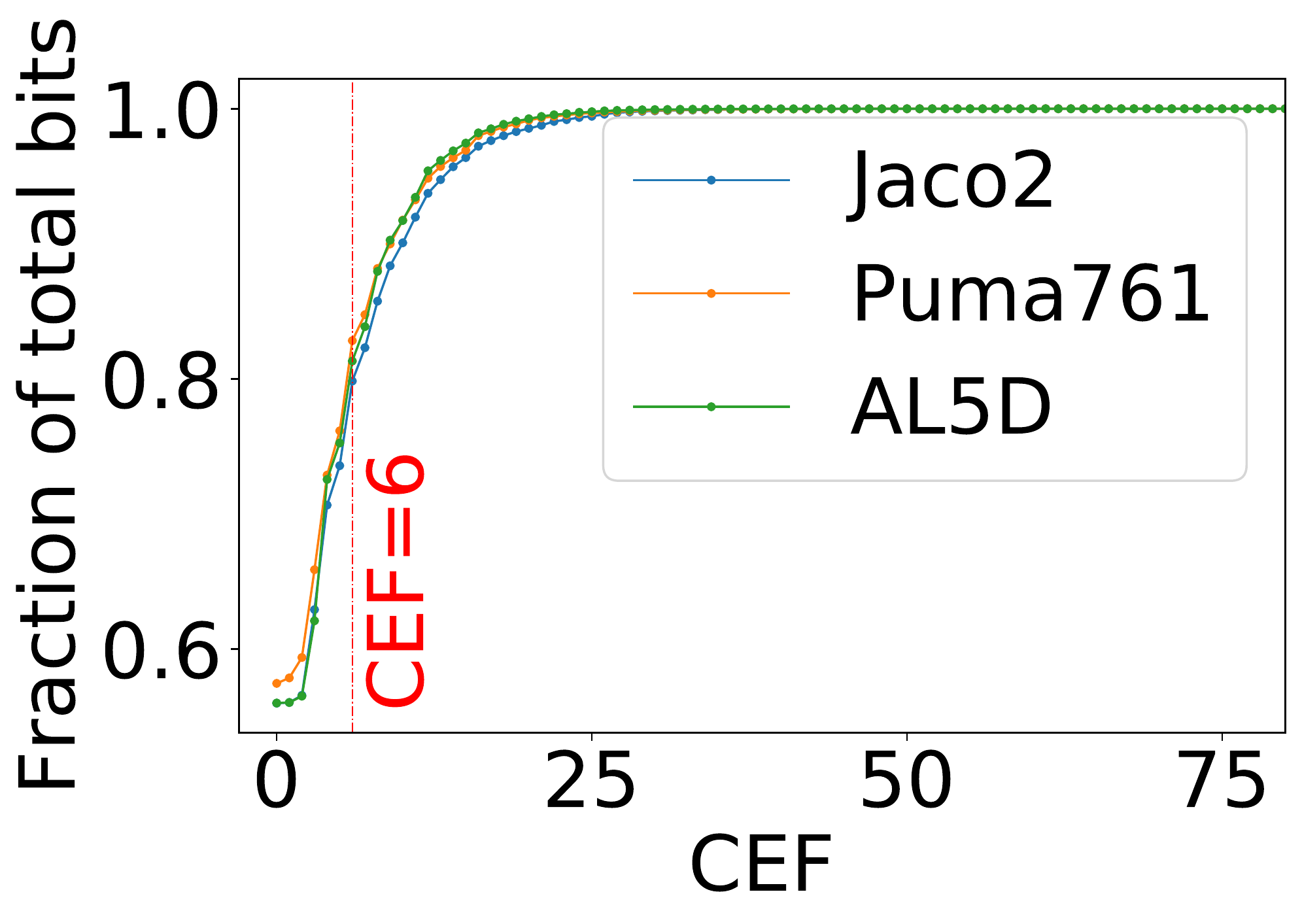}
         \caption{A2 (Box-based)}
         \label{fig:a2-cdf}
     \end{subfigure}
     \hfill
     \begin{subfigure}[b]{0.22\textwidth}
         \centering
         \includegraphics[width=0.9\textwidth]{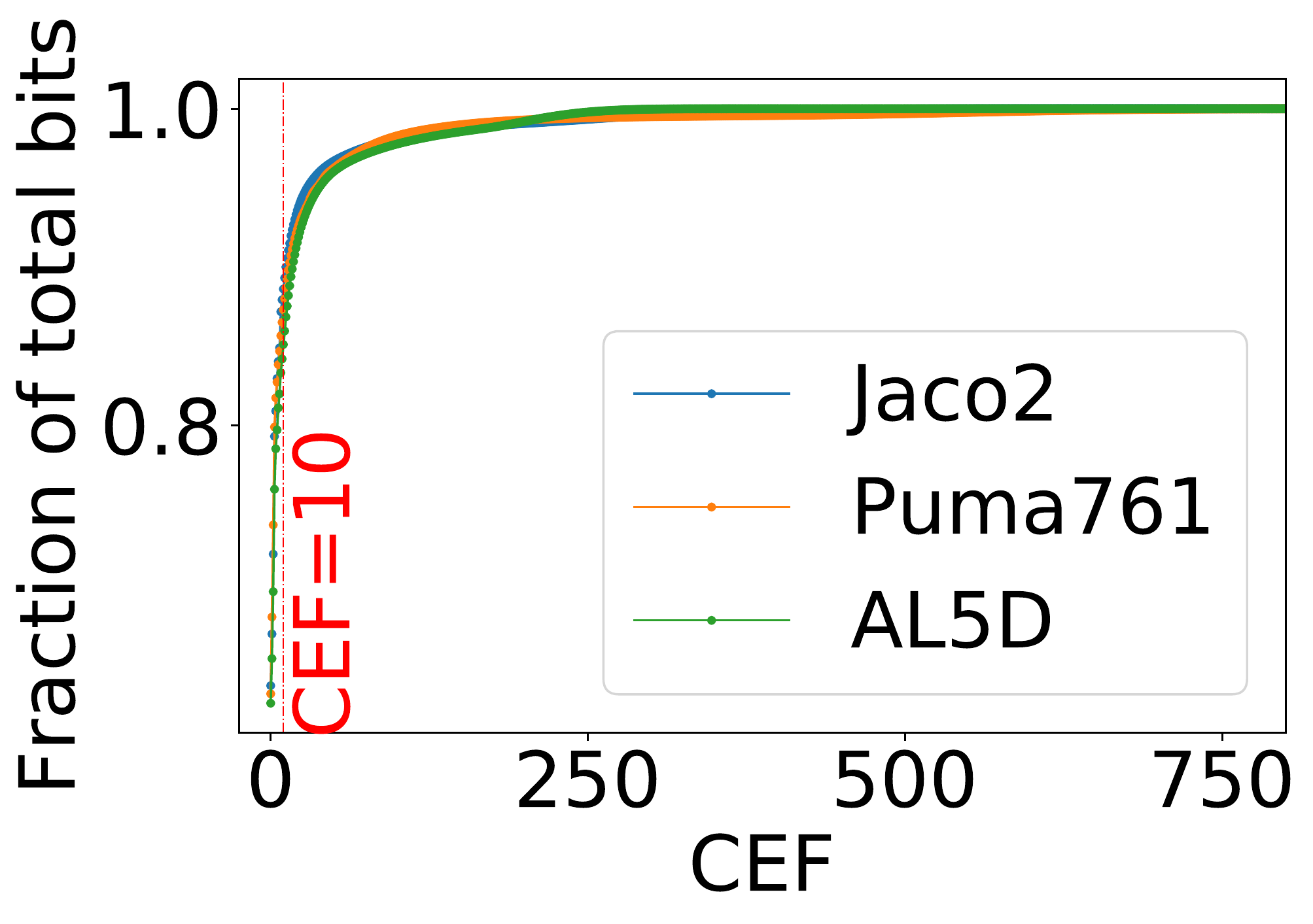}
         \caption{A3 (Octree-based)}
         \label{fig:a3-cdf}
     \end{subfigure}
     \begin{subfigure}[b]{0.22\textwidth}
         \centering
         \includegraphics[width=0.9\textwidth]{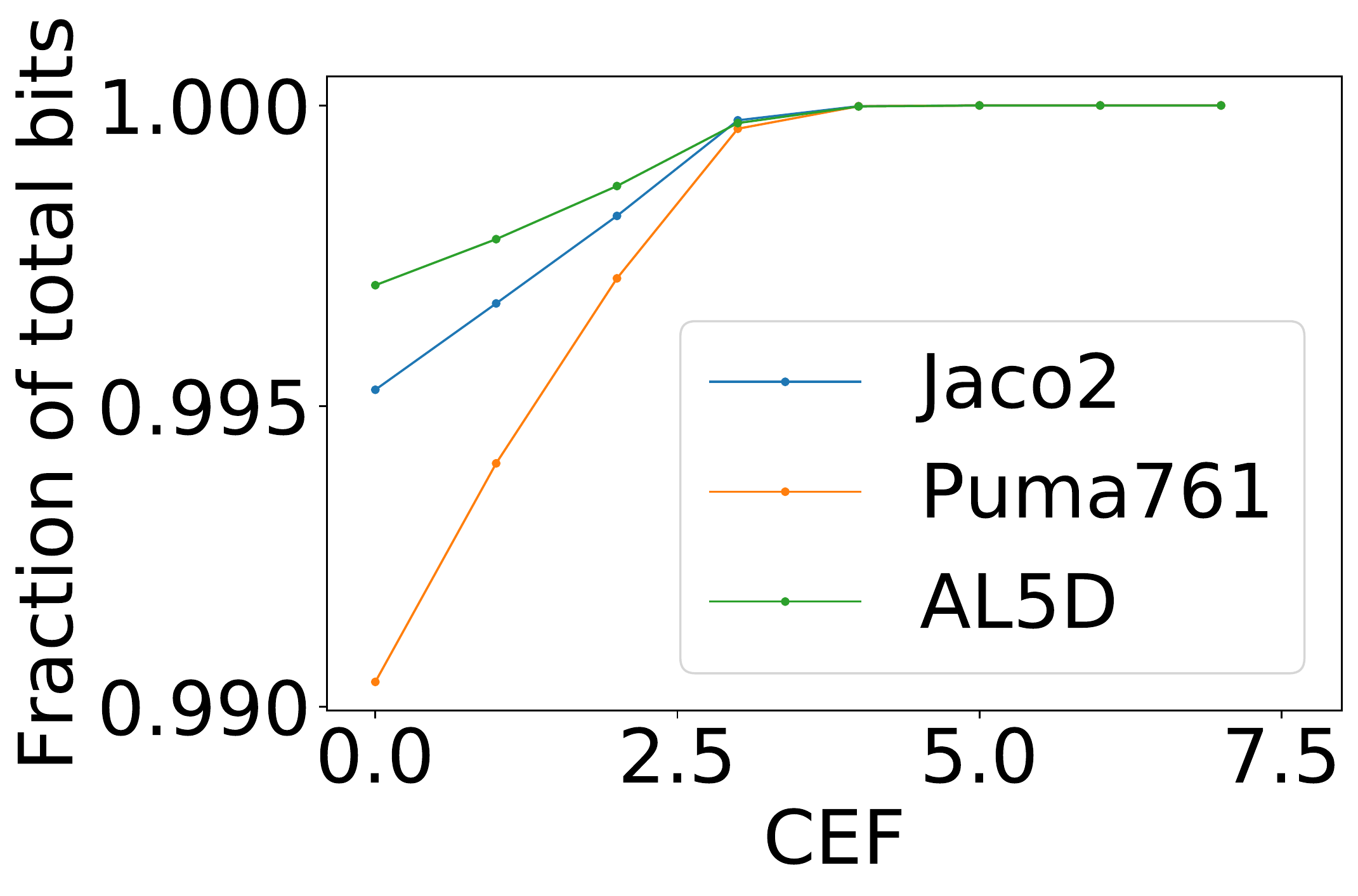}
         \caption{A4 (Flattened octree-based)}
         \label{fig:a4-cdf}
     \end{subfigure}
     \hfill
        \caption{Cumulative distribution of CEF of all bits.}
        \label{fig:fault_dist}
\end{figure}

%As described in Section~\ref{subsec:p1}, we perform Phase~1 FI using a detailed microarchitectural simulator to find the CEF and critical space of all bits. 
To further understand the contribution of individual bits to the overall \sdc{ }probability, we group the bits according to their CEFs, and plot the cumulative distribution of bits in the CDM. 
Figure~\ref{fig:fault_dist} shows that for all four CDMs, there is a high degree of asymmetry in the distribution of bits according to the CEF. %, which we can exploit to improve resilience. 
For A1 and A4, $20\%$ and $99\%$ of the total bits have CEF equal to $0$, respectively, and do not need to be protected. 
On the other hand, for A2, only $20\%$ of the total bits have CEF greater than $6$ and significantly contribute to the overall \sdc{ }probability. Thus, protecting only $20\%$ bits can reduce the FIT rate by $60\%$, as we show in Section~\ref{sec:sol}. %\karthik{So do they need protection?} \deval{added}
Similarly, in A3, only $15\%$ of the total bits have a CEF greater than $10$. 
We further examine the CEF distribution asymmetry for each accelerator. 

\begin{figure}[]
     \centering
     \begin{subfigure}[b]{0.32\linewidth}
         \centering
         \includegraphics[width=\textwidth]{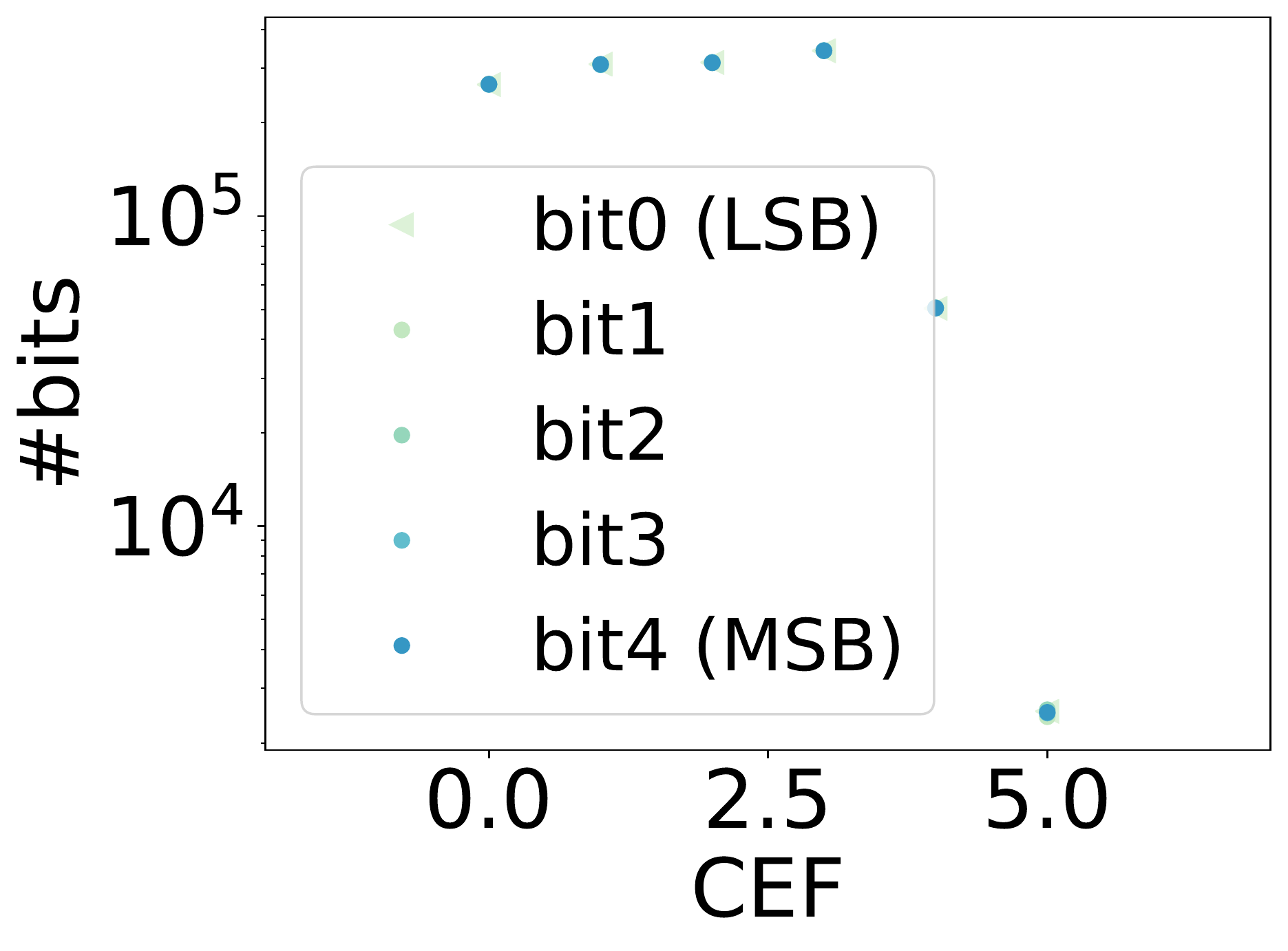}
         \caption{A1}
         \label{fig:voxel-bit}
     \end{subfigure}
     \hfill
     \begin{subfigure}[b]{0.32\linewidth}
         \centering
         \includegraphics[width=\textwidth]{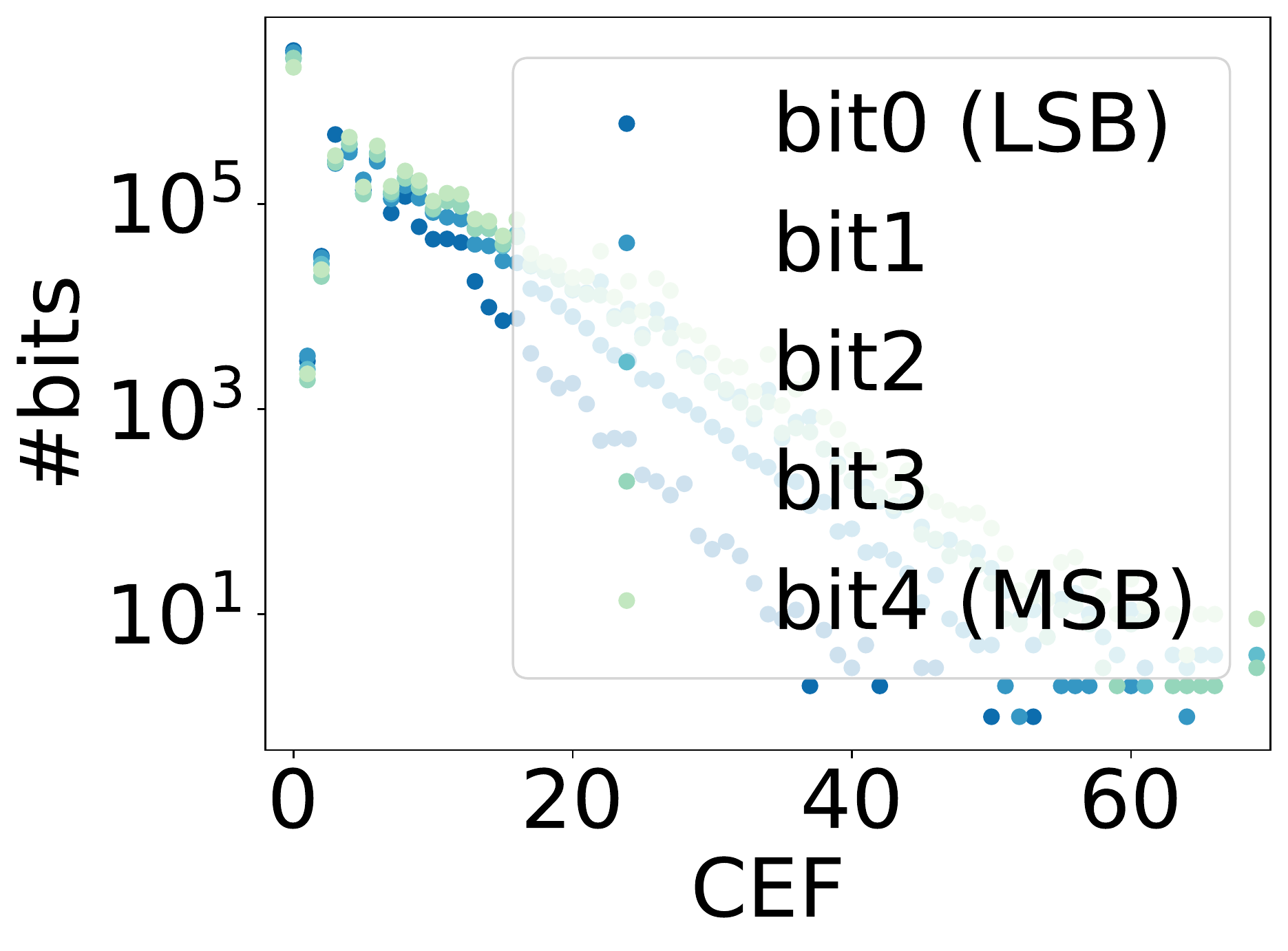}
         \caption{A2}
         \label{fig:box-bit}
     \end{subfigure}
          \begin{subfigure}[b]{0.32\linewidth}
         \centering
         \includegraphics[width=\textwidth]{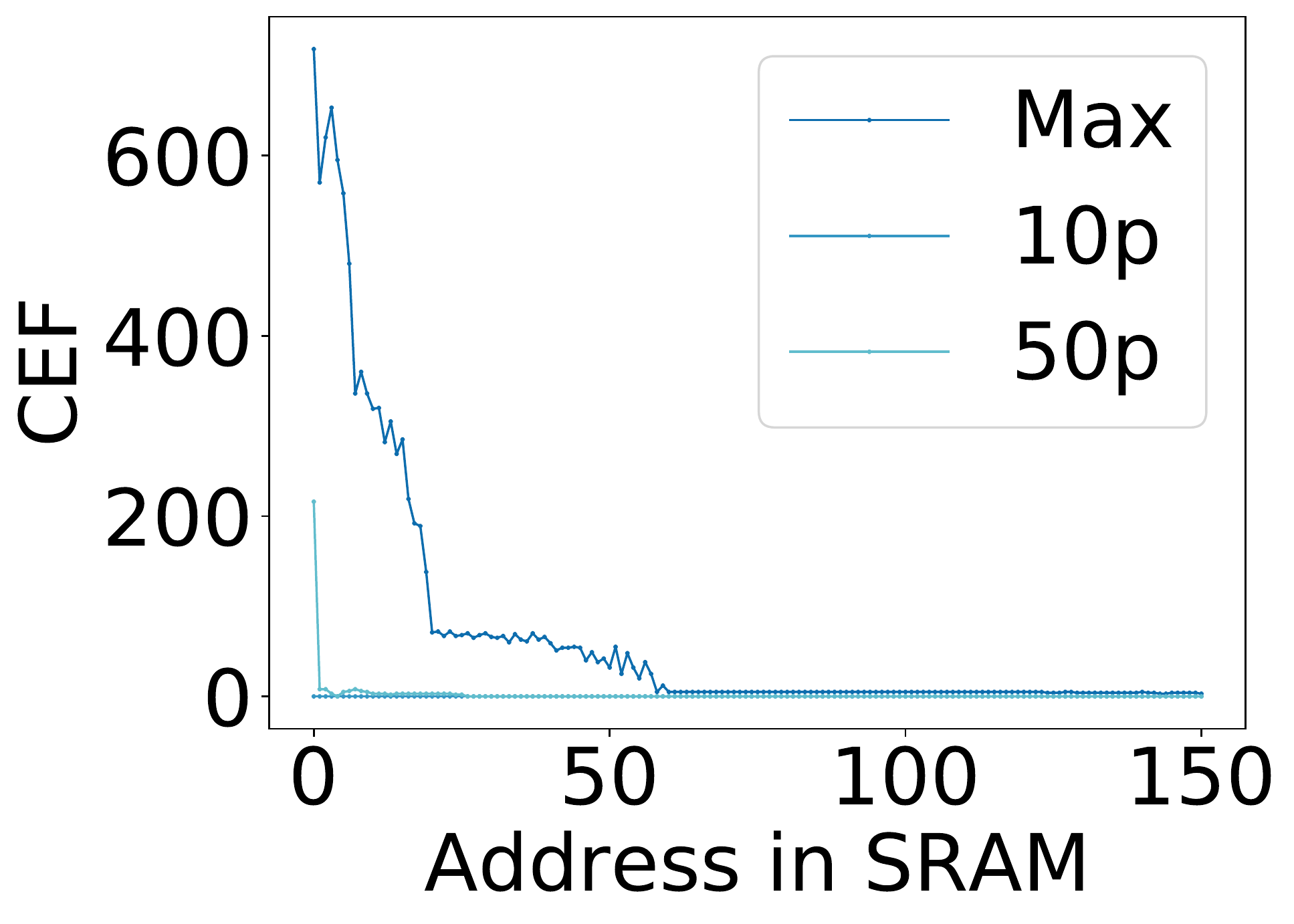}
         \caption{A3}
         \label{fig:child}
     \end{subfigure}
        \caption{CEF characterization of bit position/SRAM address.}
        \label{fig:asymm}
\end{figure}

\textbf{A1: } 
In A1, each structure/variable stores a voxel using its coordinates to represent a part of the swept space. 
A soft error in any bit of the coordinate will result in misrepresentation of that voxel, and so bit position does not affect CEF range (Figure~\ref{fig:voxel-bit}). 

\textbf{A2: }
The effect of faults in different bits of coordinates is shown in Figure~\ref{fig:box-bit}. 
We can see that the CEF range of bits decreases from the most significant bit (MSB) to the least significant bit (LSB), as a fault in the MSB is likely to result in a higher change in the box size represented by a pair of coordinates than LSB.
Thus, the number of bits with high CEF values increases from LSB to MSB. 
For example, the ratio between the number of MSBs ($bit4$) and the number of LSBs ($bit0$) with CEF $=20$ is $1000$. 

\textbf{A3: }
We calculate the range of  CEF for bits for different SRAM addresses, as shown in Figure~\ref{fig:child}. 
%In an octree, the levels near the root node are more critical as they divide the 3D space at a coarser granularity.
All nodes in the octree for a motion's swept space are stored in the continuous address space of the  SRAM; the nodes closer to the octree's root node that divide the 3D space at a coarser granularity are stored in the lower address range of SRAM. 
As can be seen, the CEFs of bits with lower SRAM addresses are much higher than those with higher SRAM addresses. 
For example, the average CEF of bits in SRAM address $0$ is $9\times$ that of the average CEF of bits in SRAM address $20$. 
%%For example, the average CEF of bits in SRAM address $0$ is $200$, which is $9\times$ that of the average CEF of bits in SRAM address $20$. 
%A3 also exhibits a much higher CEF range ($0-1000$) than A2 ($0-100$) for all the robots, as shown in Figure~\ref{fig:fault_dist} due to its different architectural design (explained in Section~\ref{sec:arch}).

\textbf{A4: }
{A flattened-octree consists of a node for each voxel in the environment. The swept space of a motion in the motion set is a small fraction of the entire environment. Thus, most of the nodes in the flattened octree are empty, and only a small fraction of nodes are occupied. }
An error in an empty node representation results in false-positive and hence has CEF equal to zero. Due to this, more than $99\%$ of the bits have CEF equal to $0$ for A4. 

\emph{In summary, for all three accelerators, the distribution of the bits as per CEF is highly asymmetric. Hence, the CEF metric facilitates finding the most \sdc-prone bits in the circuit.} 
\subsection{Error Mitigation Techniques}
\label{sec:sol}
\begin{figure*}[t!]
    \centering
    \begin{subfigure}[t]{0.20\linewidth}
        \centering
        \includegraphics[width=\textwidth]{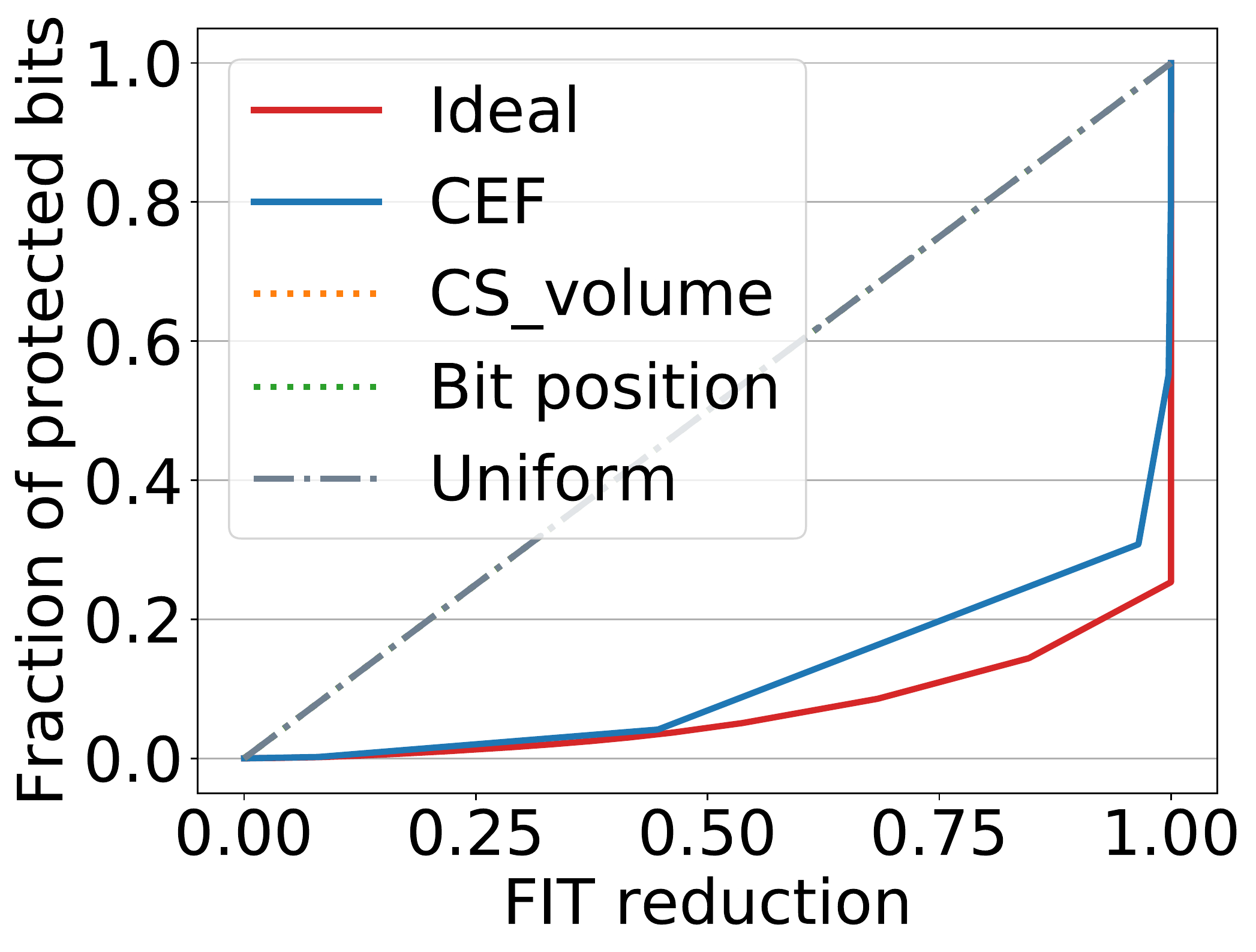}
        \caption{A1}
        \label{fig:a1frac}
    \end{subfigure}
    %\hfill
    \hspace{2mm}
    \begin{subfigure}[t]{0.20\linewidth}
        \centering
        \includegraphics[width=\textwidth]{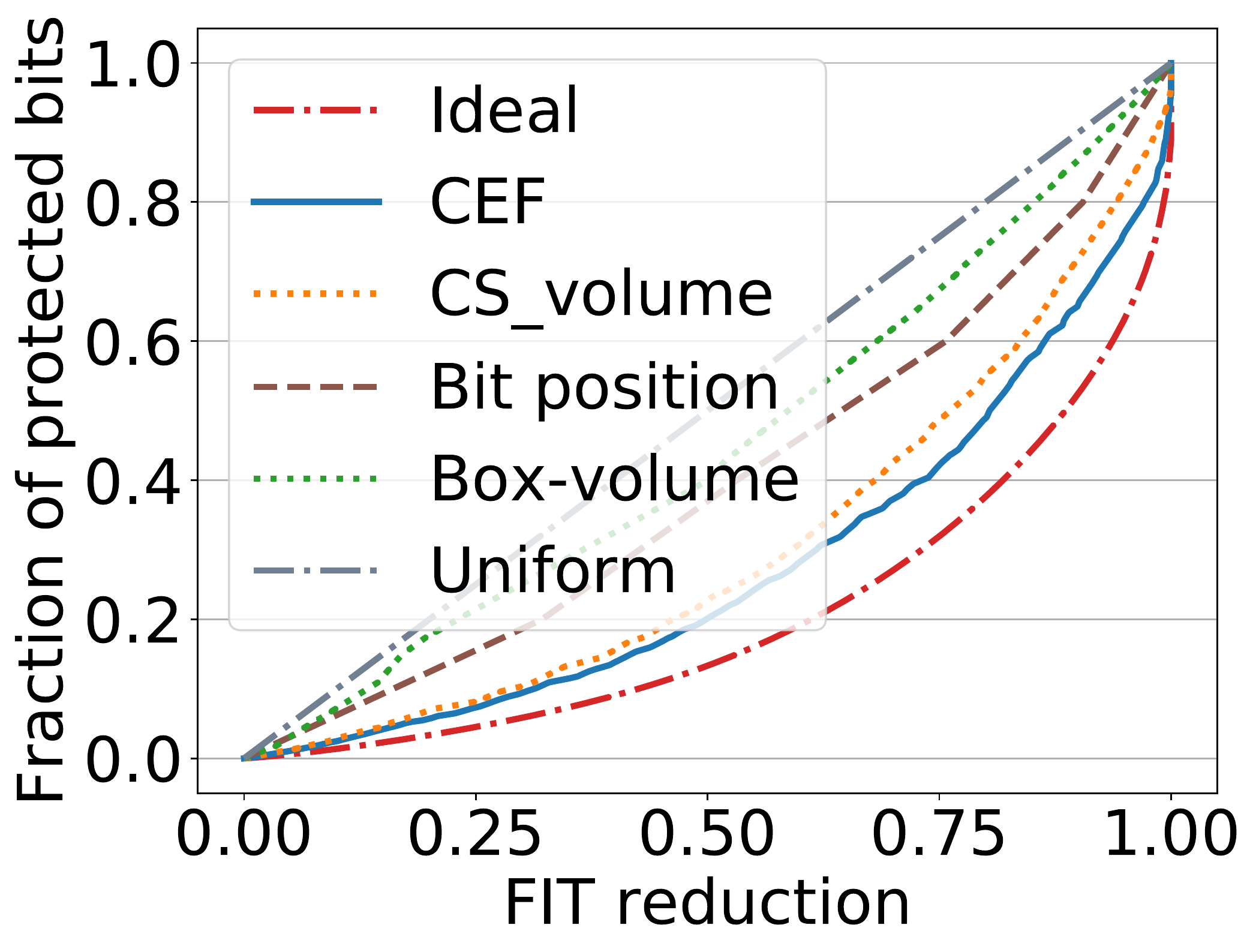}
        \caption{A2}
        \label{fig:a2frac}
    \end{subfigure}
    %\hfill
    \hspace{2mm}
    \begin{subfigure}[t]{0.20\linewidth}
        \centering
        \includegraphics[width=\textwidth]{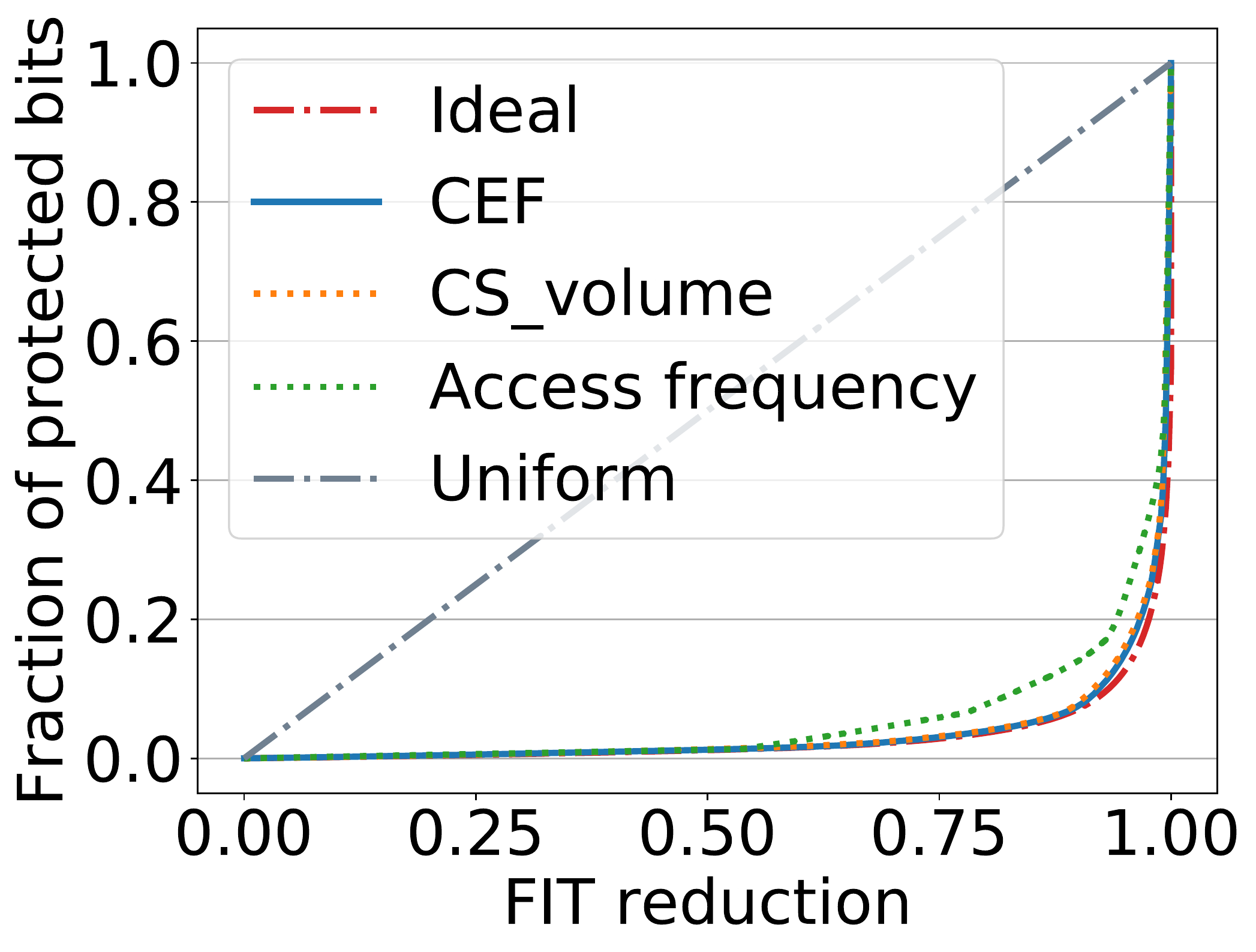}
        \caption{$\text{A3}_{\text{scaled}}$}
        \label{fig:a3frac}
    \end{subfigure}
    \hspace{2mm}
    \begin{subfigure}[t]{0.20\linewidth}
        \centering
        \includegraphics[width=\textwidth]{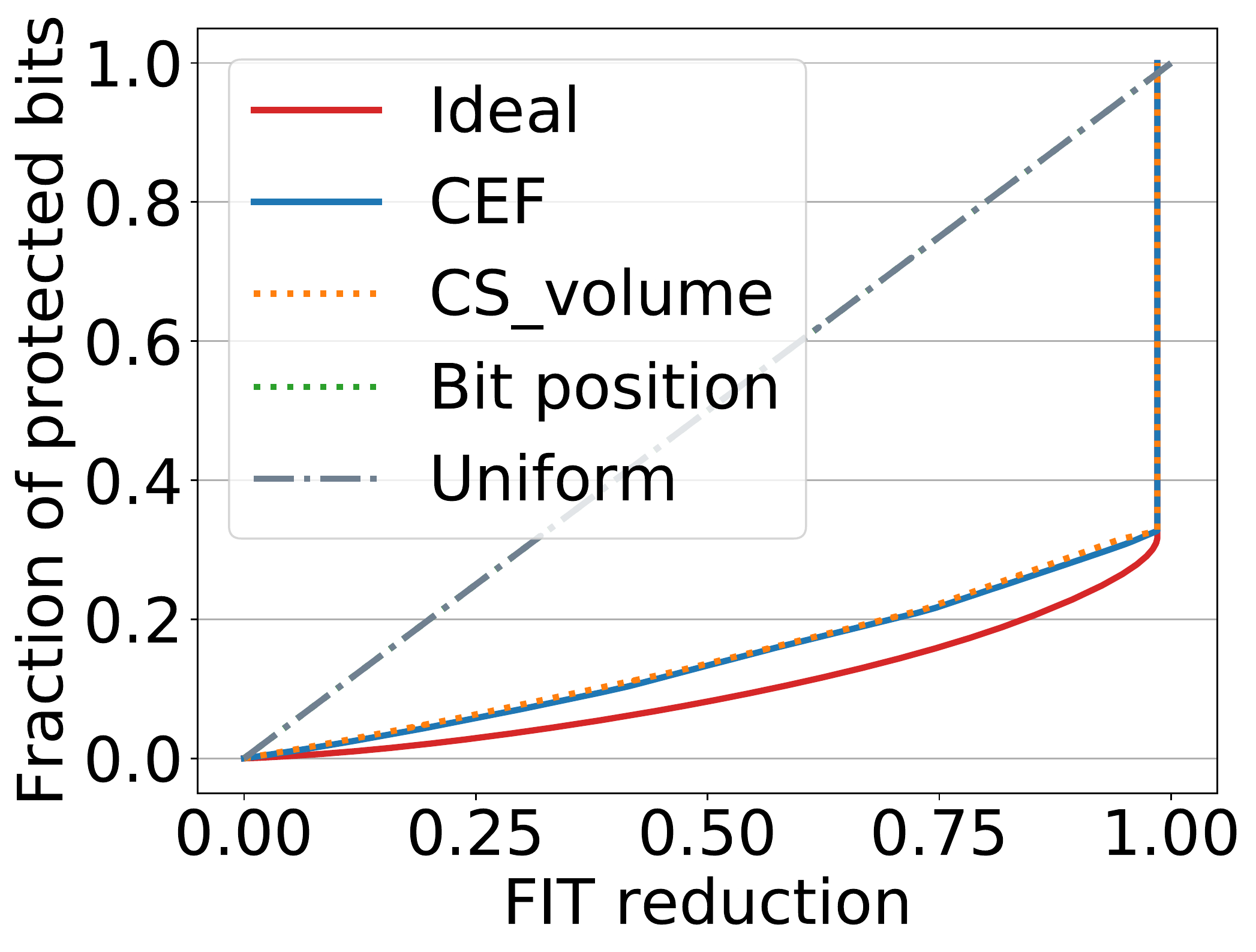}
        \caption{A4}
        \label{fig:a4frac}
    \end{subfigure}
    \hfill
       \caption{ Comparison of different selection criteria for selective error mitigation in A1, A2, $\text{A3}_{\text{scaled}}$, and A4. }
       \label{fig:selective}
\end{figure*}
As described in Section~\ref{subsec:cefem}, we propose CEF-aware selective error mitigation technique. 
We first evaluate the CEF-aware selection of structures for reliability-aware data placement at deployment time. We further evaluate the area savings achieved by the proposed techniques for accelerators A1, A2, A3$_{\text{scaled}}$, and A4 for different SIL targets at design time.  
Note that A3's FIT rate ($0.1$) is well below the highest SIL requirement, and hence it does not need error mitigation. Therefore, we do not consider A3.  

\subsubsection{CEF-aware selective error mitigation}
Figure~\ref{fig:selective} compares the FIT reduction achieved by different selection criteria for the same fraction of protected memory. 
We assume that protecting a bit reduces its FIT rate to 0 as our aim is to compare different heuristics for selecting the bits to be protected, which is independent of the underlying error mitigation technique. 
We also compare with another intuitive approach to define the reliability metric that uses the volume of the critical space (CS\_volume) instead of exposed surface area. 
We further discuss the results for A1, A2, $\text{A3}_{\text{scaled}}$, and A4. 

\textbf{A1: }
We compare our approach with ideal, uniform, bit position, and CS\_volume-aware error mitigation. 
Bit position does not require FI and can be used as a proxy for the vulnerability factor, where the vulnerability of the bits decreases from the MSB to the LSB. 
Bit position has been previously proposed for selective error mitigation in DNN accelerators~\cite{Li2017}. 
Figure~\ref{fig:a1frac} shows the fraction of protected bits versus the FIT reduction curve for CEF-aware selection and the other heuristics for A1. 
The CEF-aware selection of bits results in $52.35\times$ reduction in the FIT rate on average (geometric mean) for the same amount of protected bits compared to CS\_volume-aware, bit position-aware, and uniform selection. 
The FIT reduction for CEF-aware error mitigation is only {$1.60\times$} lower than the ideal FIT reduction that can be achieved by exhaustive FI that is $23,000\times$ slower than CEF-aware FI. 
%\karthik{Can we say something about why this is still good.}

\textbf{A2: }
We compare CEF with ideal, uniform, CS\_volume, bit position, and box-volume. 
For A2, a structure represents a box; the volume of the box can be used as the vulnerability factor of the structure. {\emph Box-volume } is an application-specific heuristic that does not require fault characterization. 
Figure~\ref{fig:a2frac} shows the fraction of protected bits versus the FIT reduction for different selection heuristics. 
The CEF-aware selection of bits results in average {$1.25\times$}, {$1.76\times$}, {$2.10\times$}, and {$2.46\times$} lower FIT rate than CS\_volume, bit position, box volume, and uniform selection for the same amount of protected bits. The FIT reduction for CEF-aware error mitigation is only $1.66\times$ lower than the ideal FIT reduction achieved by exhaustive FI. 

Intuitively, one may expect bit position to provide higher benefits than CEF, as typically MSBs are more critical. 
However, we find that CEF provides a higher reduction in the FIT rate than the bit position, due to two reasons. 
First, though the bit position captures the failure probability of bits within a structure, it does not capture the relative failure probabilities across different structures. 
For example, in A2, different boxes cover different numbers of voxels $x$ in the swept space that is not covered by other boxes; these voxels contribute to the critical space. The CEF increases as the $x$ value increases, and so structures with higher $x$ have higher failure probabilities, which is not captured by bit position. 
Second, the bit position does not consider whether the error will lead to a false-negative or a false-positive. 
Similarly, box-volume does not necessarily capture the number of critical voxels $x$. 
In contrast, the CEF captures the relative failure impact of all the structures, and hence provides a higher FIT rate reduction. % than the box-volume heuristic. 

\textbf{$\text{A3}_{\text{scaled}}$: }
Figure~\ref{fig:a3frac} shows the fraction of protected bits versus the FIT reduction for different selection criteria in A3$_{\text{scaled}}$. 
%We compare our approach with access-frequency-based selection of structures. A similar approach was used by Mehrara et al.~\cite{Mehrara2008} to selectively protect frequently accessed data.
We compare CEF with two heuristics: uniform and access-frequency-based selection. Access-frequency-based selection has been proposed for embedded applications~\cite{765955,Mehrara2008}.
We find that the CEF-aware selection of bits results in an average $1.07\times$ and {$1.90\times$} lower FIT rate for the same amount of protected bits compared to CS\_volume and access-frequency. This is because the access-frequency-based heuristic captures the failure probability of structures within a single motion, but not the relative failure probabilities across different motions. Further, CEF-aware selection achieves {$18.89\times$} and $0.79\times$ lower FIT rate than uniform and ideal error mitigation, respectively.

\textbf{A4: }
Figure~\ref{fig:a4frac} compares different selection criteria for A4. 
We find that the CEF-aware selection of bits results in an average {$1.02\times$} reduction in  FIT rate for the same amount of protected bits compared to CS\_volume. 
In the proposed accelerator, the access-frequency for all the bits of a flattened octree is the same, and hence uniform and access-frequency-based selection given the same reduction in the FIT rate.  
%%This is because the access-frequency-based heuristic captures the failure probability of structures within a single motion, but not the relative failure probabilities across different motions. 
Further, CEF-aware selection achieves {$9.41\times$} and {$0.83\times$} reduction in FIT rate than uniform/access-frequency and ideal error mitigation respectively.

\subsubsection{Area savings using CEF-aware error mitigation}
\label{subsec:lh}
Further, we measure the area/power overheads to achieve different SILs using the CEF-aware error mitigation approach. 
Latch hardening is more suitable than ECC for error mitigation in registers that are accessed in parallel, as adding an ECC encoder-decoder circuit for every  register would incur significant area and energy overheads - this is  shown in Table~\ref{tab:accel} for A1 and A2 accelerators. % 
Hence, we do not use ECC for A1 and A2 accelerators. 
We instead use the latch hardening techniques (summarized in Table~\ref{tab:overhead}) in Sullivan et al.~\cite{Sullivan} to measure the overheads for A1 and A2. 

In contrast, we use ECC (SEC-DED code) for A3$_{\text{scaled}}$ and A4 as these accelerators use SRAM for on-chip storage or DRAM (Table~\ref{tab:accel}). 
For A3$_{\text{scaled}}$, each entry in SRAM consists of 24 bits. We assume an overhead of 7 bits ($30\%$) to store the ECC bits. 
For A4, each entry in DRAM consists of $64$ bits, with {8} bits overhead for ECC bits.   
We ignore the area overhead of the error detection/correction circuits themselves. 
Note that other approaches for ECC~\cite{8671577} can also be combined with CEF-aware selection. 
\newdimen\emh
\settoheight{\emh}{%
\includegraphics[width=0.23\textwidth]{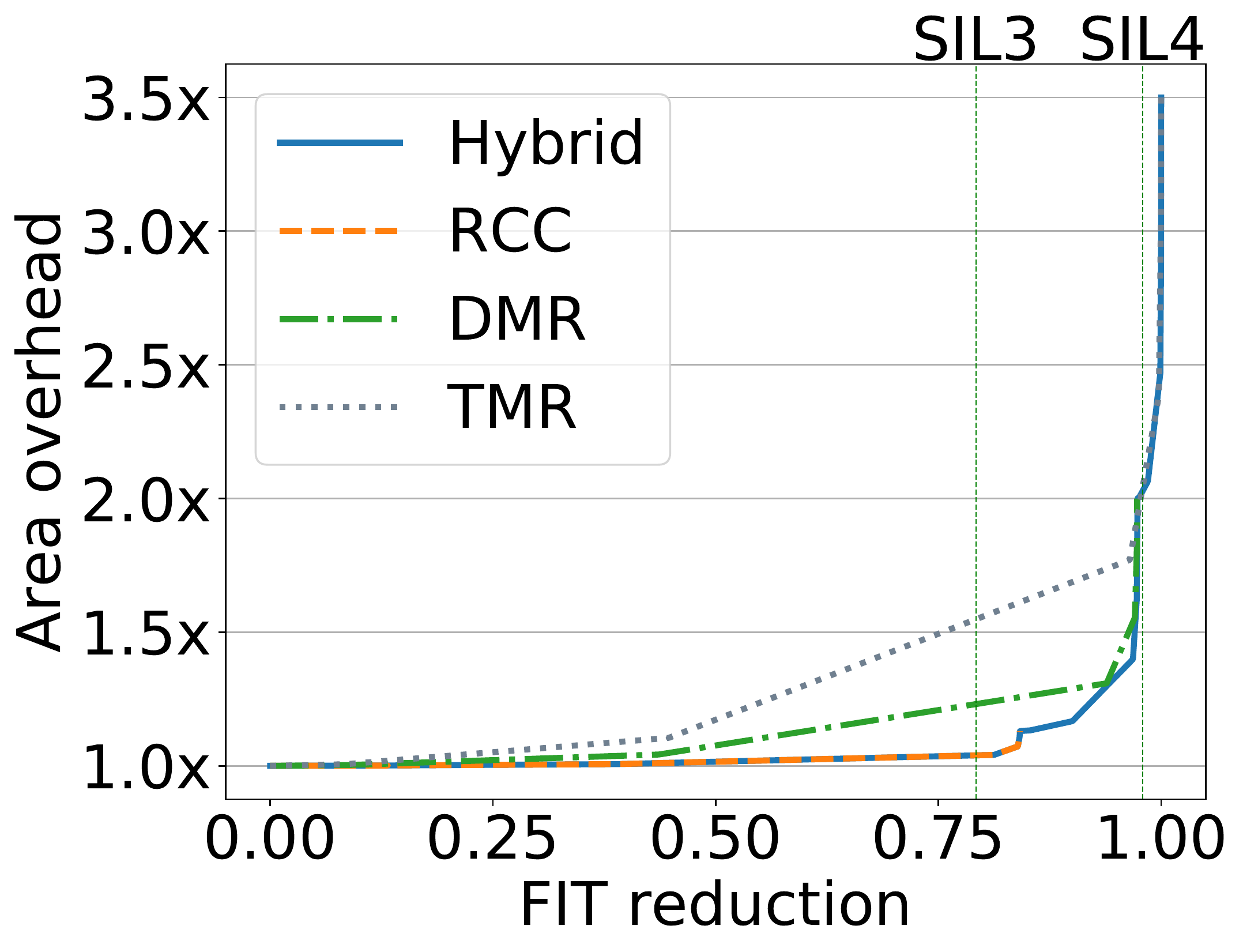}} 
\begin{figure*}[]
     \centering
     \begin{subfigure}[t]{0.23\linewidth}
         \centering
         \includegraphics[height=\emh]{figures/em_a1_lh.pdf}
         \caption{A1}
         \label{fig:slha1}
     \end{subfigure}
     %\hfill
     \hspace{2mm}
     \begin{subfigure}[t]{0.23\linewidth}
         \centering
         \includegraphics[height=\emh]{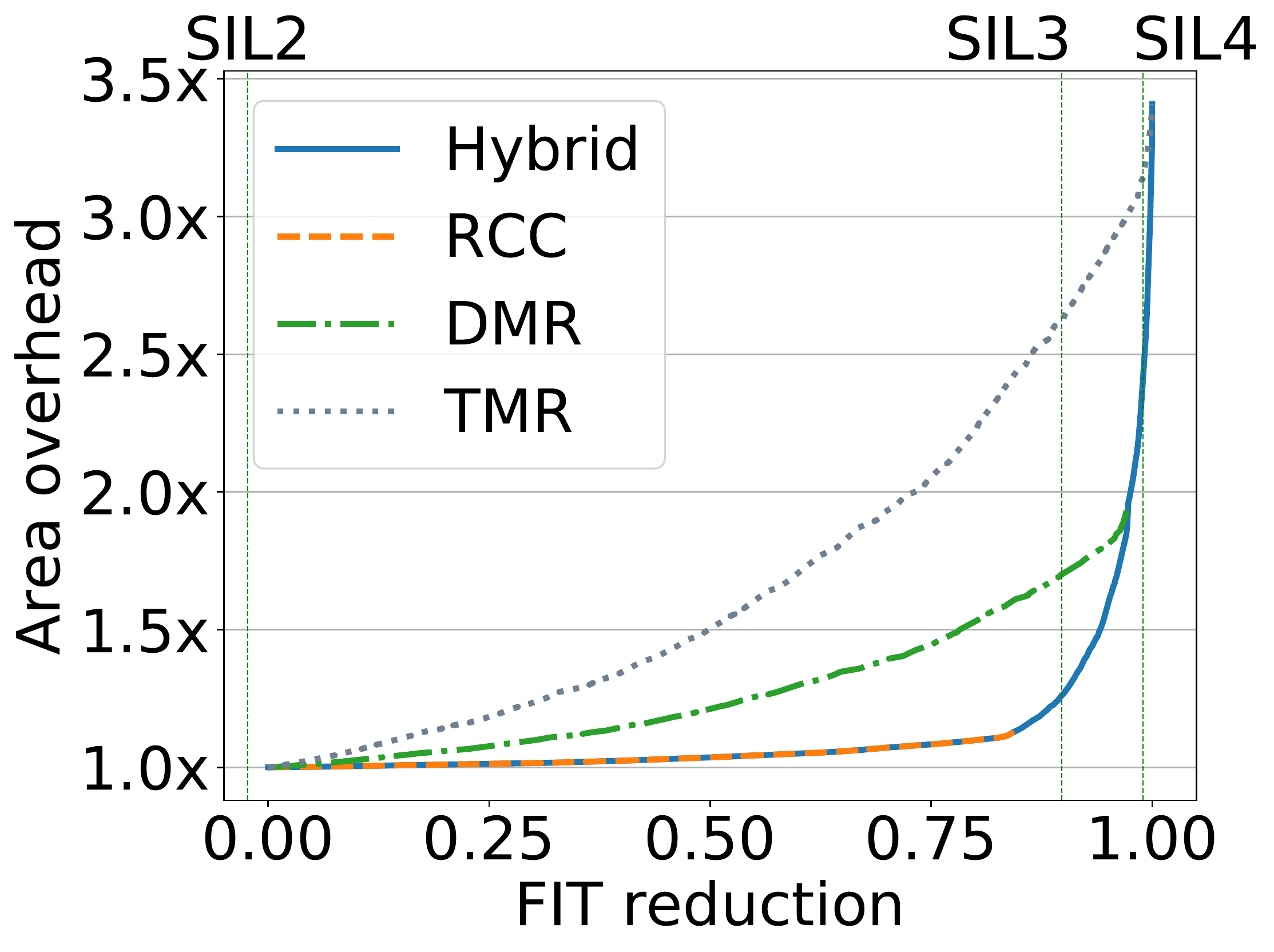}
         \caption{A2}
         \label{fig:slha2}
     \end{subfigure}
     %\hfill
     \hspace{2mm}
     \begin{subfigure}[t]{0.23\linewidth}
        \centering
        \includegraphics[height=\emh]{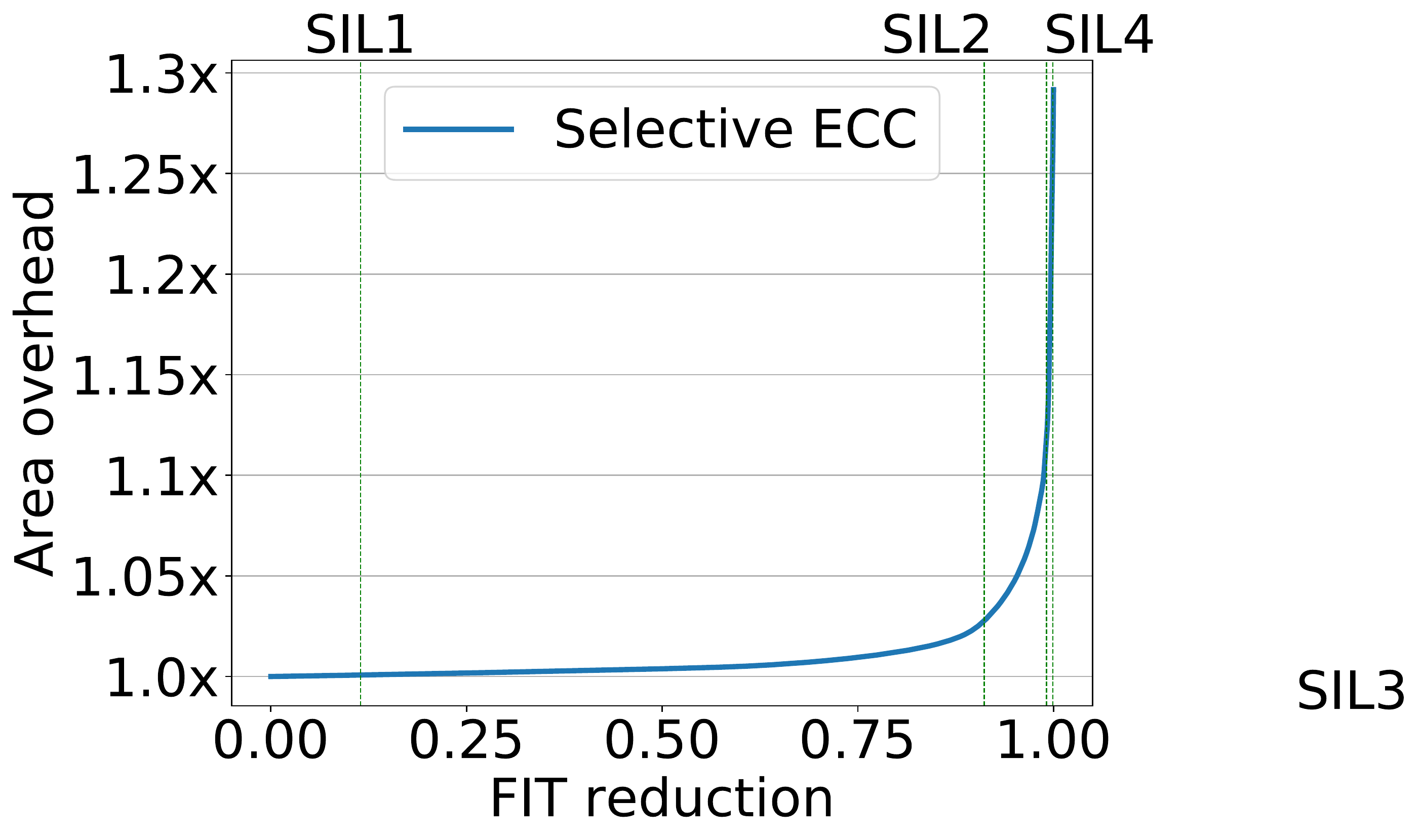}
        \caption{$\text{A3}_{\text{scaled}}$}
        \label{fig:slha3}
    \end{subfigure}
    \hspace{2mm}
    \begin{subfigure}[t]{0.23\linewidth}
       \centering
       \includegraphics[height=\emh]{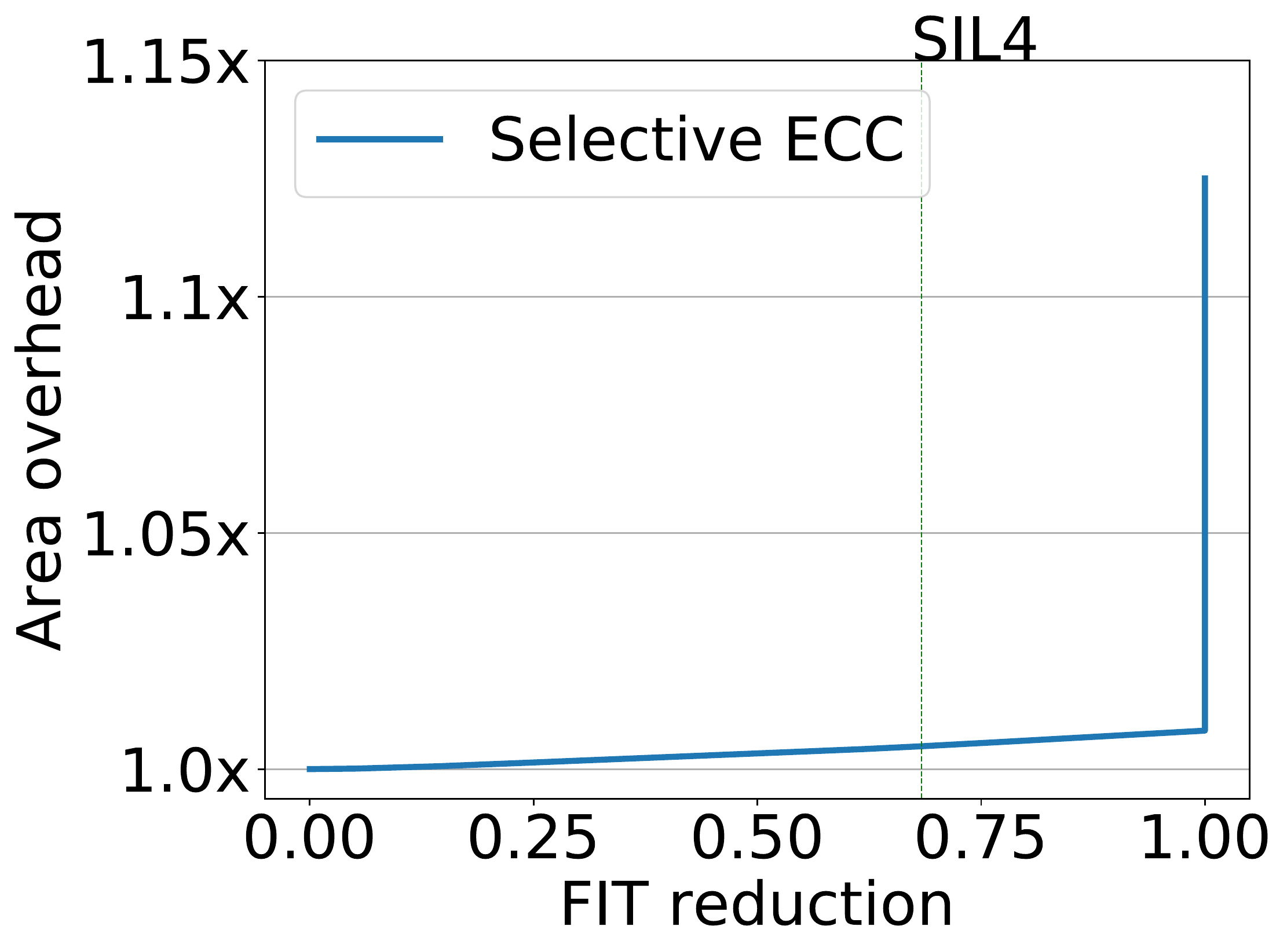}
       \caption{{A4}}
       \label{fig:slha4}
   \end{subfigure}
        \caption{Area overhead versus FIT reduction for selective error mitigation in A1, A2, $\text{A3}_{\text{scaled}}$, and A4. }
        \label{fig:select}
\end{figure*}
\begin{table}[b]
    \scriptsize
\setlength\tabcolsep{4.6pt} 
\caption{Area overhead and FIT reduction for different latch hardening techniques.}
\label{tab:overhead}
%\resizebox{0.45\textwidth}{!}{%
\begin{tabular}{|L{4.1cm}|R{1.8cm}|R{1.8cm}|}
    \hline
Latch type & Area Overhead & FIT reduction \\ \hline
Strike Supression (RCC \cite{5488831}) & 1.15x & 6.3x \\ \hline
Redundant Node (SEUT \cite{5488831}) & 2x & 37x \\ \hline
Triple Module Redundancy (TMR \cite{tmr}) & 3.5x & 1,000,000x \\ \hline
\end{tabular}%
%}
\end{table}
Figure~\ref{fig:select} shows the storage area overhead versus FIT reduction for CEF-aware error mitigation for A1, A2, A3$_\text{scaled}$, and A4. 
SIL~1-SIL~4 markers are shown for FIT achievement for the Jaco2 robot and different safety standard levels. 

In both A1 and A2, on-chip storage elements consume $\sim50\%/\sim30\%$ of the total area/power of the CDM.
Thus, applying TMR ($3.5\times$ overhead) on all the bits of the CDM incurs $125\%/75\%$ area/power overheads. 
In comparison, for A1, SIL~3 and SIL~4 can be achieved using CEF-aware hybrid latch hardening at $3\%/1.8\%$ and $52.5\%/31.5\%$ area/power overheads, respectively. %, using CEF-aware hybrid latch hardening. 
Similarly, for A2, SIL~2, SIL~3, and SIL~4 can be achieved at {$0.4\%/0.24\%$}, {$18\%/10.8\%$}, and {$76\%/45.6\%$} area/power overheads. %, respectively, using CEF-aware hybrid latch hardening. 

For A3$_\text{scaled}$, the SRAM contributes to $\sim40\%/\sim30\%$ of the total area/power of the CDM, and the complete ECC results in a $12\%/9\%$ area/power overhead. 
However, with CEF-aware selection, SIL~1, SIL~2, SIL~3, and SIL~4 incur $0.004\%/0.003\%$,  $1\%/0.75\%$, $4.4\%/3.3\%$, and $10.5\%/7.62\%$ area/power overheads, respectively. 
For A4, the storage elements of the DRAM contributes to $\sim98\%$ of the total area/power of the CDM, and the blanket application of ECC results in a $12\%$ area/power overhead. 
In comparison, CEF-aware selection incurs $0.6\%$ area/power overheads for SIL~4, which is significantly lower than uniform application. 
\section{Architectural Implications} 
\label{sec:arch}
In this section, we compare the reliabilities of the four CDMs to draw lessons for resilience-aware CDM design. 
The CDM area is typically dominated by the sequential elements~\cite{Murray, Lian2018}, which are more prone to soft errors compared to combinational elements~\cite{5173251,6338321, 1028924 }. 
Error mitigation can incur significant performance, area, and energy overheads (e.g., $3.5\times$ area and energy for TMR), depending upon the CDM architecture. 
Hence, we need to consider the overheads of error mitigation for making architectural decisions. 

{Figure~\ref{fig:fit_calc_new}} compares the FIT rates of accelerators A1, A2, A3, $\text{A3}_{\text{scaled}}$, and A4. 
Figure~\ref{fig:overall} compares the error mitigation overhead for these accelerators for different FIT rates. 
Even though $\text{A3}_{\text{scaled}}$ has the highest FIT rate (Figure~\ref{fig:fit_calc_new}), ECC can be used for low overhead error mitigation in $\text{A3}_{\text{scaled}}$ as it uses SRAM. 
%As a result, it exhibits a steeper area versus FIT rate curve (Figure~\ref{fig:overall}),  and hence consumes the least area below the FIT rate of $\sim100$. 
%We further discuss the design parameters and their effects on this trend.
\newdimen\arch
\settoheight{\arch}{%
\includegraphics[width=0.23\textwidth]{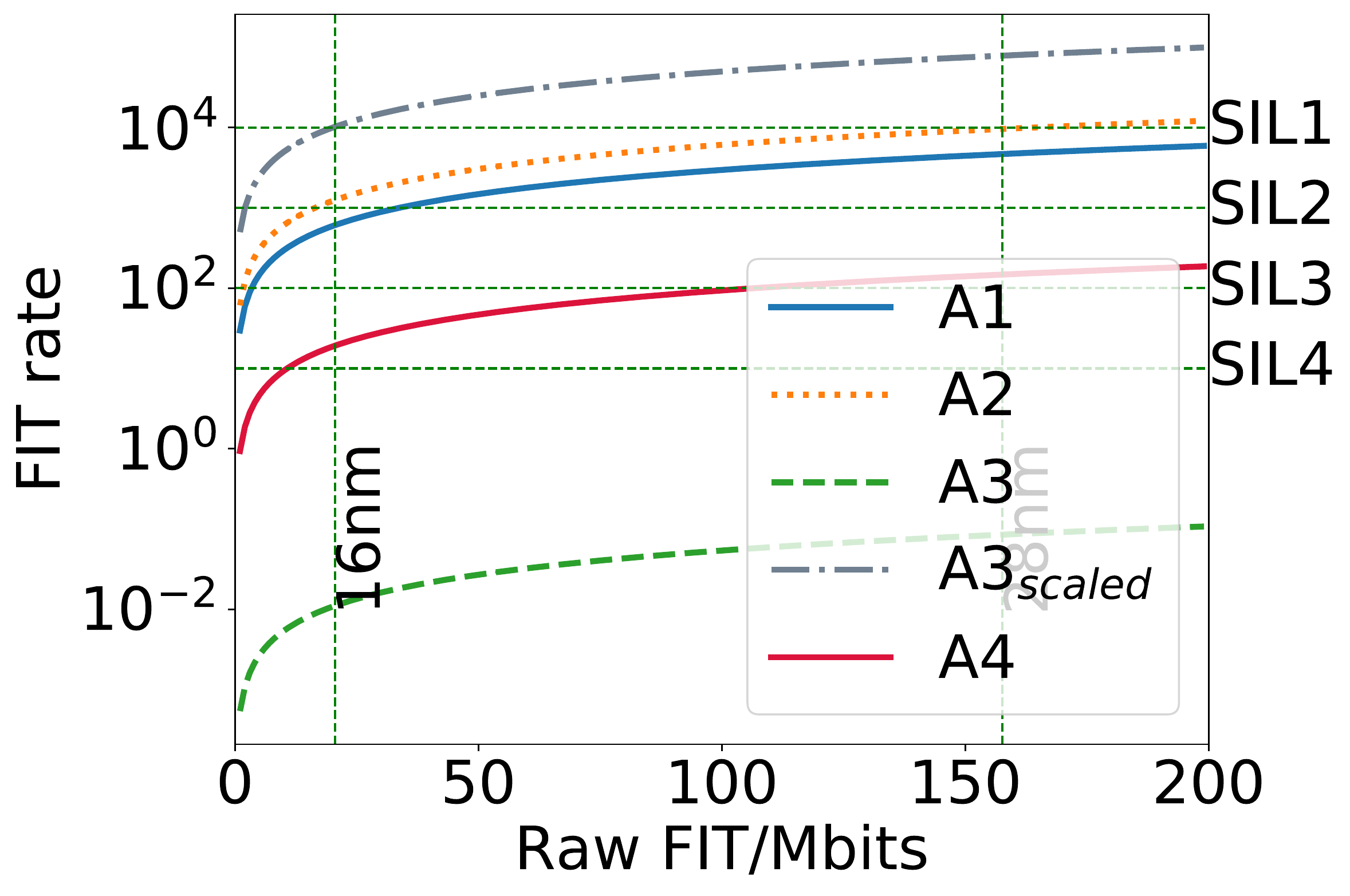}} 
\begin{figure*}[t]
    \centering
    \begin{subfigure}[t]{0.23\textwidth}
     \centering
    \includegraphics[height=\arch]{figures/FIT_rate.pdf}
    \caption{Calculated FIT rate}
    \label{fig:fit_calc_new}
    \end{subfigure}
     \hfill
        \begin{subfigure}[t]{0.23\textwidth}
      \centering
    \includegraphics[height=\arch]{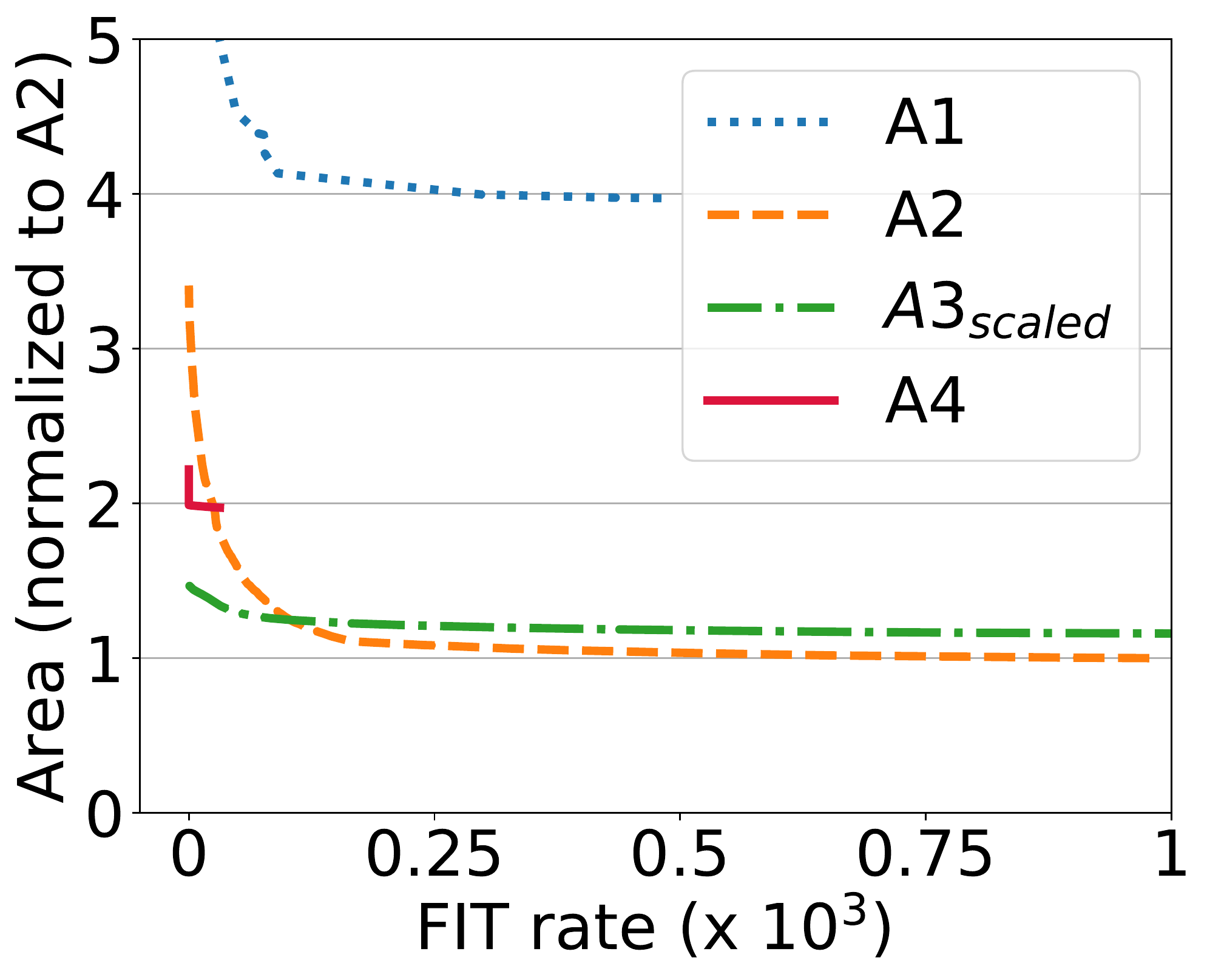}
    \caption{Area versus FIT rate}
    \label{fig:overall}
    \end{subfigure}
     \hfill
      \begin{subfigure}[t]{0.23\textwidth}
        \centering
      \includegraphics[height=\arch]{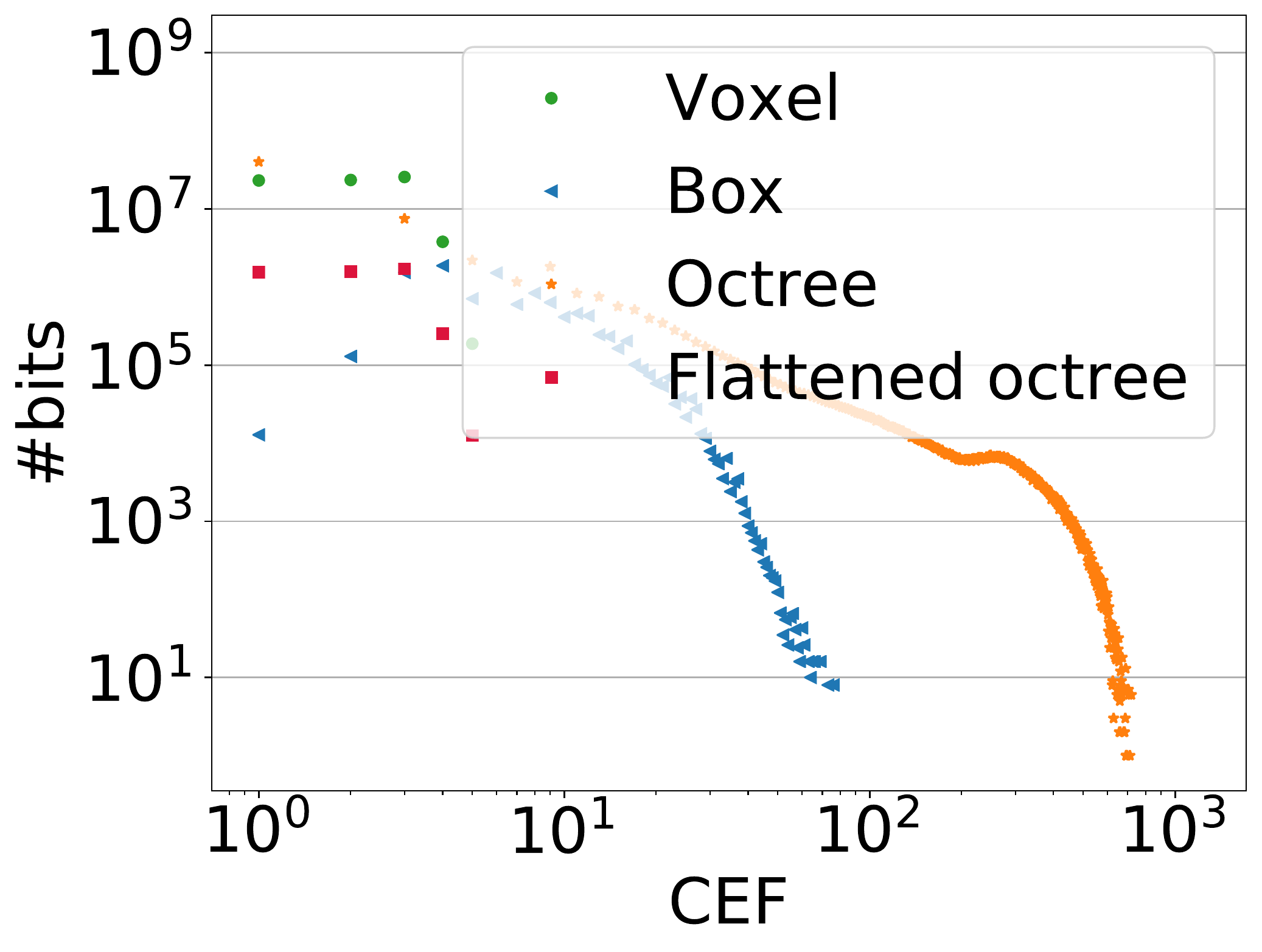}
      \caption{CEF distribution}
      \label{fig:geomcef}
      \end{subfigure}
       \hfill
          \begin{subfigure}[t]{0.23\textwidth}
        \centering
      \includegraphics[height=\arch]{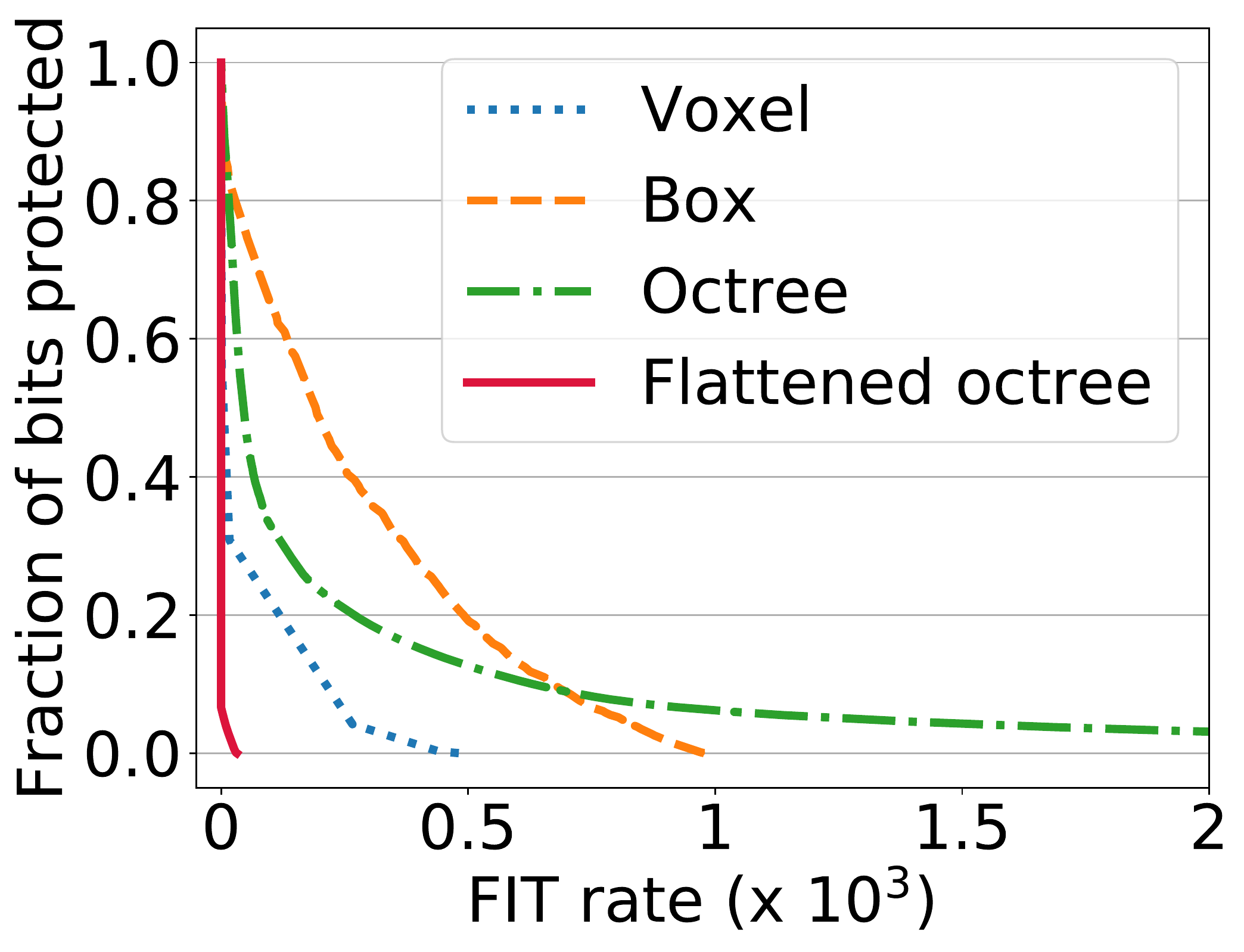}
      \caption{Fraction of protected bits vs. FIT rate}
      \label{fig:geom}
      \end{subfigure}
       \hfill
       \caption{Comparison of different CDM architectures geometric representations.}
  \end{figure*} 

\textbf{Geometric representation of swept space: }
%The geometric representation is a crucial design parameter in a CDM as it directly affects the datapath design, storage requirement, and collision detection latency~\cite{Ericson}. 
%The CDM architectures studied in this work cover four geometric representations: Voxel, Box, Octree, and Flattened octree. 
%All geometric modeling methods exhibit different storage requirements and collision detection logic. 
{Figure~\ref{fig:geomcef}} compares the CEF distribution of different geometric representations used in the CDMs studied for a motion set of the Jaco2 robot. 

We find that two aspects of the geometric representation mainly affect the CEF distribution and range: (1) redundancy, and the (2) volume covered by a structure.
In the box-based representation, to take advantage of spatial locality, the optimization process converts the swept space into a set of boxes. Each box covers the maximum possible number of voxels in the swept space. 
This adds redundancy as some voxels are covered by multiple boxes, and are hence not included in critical space. %
In the octree-based representation, nodes near the root node divide the space at a coarser level and represent a much larger volume. 
 %A fault in these nodes can result in high CEF values. %, as seen in A3 (Section~\ref{sec:part1}). 
Also, there is no redundancy in the representation. 
These factors result in an overall higher CEF for octree. %\karthik{Please check}
Conversely, in voxel- and flattened-octree-based representation, each structure/bit represents a single voxel, and so their CEF is less than $6$ (maximum $6$ surfaces of a voxel). 
%Similarly, for flattened-octree, more than $99\%$ bits represent empty voxels and have CEF equal to 0. 
%compared to the box and voxel representations. 
%The CEF range of bits directly affects the FIT rate of accelerators, as shown in Figure~\ref{fig:fit_calc_new}.
% for all bits.
%The CEF of bits in the voxel-based representation results in a lower FIT rate for A1 than A2 and $\text{A3}_{\text{scaled}}$. Similarly, the lower CEF value for box-based representation results in lower FIT rate of A2 than $\text{A3}_{\text{scaled}}$ (Figure~\ref{fig:fit_calc_new}). 

%High asymmetry in the CEF distribution can be exploited to balance the FIT reduction versus error mitigation overhead. 
Further, geometric representation determines the suitable storage structures and the error mitigation approach.  
For example, using SRAM/DRAM banks and ECC error mitigation results in lower costs of error mitigation for stringent FIT rate requirements as shown in Figure~\ref{fig:geom} (A3$_\text{scaled}$ and A4). 
%Figure~\ref{fig:geom} plots the FIT reductions for protecting different fractions of bits for the three representations. 

\textbf{Data reuse: }
Many accelerators use on-chip buffers to store frequently used data, and exploit data reuse to reduce memory accesses. 
As a result, a soft error-induced bit flip persists until it is overwritten by reloading the swept space data as explained in Section~\ref{subsec:fitN}. 
%\karthik{We said this earlier, didn't we ? Point to the section.} \deval
Therefore, if the erroneous data is used for $N$ collision queries, the FIT rate increases by $N$ times (Equation~\ref{eq:2}).  
%For example, A1, A2, and A3\_scaled use a dedicated CDC for each edge and configure only at the start, which increases the FIT rate. 
However, frequent reloading of on-chip data increases DRAM accesses and incurs performance and energy overheads. Thus there is a trade-off between the overhead of error mitigation and DRAM accesses. 
\section{Related work}
\label{sec:rw}
% \textit{Fault characterization:}
% \red{Mukherjee et al. \cite{Mukherjee2003} defined the Architectural Vulnerability Factor (AVF) of a structure as the probability that a fault in this structure will result in incorrect output and proposed Architectural Correct Execution analysis (ACE-analysis) to approximate AVF for microprocessor structures.   
% Sridharan et al. \cite{4798243} proposed the Program Vulnerability Factor (PVF) to decouple the architectural fault-masking effect from the microarchitecture and quantify the vulnerability of a program. 
% Fang et al. further extended it in their work ePVF ~\cite{7579739} to consider only SDC-causing bits in ACE-analysis for CPUs. 
% The architectural or microarchitectural ACE-analysis is challenging to apply to accelerators. Also,  not all architecturally visible faults would lead to safety violations with equal probability in motion planning, and ACE-analysis in such cases would be very conservative. 
% In comparison, CEF gives a measure of exposed space due to a faulty bit and can be used to estimate the probability of safety violation from a  fault in the bit.
\textit{Fault Characterization: } Several FI tools focus on CPUs and GPUs~\cite{Tselonis2016, Li2018, Hari2017, 8809540, Bo2016, Nie2018}. 
{
    %Tselonis2016- GUFI
    %Li2018 - Trident
    %Hari2017- SASSIFI
    %8809540- gem5-Approxilyzer
    %Bo2016- GPU-Qin
    %Nie2018-  GPU thread fault site pruning
    TRIDENT~\cite{Li2018} uses compiler information to estimate SDC probability without performing FI for CPU. 
    gem5-Approxilyzer~\cite{8809540} uses gem5 simulator~\cite{gem5} for FI, and proposes a methodology for fault site pruning. 
    GPU FI tools~\cite{Bo2016,Nie2018} use GPGPU-Sim~\cite{4919648} for FI, and propose fault site pruning for GPU SIMT execution model. 
    SASSIFI~\cite{Hari2017} proposes an instrumentation-based FI tool for NVIDIA GPUs. 
%Trident uses compiler information to estimate SDC without any FI. SASSIFI is an instrumentation based FI framework for NVIDIA GPU. GPU-Qin uses GPGPU-Sim and thread level information for fault site pruning-Nie2018 also. GUFI doed not speed up- it is just a tool that injects fault at ptx or  sass level and compare. Gem5-approxilyzer is a tool with gem5, for fault site pruning it uses program speciifc analysis and prpose some simple heuristics that can not be applied directly to MPAs.
}
These tools are useful for fault characterization of general purpose applications, but they cannot be easily applied to robotics accelerators that use specialized microarchitectures and instruction sets. 
%%\red{However, these tools lack support for studying the resilience properties of accelerators. }
More recent papers have studied the resilience of DNN accelerators~\cite{Li2017, Reagan2018}, and autonomous vehicle systems~\cite{Banerjee2018, Jha2019, Jha2018}. We focus, instead, on robotics accelerators processing spatial information.
Due to the differences between the datapath and information processed by the DNN accelerators and motion planning accelerators, the fault propagation to the output is significantly different between them.
%\red{Trident uses compiler information to estimate SDC without any FI. SASSIFI is an instrumentation based FI framework for NVIDIA GPU. GPU-Qin uses GPGPU-Sim and thread level information for fault site pruning-Nie2018 also. GUFI doed not speed up- it is just a tool that injects fault at ptx or  sass level and compare. Gem5-approxilyzer is a tool with gem5, for fault site pruning it uses program speciifc analysis and prpose some simple heuristics that can not be applied directly to MPAs.   }\deval{ToDo: explain it properly}
%\karthik{in what way? Their architecture?} \deval{added}

Another body of work has proposed metrics to characterize the vulnerability of structures to faults~\cite{Mukherjee2003, 7579739, Sridharan2009, Sridharan2010}. Mukherjee et al. \cite{Mukherjee2003} defined the Architectural Vulnerability Factor (AVF), and proposed Architectural Correct Execution analysis (ACE-analysis) to approximate AVF for microprocessor structures~\cite{Mukherjee2003}. 
Sridharan et al. \cite{Sridharan2009} proposed the Program Vulnerability Factor (PVF) to decouple the program's fault-masking effect from that of the microarchitecture, and quantify the vulnerability of a program. 
Fang et al. extended this work in ePVF~\cite{7579739} to consider only SDC-causing bits. %\karthik{Can we also cite the PVF paper by Vilas Sridharan here ?}
%Sridharan et al. \cite{Sridharan2010} also proposed Hardware Vulnerability Factor (HVF) to study the vulnerability solely from the microarchitecture level. 
%
Although useful for CPUs and GPUs, architectural or microarchitectural ACE-analysis is challenging to apply to accelerators due to differences in the ISA and workload. 
For MPAs in particular, whether a bit is ACE heavily depends upon the position of obstacles, and requires multiple simulations to estimate. The CEF gives a measure of the fraction of runtime for which a bit is ACE without such simulations. 
%Further, not all architecturally visible faults will lead to safety violations in motion planning, and hence ACE-analysis would be overly conservative. %\karthik{Rephrased this sentence} 
%%In contrast, the CEF estimates the probability of safety violations due to faults.

\textit{Error Mitigation Techniques: } There has been significant work on selective error mitigation techniques for CPUs~\cite{Reis2005} and GPUs~\cite{Mittal2017, Palframan2014, Mahmoud2018}. Mittal et al.~\cite{Mittal2017} proposed compressing similar values in GPUs. Palframan et al.~\cite{Palframan2014} analyzed GPGPU applications and proposed architectural modifications to reduce the magnitude of errors. Unfortunately, these methods are difficult to apply to MPAs due to differences in the ISA and microarchitecture. 

Reis et al.~\cite{Reis2005} and Mahmoud et al.~\cite{Mahmoud2018} proposed software-level instruction replication to improve the resilience of CPUs and GPUs respectively. However, accelerators typically use complex custom instructions and perform more computations per instruction~\cite{eyeriss, Lian2018, Murray2016ori}, and hence software-level duplication would result in significant overheads. 
Li et al.~\cite{Li2018} examined the resilience properties of DNN accelerators and proposed a method to protect vulnerable bits.  Guan et al.~\cite{Guan2019} proposed leveraging application-specific data properties in CNNs to minimize the error correction overhead. In contrast, we focus on accelerators processing spatial information. These are very different from DNN/CNN accelerators. 
%, and propose a novel reliability metric for protecting them.  %\karthik{Can we say how our work differs from these 2 ?}
%Our work focuses on finding important bits in motion planning accelerators. %, where resilience techniques used in neural network accelerators perform poorly. 

 \textit{Motion Planning: } Many techniques have been proposed to accelerate motion planning on CPUs and GPUs~\cite{Bialkowski2011, Gayle, Atay2006, Shi2018}. However, these do not meet the energy and performance requirements of autonomous robots~\cite{Murray2016, Lian2018}.  %\karthik{Provide citation}\deval{added}
 %performance and energy consumption of these techniques are ill-suited for autonomous robots.  
ASIC-based and FPGA-based MPAs~\cite{Murray2016, sorin, Lian2018, daducd} meet real-time constraints, but they focus on performance and energy optimizations, rather than resilience. Other work~\cite{Sun2019, Christensen2008} has studied resilient motion planning under sensor and communication faults, but not soft errors. Note that many of these techniques use the term {\em roadmap} instead of {\em motion set} to denote the same idea. %\karthik{Why do we need to say this?} \deval{The term ``roadmap'' is widely used by the robotics community. We replaced it with ``motion set'' because the reviewers were confused about environment dependence due to the use of this term. But if someone from a robotics background reads the paper, they will understand immediately, but I feel  it would be better to mention it explicitly}
%Most research works on motion planning use the term {\em{roadmap}} for the motion set used in this work.

%%%%%%%%%%%%%%%%%%%%%%%%%%%%%%%%%%%%%%%%%%%%%%%%%%%%%%%%%%%%%%%%%%%%%%%%%%%%%%%%
%%%%%%%%%%%%%%%%%%%%%%%%%%%%%%%%%% Conclusion %%%%%%%%%%%%%%%%%%%%%%%%%%%%%%%%
%%%%%%%%%%%%%%%%%%%%%%%%%%%%%%%%%%%%%%%%%%%%%%%%%%%%%%%%%%%%%%%%%%%%%%%%%%%%%%%%
%\section{Discussion and Future Work}
%\label{sec:diss}
%%We have used single-bit fault model for our evaluation and characterization, which is in line with related work
 
\section{Conclusions and Future Work}
\label{sec:concl}

Motion planning is a critical task in autonomous robots, and motion planning accelerators (MPAs) have been proposed to speed it up significantly.
Collision detection is the most resource-consuming and safety-critical module in MPAs.
In this work, we propose a spatially-aware reliability metric (CEF) for MPAs, based on the exposed surface area of critical space.
We propose a CEF-aware mitigation strategy and Fault Injection (FI) method based on this metric.
We also find that CEF-aware error mitigation achieves significant FIT (Failures in Time) rate reduction, even while incurring low area and energy overheads. 
We find that CEF-aware FI results in  $\sim23,000\times$ speedup over exhaustive FI to identify the critical bits.   
Finally, we identify the architectural design parameters affecting the resilience and error mitigation overheads in MPAs. 

%We use a single bit fault model for our evaluation and characterization, which is consistent with most other papers studying the effects of soft errors \cite{Ayatolahi2013, 8809532, 8416519,  Bo2016}.
There are several possible directions for future work. 
First, while we focus on single bit flips, both the reliability metric CEF, and the proposed error mitigation and FI methods can be extended to a multi-bit fault model. 
In particular, Phase~1 of FI, CEF measurement needs to be modified to include multi-bit fault model, where CEF of a bit can be calculated as an average of CEF for different multi-bit fault combinations. %We believe that
%Phase~2  of CEF-aware FI or error mitigation does not need to be changed. 
Second, while we focused on MPAs, the underlying ideas can be applied to other robotics accelerators that process spatial information.
%to motivate resilience-aware design space exploration of accelerators.
%
%We also analyzed the effect of different design parameters such as geometric modeling,  data reuse, workloads, and data access pattern.  
Finally, our observations on the effect of the different design parameters on the resilience and error mitigation overhead open up the direction of ``resilience-aware'' algorithm-hardware co-design. 
\ifCLASSOPTIONcaptionsoff
  \newpage
\fi



%\bibliographystyle{IEEEtran}
\bibliographystyle{IEEEtranS}
\bibliography{ref}


%\end{IEEEbiography}

%%\begin{IEEEbiographynophoto}{John Doe}
%%Biography text here.
%%\end{IEEEbiographynophoto}

%\newpage

%%\begin{IEEEbiographynophoto}{Jane Doe}
%%Biography text here.
%%\end{IEEEbiographynophoto}


%\vfill


\end{document}